%% file: DNS_ZPG_TBL.tex
\documentclass[lineno]{jfm}

\usepackage{graphicx}
\usepackage{newtxtext}
\usepackage{newtxmath}
\usepackage{natbib}
\usepackage{hyperref}
\hypersetup{
    colorlinks = true,
    urlcolor   = blue,
    citecolor  = black,
}

\newcommand{\RomanNumeralCaps}[1]
\linenumbers

\usepackage{mathtools}
\usepackage[dvipsnames]{xcolor}

\definecolor{colorA}{rgb}{0.12156862745098039,0.4666666666666667,0.7058823529411765}
\definecolor{colorB}{rgb}{1.0,0.4980392156862745,0.054901960784313725}
\definecolor{colorC}{rgb}{0.17254901960784313,0.6274509803921569,0.17254901960784313}
\definecolor{colorD}{rgb}{0.8392156862745098,0.15294117647058825,0.1568627450980392}
\definecolor{colorE}{rgb}{0.5803921568627451,0.403921568627451,0.7411764705882353}
\definecolor{colorF}{rgb}{0.5490196078431373,0.33725490196078434,0.29411764705882354}
\definecolor{colorG}{rgb}{0.5019607843137255,0.5019607843137255,0.5019607843137255}
\definecolor{colorH}{rgb}{0.0,0.0,0.0}

\usepackage{tikz}
\usepackage{pgfplots}
\newcommand{\plotA}{\raisebox{2pt}{\tikz{\draw[colorA,solid,line width=0.9pt](0,0) -- (5mm,0);}}}
\newcommand{\plotB}{\raisebox{2pt}{\tikz{\draw[colorB,solid,line width=0.9pt](0,0) -- (5mm,0);}}}
\newcommand{\plotC}{\raisebox{2pt}{\tikz{\draw[colorC,solid,line width=0.9pt](0,0) -- (5mm,0);}}}
\newcommand{\plotD}{\raisebox{2pt}{\tikz{\draw[colorD,solid,line width=0.9pt](0,0) -- (5mm,0);}}}
\newcommand{\plotE}{\raisebox{2pt}{\tikz{\draw[colorE,solid,line width=0.9pt](0,0) -- (5mm,0);}}}
\newcommand{\plotF}{\raisebox{2pt}{\tikz{\draw[colorF,solid,line width=0.9pt](0,0) -- (5mm,0);}}}
\newcommand{\plotgrey}{\raisebox{2pt}{\tikz{\draw[colorG,solid,line width=0.9pt](0,0) -- (5mm,0);}}}

\newcommand{\plotAdashed}{\raisebox{2pt}{\tikz{\draw[colorA,dashed,line width=0.9pt](0,0) -- (5mm,0);}}}

\newcommand{\plotAdotted}{\raisebox{2pt}{\tikz{\draw[colorA,dotted,line width=0.9pt](0,0) -- (5mm,0);}}}
\newcommand{\plotBdotted}{\raisebox{2pt}{\tikz{\draw[colorB,dotted,line width=0.9pt](0,0) -- (5mm,0);}}}
\newcommand{\plotCdotted}{\raisebox{2pt}{\tikz{\draw[colorC,dotted,line width=0.9pt](0,0) -- (5mm,0);}}}
\newcommand{\plotDdotted}{\raisebox{2pt}{\tikz{\draw[colorD,dotted,line width=0.9pt](0,0) -- (5mm,0);}}}

\newcommand{\plotAdashdot}{\raisebox{2pt}{\tikz{\draw[colorA,dash pattern={on 8pt off 2pt on 1pt off 4pt},line width=0.9pt](0,0) -- (5mm,0);}}}

\newcommand{\plotblacksolid}{\raisebox{2pt}{\tikz{\draw[black,line width=0.9pt](0,0) -- (5mm,0);}}}
\newcommand{\plotblackdashed}{\raisebox{2pt}{\tikz{\draw[black,dashed,line width=0.9pt](0,0) -- (5mm,0);}}}
\newcommand{\plotblackdotted}{\raisebox{2pt}{\tikz{\draw[black,dotted,line width=0.9pt](0,0) -- (5mm,0);}}}
\newcommand{\plotblackdashdot}{\raisebox{2pt}{\tikz{\draw[black,dash pattern={on 8pt off 2pt on 1pt off 4pt},line width=0.9pt](0,0) -- (5mm,0);}}}
\newcommand{\plotblackmarker}{\tikz[baseline=-0.6ex,thick]\node[black,mark size=0.5ex]{\pgfuseplotmark{o}};}

\newcommand{\plotAmarker}{\tikz[baseline=-0.6ex,thick]\node[colorA,mark size=0.5ex]{\pgfuseplotmark{o}};}
\newcommand{\plotBmarker}{\tikz[baseline=-0.6ex,thick]\node[colorB,mark size=0.5ex]{\pgfuseplotmark{o}};}
\newcommand{\plotCmarker}{\tikz[baseline=-0.6ex,thick]\node[colorC,mark size=0.5ex]{\pgfuseplotmark{o}};}
\newcommand{\plotDmarker}{\tikz[baseline=-0.6ex,thick]\node[colorD,mark size=0.5ex]{\pgfuseplotmark{o}};}
\newcommand{\plotAmarkers}{\tikz[baseline=-0.6ex,thick]\node[colorA,mark size=0.5ex]{\pgfuseplotmark{square}};}
\newcommand{\plotkmarkersq}{\tikz[baseline=-0.6ex,thick]\node[colorH,mark size=0.5ex]{\pgfuseplotmark{square}};}
\newcommand{\plotkmarkercr}{\tikz[baseline=-0.6ex,thick]\node[colorH,mark size=0.5ex]{\pgfuseplotmark{x}};}
\newcommand{\plotAmarkerut}{\tikz[baseline=-0.6ex,thick]\node[colorA,mark size=0.5ex]{\pgfuseplotmark{triangle}};}
\newcommand{\plotAmarkerdt}{\tikz[baseline=-0.6ex,thick]\node[colorA,mark size=0.5ex,rotate=180]{\pgfuseplotmark{triangle}};}

\usepackage[normalem]{ulem}
\usepackage{color}

\title{Direct numerical simulation of a zero-pressure-gradient thermal turbulent boundary layer up to \boldmath{$Pr = 6$}}

\author{
Arivazhagan~G.~Balasubramanian\aff{1,2}\corresp{\email{argb@kth.se}},
Luca~Guastoni\aff{1,2},
Philipp~Schlatter\aff{1,2}, 
\and Ricardo~Vinuesa\aff{1,2}\corresp{\email{rvinuesa@mech.kth.se}}
}

\affiliation{
\aff{1}FLOW, Engineering Mechanics, KTH Royal Institute of Technology, Stockholm, Sweden
\aff{2}Swedish e-Science Research Centre (SeRC), Stockholm, Sweden
}

\begin{document}

\maketitle

\begin{abstract}
The objective of the present study is to provide a numerical database of thermal boundary layers and to contribute to the understanding of the dynamics of passive scalars at different Prandtl numbers. In this regard, a direct numerical simulation (DNS) of an incompressible zero-pressure-gradient turbulent boundary layer is performed with the Reynolds number based on momentum thickness $Re_{\theta}$ up to $1080$. Four passive scalars, characterized by the Prandtl numbers $Pr = 1,2,4,6$ are simulated with constant Dirichlet boundary conditions, using the pseudo-spectral code SIMSON~\citep{simson}. To the best of our knowledge, the present direct numerical simulation provides the thermal boundary layer with the highest Prandtl number available in the literature. It corresponds to that of water at $\sim$24$^{o}C$, when the fluid temperature is considered as a passive scalar. Turbulence statistics for the flow and thermal fields are computed and compared with available numerical simulations at similar Reynolds numbers. The mean flow and temperature profiles, root-mean squared (RMS) velocity and temperature fluctuations, turbulent heat flux, turbulent Prandtl number and higher-order statistics agree well with the numerical data reported in the literature. Furthermore, the pre-multiplied two-dimensional spectra of the velocity and of the passive scalars are computed, providing a quantitative description of the energy distribution at the different lengthscales for various wall-normal locations. The energy distribution of the heat flux fields at the wall is concentrated on longer temporal structures and exhibits different footprint at the wall, with increasing Prandtl number. 

\end{abstract}


\input{introduction}

\input{sec2_methodology}

\input{sec3_comparison}

\input{sec4_PSD}

\input{sec5_conclusions}

\input{Appendix}

\backsection[Acknowledgements]{The simulations were run with the computational resources provided by the Swedish National Infrastructure for Computing (SNIC). The authors of the various data bases used in this study are acknowledged for sharing the simulation data. All the data to reproduce the results obtained in the present study will be available on GitHub upon publication.}

\backsection[Funding]{This work is supported by the funding provided by the Swedish e-Science Research Centre (SeRC), ERC grant no.~"2021-CoG-101043998, DEEPCONTROL" to RV and the Knut and Alice Wallenberg (KAW) Foundation.}

\backsection[Declaration of interests]{The authors report no conflict of interest.}

\bibliographystyle{jfm}
\bibliography{literature}

\end{document}

%% file: introduction.tex
\section{Introduction}
Wall-bounded turbulence is a phenomenon of huge technological importance in many industrial and environmental applications. The study of turbulent flow in complex geometries is quite challenging both from the numerical and experimental points of view, but it is relevant for various practical applications. For this reason, simpler geometries are chosen when the fundamental physics of the flow is studied. One canonical flow case widely used in the literature is the boundary layer developing on a flat surface~\citep{schlatter09,sillero}. The spatially-evolving fully-turbulent boundary layer has been studied using different experimental techniques~\citep{osterlund,rah,shehzad} with the researchers constantly improving the measurement techniques~\citep{orlu01,bailey,vinuesaexp} to obtain reliable measurements at high Reynolds number~\citep{samie}. 
At the same time,  direct numerical investigations of turbulent boundary layer have been performed in several studies~\citep{spalart,ferrante,wu,simens,schlatter4300} implementing different solution methods for an increasing range of Reynolds number.
Since the experimental techniques have resolution limitations in the near-wall region of boundary-layer flow, direct numerical simulations (DNSs) have been helpful to study the relevant transport phenomena~\citep{araya12}. On the other hand, DNSs are limited to low-Reynolds-number flows owing to high computational cost. The direct numerical simulation of turbulent boundary layer is not only limited to canonical flows but also extended to more complex geometries like airfoils~\citep{vinuesaFTC}.

Many engineering applications involve heat and mass transfer, turbulent mixing, combustion, etc.~\citep{kozuka}. For this reason, scalar quantities like temperature also become important to be simulated. 
Understanding and predicting the dynamics of passive scalars like air and water pollutants play an important role in local and global environmental problems~\citep{kasagi99,lazpita}, as well as in the design of transport and energy systems~\citep{straub}.

Several experimental studies~\citep{kays72,perry76,simonich78,subramanian,krishnamoorthy} have analyzed different aspects of heat transfer of passive scalars in turbulent boundary layers. The investigation by~\cite{kays72} presented the variation of skin-friction coefficient and Stanton number in boundary layer over a transpiring wall, with different blowing and suction conditions for constant free-stream velocity condition. They have also discussed and proposed theoretical models to enable the prediction of heat transfer in a turbulent boundary layer. A zero-pressure gradient turbulent boundary layer with constant wall-temperature conditions was setup by~\cite{perry76}, which enabled them to test similarity relations between instantaneous heat and momentum fluxes.~\cite{simonich78} investigated the effects of the free-stream on heat transfer in a turbulent boundary layer and reported the increase in Stanton number with respect to free-stream turbulence.~\cite{subramanian} studied the effects of Reynolds number in a turbulent boundary layer and reported the constants in the logarithmic law for velocity and temperature to be independent of Reynolds number. Further,~\cite{krishnamoorthy} were able to measure the three components of average temperature dissipation very close to the wall in a turbulent boundary layer, in the effort to model turbulence for the temperature fields computation. This is only a small summary of the recent investigations on this topic, as there have been a continual investigation of the heat transfer behaviour both from engineering and numerical-modelling aspects.

In numerical investigations, the fluid temperature can be considered as a passive scalar, provided that the buoyancy effects and the temperature dependence of fluid properties are considered as negligible~\citep{araya12}. Many DNS studies of turbulent scalar transport have been performed to analyze the convective heat transfer between the fluid and solid walls in spatially-developing flows. ~\cite{ferziger} first performed the DNS for a turbulent thermal boundary layer. Later,~\cite{kong} performed a DNS at a Prandtl number of~$Pr=0.71$ with different boundary conditions including isothermal (Dirichlet) and isoflux (Neumann), for $Re_\theta$ ranging between 300 and 420 (note that $Re_{\theta}$ is the Reynolds number based on momentum thickness). The Reynolds-number range was extended in the studies by~\cite{hattori}, who simulated $Re_\theta$ from $1000$ to $1200$, at $Pr = 0.71$. At the same time, the numerical investigations for Prandtl numbers up to $Pr=2$ were performed by~\cite{tohdoh} for a relatively lower Reynolds-number range, up to $Re_\theta=420$. The effect of different boundary conditions at $Pr=0.2,\, 0.71$ and $2.0$ in the Reynolds-number range of $Re_\theta \in \left[350,830\right]$ was reported by~\cite{qiang}. Differently from thermal channel-flow simulations, which have been conducted at higher Prandtl numbers of 49 and low $Re$ by~\cite{schwertfirm} and at a $Pr$ of 10 and high Reynolds number by~\cite{alcantra21}, the thermal turbulent boundary layers have been only partially explored at a medium Prandtl number of 2 by~\cite{qiang} owing to the significant computational cost associated with higher $Pr$. In this study, we consider higher Prandtl numbers in a turbulent thermal boundary layer, reporting analysis that are currently not available in the literature, according to the authors' knowledge. Thereby, the passive scalars at $Pr = 1,2,4$ and $6$ are simulated for $Re_\theta$ up to 1070 in a zero-pressure gradient turbulent boundary layer using an isothermal wall boundary condition.

The details of the simulation setup are provided in~$\S$\ref{sec_2}. The statistics obtained from the simulation at different Prandtl numbers are compared with the data available in the literature for the fully-developed thermal turbulent boundary layer.  Since the thermal channel flow and thermal boundary layer exhibit a similar behaviour in the near-wall region, the statistical quantities of the channel flow reported in the literature are compared with the thermal boundary layer quantities at similar Reynolds number. Both comparisons are presented in~$\S$\ref{sec_3}.
In~$\S$\ref{sec_4} we analyze the distribution of energy in different scales for the wall-heat flux field and wall-parallel fields at $y^+ = 15, 30$ and $50$ (where the superscript `+' denotes scaling in terms of friction velocity $u_{\tau}$, see below). The premultiplied two-dimensional power-spectral density provide additional insight into the scalar transport at different Prandtl numbers. Finally, a short summary of the observations discussed in this work is reported in~$\S$\ref{sec_5}. 

%% file: sec2_methodology.tex
\section{Methodology}\label{sec_2}

\subsection{Governing equations}
A DNS of the zero-pressure-gradient (ZPG) turbulent boundary layer (TBL) is performed using the pseudo-spectral code SIMSON~\citep{simson}. The code solves the governing equations in  non-dimensional form (here, written in index notation), in particular the flow and scalar variables are non-dimensionalized as
\begin{align}
     \tilde{{x}_{i}} = \frac{x_{i}}{\delta_{0}^{*}}\,, \; \tilde{{U}_{i}} = \frac{U_{i}}{U_{\infty}}\,, \; \tilde{t} = \frac{tU_{\infty}}{\delta_{0}^{*}}\,, \; \tilde{P} = \frac{P}{\rho U_{\infty}^2}\,, \; \tilde{{\theta}_{i}} = \frac{\theta_{i}-\theta_{i,\infty}}{\theta_{i,w}-\theta_{i,\infty}}\,, \label{eqn_non_D}
\end{align}
where $\left(x_1,x_2,x_3\right) = \left(x,y,z\right)$ are the Cartesian coordinates in the streamwise, wall-normal and spanwise directions, respectively and $t$ denotes the time. The length scale used for the non-dimensionalization is the displacement thickness at $x=0$ and $t=0$, denoted by $\delta_{0}^{*}$. The corresponding instantaneous velocity components are denoted by $\left(U_1,U_2,U_3\right)=\left(U,V,W\right)$ with the mean quantities identified by $\left(\left<U\right>,\left<V\right>,\left<W\right>\right)$ and the fluctuations by $\left(u,v,w\right)$. Here $U_{\infty}$ is the undisturbed laminar free-stream velocity at $x=0$ and time $t=0$. The total pressure is denoted by $P$ and the density and kinematic viscosity of the fluid is represented by $\rho$ and $\nu$, respectively. In this study, four different passive scalars $\left(\theta_1,\theta_2,\theta_3,\theta_4\right)$ are simulated at different Prandtl numbers $\left(Pr=1,2,4,6\right)$, respectively. Here, $\theta_{i,\infty}, \theta_{i,w}$ correspond to the $i^{th}$ scalar concentration in the free-stream and at the wall, respectively with the mean quantities indicated by~$\left<\theta_i\right>$ and the corresponding fluctuations by~$\theta_i^{\prime}$. The superscript $\tilde{\cdot}$ introduced in equation~\ref{eqn_non_D} identifies a non-dimensional variable and it shall be dropped in the non-dimensional quantities for the rest of the sections for simplicity.

The non-dimensional form of the incompressible Navier--Stokes equation and the transport equation for passive scalars are given by
\begingroup
\allowdisplaybreaks
\begin{align}
    \label{nd_continuity} \frac{\partial U_{i}}{\partial x_{i}} &= 0\,,\\
    \label{nd_momentum} \frac{\partial U_{i}}{\partial t} + U_{j}\frac{\partial U_{i}}{\partial x_{j}} &= -\frac{\partial P}{\partial x_{i}} + \frac{1}{Re_{\delta_0^*}}\frac{\partial^2U_{i}}{\partial x_{j}\partial x_{j}} + F_{i}\,,\\
    \label{nd_scalar} \frac{\partial \theta_{i}}{\partial t} + U_{j}\frac{\partial \theta_{i}}{\partial x_{j}} &= \frac{1}{Re_{\delta_0^*}Pr}\frac{\partial^2\theta_{i}}{\partial x_{j}\partial x_{j}} + F_{\theta_i}\,,
\end{align}
\endgroup
where $Re_{\delta_0^*}$ identifies the Reynolds number based on the free-stream velocity $\left(U_{\infty}\right)$ and the displacement thickness at the inlet $\left(\delta_0^*\right)$.  The product of Reynolds number $\left(Re_{\delta_0^*}\right)$ and Prandtl number $\left(Pr\right)$ results in another non-dimensional number called P\'eclet number ($Pe =  Re_{\delta_0^*}Pr$), which measures the ratio between the scalar convective transport and the scalar molecular diffusion. Here, $F_i$ and $F_{\theta_i}$ correspond to the volume force terms for the velocity and passive scalars, respectively. The velocity-vorticity formulation of the incompressible Navier--Stokes equation is implemented in the solver as the divergence-free condition is implicitly satisfied by the formulation.

\subsection{Boundary conditions}
Having defined the governing equations, the problem definition is completed by providing appropriate boundary conditions. At the wall, the velocity of the fluid is the same as that of the solid surface and is given by the following no-slip and no-penetration boundary conditions
\begin{align}
    U\rvert_{y=0} = 0\,,\; V\rvert_{y=0} = 0\,, \; W\rvert_{y=0} = 0\,.
\end{align}

From the continuity equation, we also obtain
\begin{align}
    \left. \frac{\partial V}{\partial y}\right\rvert_{y=0} = 0\,.
\end{align}

The flow is assumed to extend to an infinite distance perpendicular to the plate, but discretizing an infinite domain is not feasible. Hence, a finite domain has to be considered, for which artificial boundary conditions have to be applied. A simple Dirichlet condition can be considered; however, the desired flow solution generally contains a disturbance that would be forced to zero. This would introduce an error due to the increased damping of the disturbances in the boundary layer~\citep{lundbladh}. An improvement to the aforementioned boundary condition can be made by using the Neumann boundary condition given by
\begin{equation}
    \left.\frac{\partial U_{i}}{\partial y}\right\rvert_{y=y_{L}} = \left.\frac{\partial \mathcal{U}_{i}}{\partial y}\right\rvert_{y=y_{L}}\,.
\end{equation}
where $y_{L}$ is the height of the solution domain in the wall-normal direction in physical space and $\mathcal{U}_{i}$ is the laminar base flow that is chosen as the Blasius flow for the present study. For the passive scalars an isothermal wall boundary condition is applied, as given by
\begin{equation}
    \theta_{i}\rvert_{y=0}= 0\,,
\end{equation}
which corresponds to a vanishing thermal-activity ratio $K$. The thermal-activity ratio defines the ratio between the fluid density, thermal conductivity and specific heat capacity and the same properties of the boundary surface as defined below
\begin{equation}
    K = \sqrt{\frac{\rho k C_{p}}{\rho_{w} k_{w} C_{p,w}}}\,.
\end{equation}
Here, $\rho_{w}$, $k_{w}$ and $C_{p,w}$ correspond to the density, thermal conductivity and specific heat capacity of the wall. The isothermal wall boundary condition corresponds to the fluid that exchanges heat with the boundary surface, without modifying the wall temperature.
The boundary condition in the free-stream is given by
\begin{equation}
    \theta_{i}\rvert_{y=y_{L}} = 1\,.
\end{equation}

\subsection{Numerical scheme}
The direct numerical simulation is performed with a pseudo-spectral method, where Fourier expansions are used in the streamwise and spanwise directions and Chebyshev polynomials $T_{k} (\xi)$ (on $-1\le\xi\ge1$) are used in the wall-normal direction employing the Chebyshev-tau method for faster convergence rates. 
The time advancement is performed using the second-order Crank--Nicholson scheme for linear terms and the third-order Runge--Kutta method for non-linear terms with a constant time step $\Delta t$. The non-linear terms are calculated in physical space and the aliasing errors in the evaluation of non-linear terms are removed by the 3/2 rule.

Since the TBL is developing in $x$, the periodic boundary condition cannot be directly used in this particular direction, which requires specific numerical treatment. In this regard, one approach is to impose an appropriate instantaneous velocity and scalar profile at the inlet for every time step. Assuming self-similarity of the flow in the streamwise direction, \cite{lund} proposed a rescaling-recycling method to generate the required inlet profiles based on the solution downstream. An alternative approach is the addition of the fringe region downstream of the physical domain to retain the periodicity in the streamwise direction as described by~\citep{bertolotti,nordstrom}. In this method, the disturbances are damped, and the flow is forced from the outflow of the physical domain to the same profile as the inflow. The fringe technique is used in the present study, as the inflow conditions from a laminar profile at followed by the tripping produce natural instantaneous fluctuations for the velocity and the scalar fields~\citep{araya12}. The fringe region is implemented by adding a volume force $\left(F_i; F_{\theta_i}\right)$ to the momentum and scalar transport equation~$\left(\ref{nd_momentum},\ref{nd_scalar}\right)$, respectively. The forcing term is given by
\begin{align}
    F_{i} = \lambda(x)\left(\mathcal{U}_{i} - U_{i}\right)\,,\\
    F_{\theta_i} = \lambda(x)\left(\Breve{\theta}_{i} - \theta_{i}\right)\,,
\end{align}
where $\lambda (x)$ is the strength of the forcing which is non-zero only in the fringe region. The flow field at the inlet is the laminar Blasius profile $\mathcal{U}_{i}$ and for the scalar $\theta_i$ it is the linear variation with $y$ ranging from 0 to 1, denoted by $\Breve{\theta}_i$.

\subsection{Computational domain and numerical setup}
The laminar base flow is tripped by a random volume force strip $\left(\text{at }x/\delta_0^* = 10\right)$ to trigger transition of the flow to a turbulent state. For this simulation, a three-dimensional cuboid is considered with length, height and width equal to $x_{L}, y_{L}, z_{L}$, respectively. The lower surface of the cuboid is considered as a flat plate with no-slip boundary conditions. The boundary layer grows in the considered computational domain with initial thickness denoted by $\delta_{0}^{*}$. In the streamwise direction, the computational domain terminates the fringe region. The vertical extent of the computational domain includes the whole boundary layer and the domain height is chosen based on the free-stream boundary condition. In particular, the problem setup in this work is similar to that studied by~\cite{qiang}.

The computational domain is discretized by $N_{x}, N_{y}$ and $N_{z}$ grid points in the streamwise, wall-normal and spanwise directions, respectively. The grid spacing is uniform in the streamwise and spanwise directions. For the wall-normal direction, the collocation points follow the Gauss--Lobatto (GL) distribution given by
\begin{align}
    y_{i} = \cos\left(\pi\frac{i}{N_{y}}\right) \quad i = 0,1,2...,N_{y}\,.
\end{align}
The computational box has a dimension of 1000$\,\delta_0^*\,\times\,$40$\,\delta_0^*\,\times\,$50$\,\delta_0^*$ in the streamwise, wall-normal and spanwise directions, respectively. The number of grid points in each direction corresponds to 3200$\,\times\,$385$\,\times\,$320. Considering the friction velocity $u_{\tau}$ at the middle of the computational domain $\left(x/\delta_0^*=500\text{, corresponding to }Re_\theta=794\right)$, the grid resolution in viscous units is $\Delta x^+=6.6$ and $\Delta z^+=3.3$ in the wall-parallel directions. In the wall-normal direction we have an irregular distribution 
of collocation points, hence $\Delta y^+$ varies between $0.01$ and $3.5$.
Note that the smallest scale in the scalar fluctuation is inversely proportional to $Pr^{1/2}$~\citep{lumley} and hence, the Batchelor length scale $\eta_{\theta_i}$ (which is analogous to the smallest scale in turbulent flow, Kolmogorov scale $\eta$) is estimated as $\eta Pr^{-1/2}$~\citep{kozuka}. Similarly, the ratio of largest to the smallest scales is proportional to $~Re^{1.5}Pr^{0.5}$ at high $Pr$~\citep{batchelor,lumley}. In this study, an adequate grid resolution is adopted to resolve all the physically-relevant scales. The Reynolds number based on free-stream velocity and displacement thickness at the inlet is $Re_{\delta_0^*} = 450$ and the friction Reynolds number based on local friction velocity $\left(u_\tau\right)$ and boundary-layer thickness $\left(\delta_{99}\right)$ is $Re_\tau = 46$. At outlet, the Reynolds number based on displacement thickness is $Re_{\delta_0^*} = 1,580$ and $Re_\tau = 396$.

In this study, five different realizations of ZPG TBL are performed by introducing different trip forcings through modification of the random seed parameter, to obtain an ensemble average of the statistical quantities. All the different realizations are run for about 2,400 time units $\left({\delta_{0}^{*}}/{U_{\infty}}\right)$ after one flow-through of initial transience, which corresponds to 1,000 time units $\left({\delta_{0}^{*}}/{U_{\infty}}\right)$. The converged statistics are obtained with the data corresponding to 12,000 time units $\left({\delta_{0}^{*}}/{U_{\infty}}\right)$.

%% file: sec3_comparison.tex
\section{Comparison with data in the literature}\label{sec_3}

\subsection{Mean velocity and scalar profiles}
\begin{figure}
\centering
\resizebox*{0.7\linewidth}{!}{\includegraphics{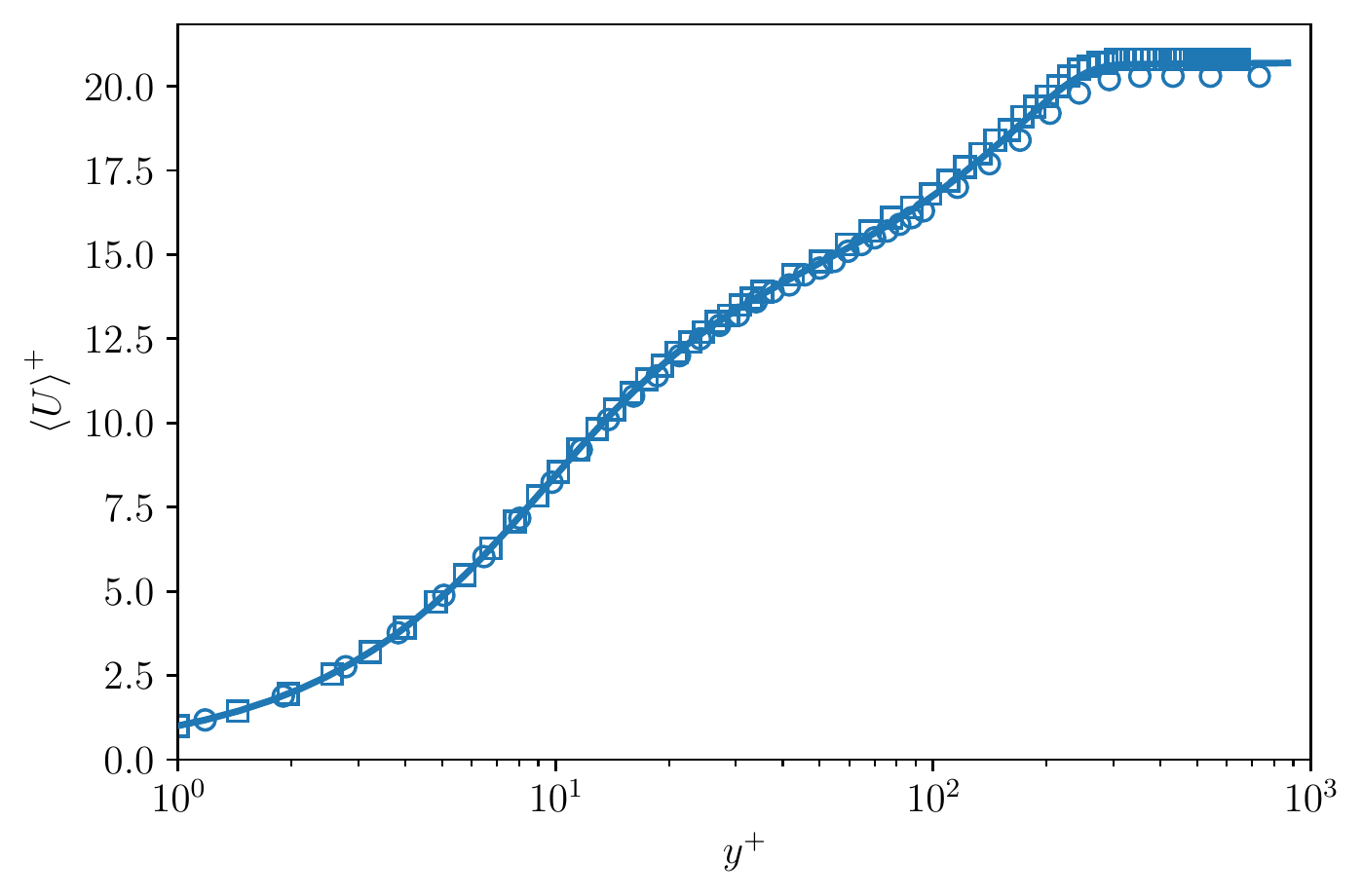}}
\caption[]{Inner-scaled mean streamwise velocity profile at $Re_{\theta} = 670$. {\plotA}~Present DNS, {\plotAmarker}~\cite{spalart}, {\plotAmarkers}~\cite{skote} at $Re_{\theta} = 666$.} 
\label{fig_mean_vel_profile_}
\end{figure}

\begin{figure}
\centering
\resizebox*{0.7\linewidth}{!}{\includegraphics{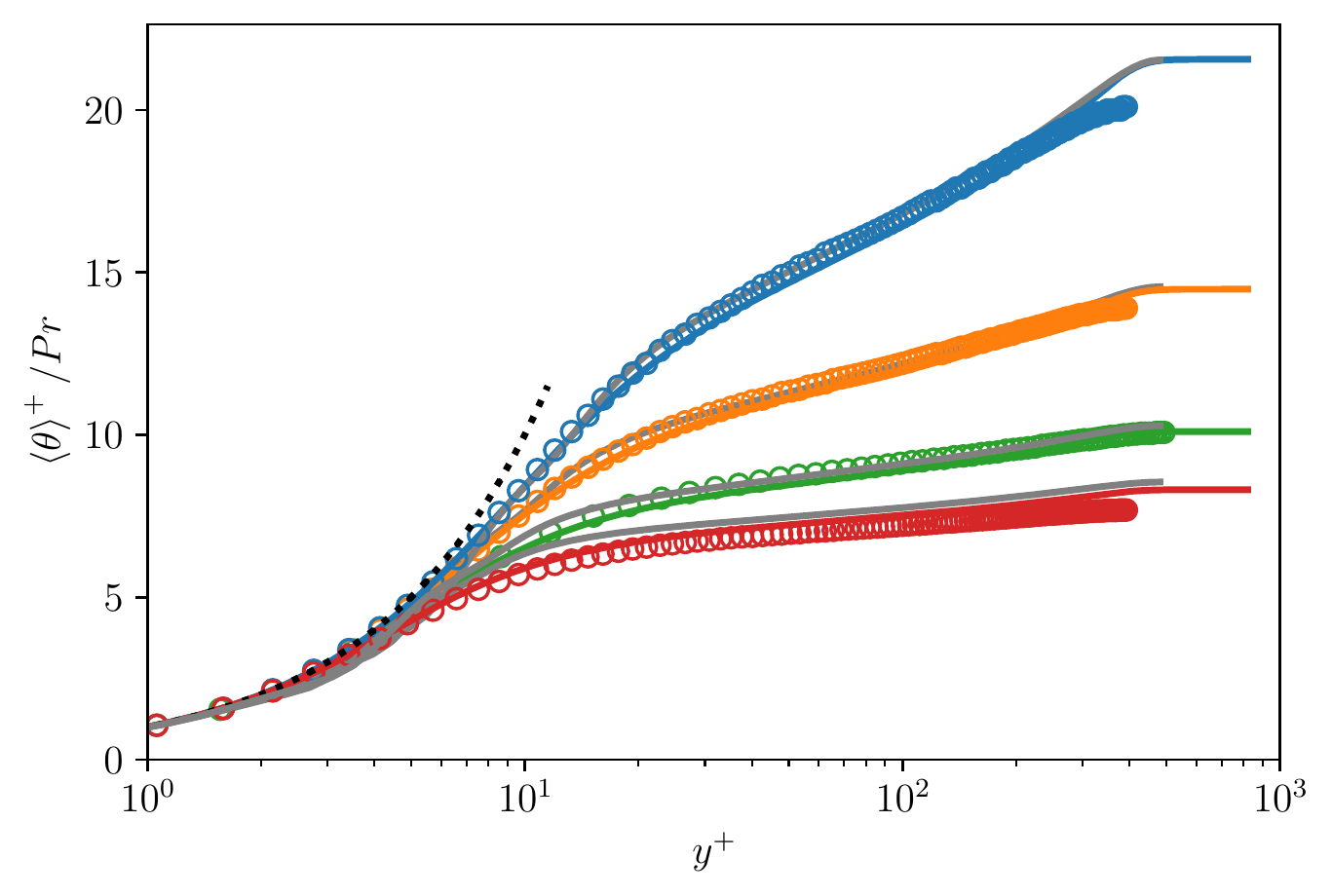}}
\caption[]{Normalized mean profiles of the passive scalars at $Re_{\theta} = 1070$ corresponding to $Re_{\tau} = 395$. {\plotA}~$Pr=1$, {\plotB}~$Pr=2$, {\plotC}~$Pr=4$, {\plotD}~$Pr=6$, channel DNS data by~\cite{kozuka} at $Re_{\tau} = 395$ and {\plotAmarker}~$Pr=1$, {\plotBmarker}~$Pr=2$, {\plotDmarker}~$Pr=7$, {\plotCmarker}~channel DNS data by~\cite{alcantra18} at $Re_{\tau} = 500$ and $Pr = 4$, with {\plotblackdotted}~conductive sub-layer relation, $\theta_{i}^{+}=Pr y^{+}$ and {\plotgrey}~fit by~\cite{kader81}. } 
\label{fig_mean_scalar_profile}
\end{figure}

The mean velocity profile obtained at the streamwise location corresponding to $Re_{\theta} = 670$ is shown in figure~\ref{fig_mean_vel_profile_}. The streamwise velocity profile is compared with the DNS data from~\cite{spalart} at $Re_{\theta} = 670$ and \cite{skote} at $Re_{\theta} = 666$. The comparison of the present data with the existing DNS results shows a good agreement in the inner region. There is a slight deviation of the mean velocity profile reported by~\cite{spalart} in the wake region with respect to the present data but it agrees well with the data provided by~\cite{skote}.
The mean profiles of the various passive scalars $\theta_{i}$ at different Prandtl numbers are normalized with the respective Prandtl numbers and with the friction scalar $\theta_{i,\tau}$ defined as
\begin{equation}
    \theta_{i,\tau} = \frac{q_{i,w}}{\rho C_{p} u_{\tau}}\,,
\end{equation}
where $C_{p}$ is the heat capacity of the fluid and $q_{i,w}$ is the rate of heat transfer from the wall to the fluid and is defined by
\begin{equation}
    q_{i,w} = \left.-k\frac{{\rm d}\left<\theta_{i}\right>}{{\rm d}y}\right\rvert_{y=0}\,.
    \label{fourier_hc}
\end{equation}
where $k$ is the thermal conductivity of the fluid. The normalized mean scalar profiles are plotted at $Re_{\theta} = 1070$, corresponding to $Re_{\tau} = 395$, as shown in figure~\ref{fig_mean_scalar_profile}. The mean scalar profiles follow the conductive sub-layer relation ($\theta_{i}^{+}=Pr y^{+}$) and is clearly identified in the plots for $y^{+} < 5$. The profiles of the passive scalars at $Pr=1,2$ are compared against the channel DNS data provided by~\cite{kozuka} at the same Prandtl numbers. The comparison shows a good agreement in the near-wall and overlap regions. It should be noted that the boundary condition used by~\cite{kozuka} is the uniform heat flux condition as opposed to the uniform temperature boundary condition applied in this study. Note that the study by~\cite{kawamura2000} concluded that the mean scalar profile should be different between the Dirichlet and Neumann boundary conditions based on low-Reynolds-number simulations and associated the difference with the differing scalar von K\'arm\'an coefficient $k_{\theta_i}$ as given in
\begin{equation}
    \frac{\left<\theta_{i}\right> - \left<\theta_{i,w}\right>}{\theta_{i,\tau}} = \frac{1}{k_{\theta_i}} \rm{log}\; y^+ + B_{\theta_i}\left(Pr\right) \,,
    \label{eqn_log_var}
\end{equation}
with  $B_{\theta_i}$ denoting the additive constant for the scalar $\theta_i$. However, based on higher-Reynolds-number simulations, \cite{pirozzoli} reported that the difference in boundary condition affects the mean passive scalar profiles only in small magnitudes in the overlap layer although the effect is evident in the scalar fluctuation profiles reported later.

The scalar profile at $Pr = 4$ is compared against the channel DNS data reported by~\cite{alcantra18}. The channel DNS data reported by~\cite{kozuka} at $Pr = 7$ is used for comparison of the passive scalar at $Pr = 6$, since there are no simulations reported in the literature at exactly the same Prandtl number. This gives us the opportunity to highlight the difference between the profiles at these high Prandtl numbers. Due to the difference in the considered Prandtl numbers, we observe a discrepancy in the mean velocity profile for $y^+ > 40$ in the overlap region. Since the channel data is used for the comparison, there is a difference observed near the wake region for all the cases. However, a good agreement of the profiles is observed for the inner region. Based on experimental data, semi-empirical fits were provided by~\cite{kader81} for a boundary layer with constant heat flux. In this study it was assumed that the overlap layer exhibited logarithmic variation as given in equation~(\ref{eqn_log_var}), and an empirical relation was provided to determine the additive constant $B_{\theta_i}$. The comparison of the mean scalar profiles against the relationships provided by~\cite{kader81} shows that the value at the wake is slightly overestimated with respect to the DNS data. The deviation for scalar $\theta_4$ corresponding to $Pr = 6$ is about 3\% for $y^+ \in \left[100,500\right]$. One possible reason for this small deviation could also be the constant-temperature boundary condition imposed in our simulations as opposed to the constant-heat-flux boundary conditions considered by~\cite{kader81}.

\subsection{Velocity and scalar fluctuations}
\begin{figure}
\centering
\resizebox*{0.45\linewidth}{!}{\includegraphics{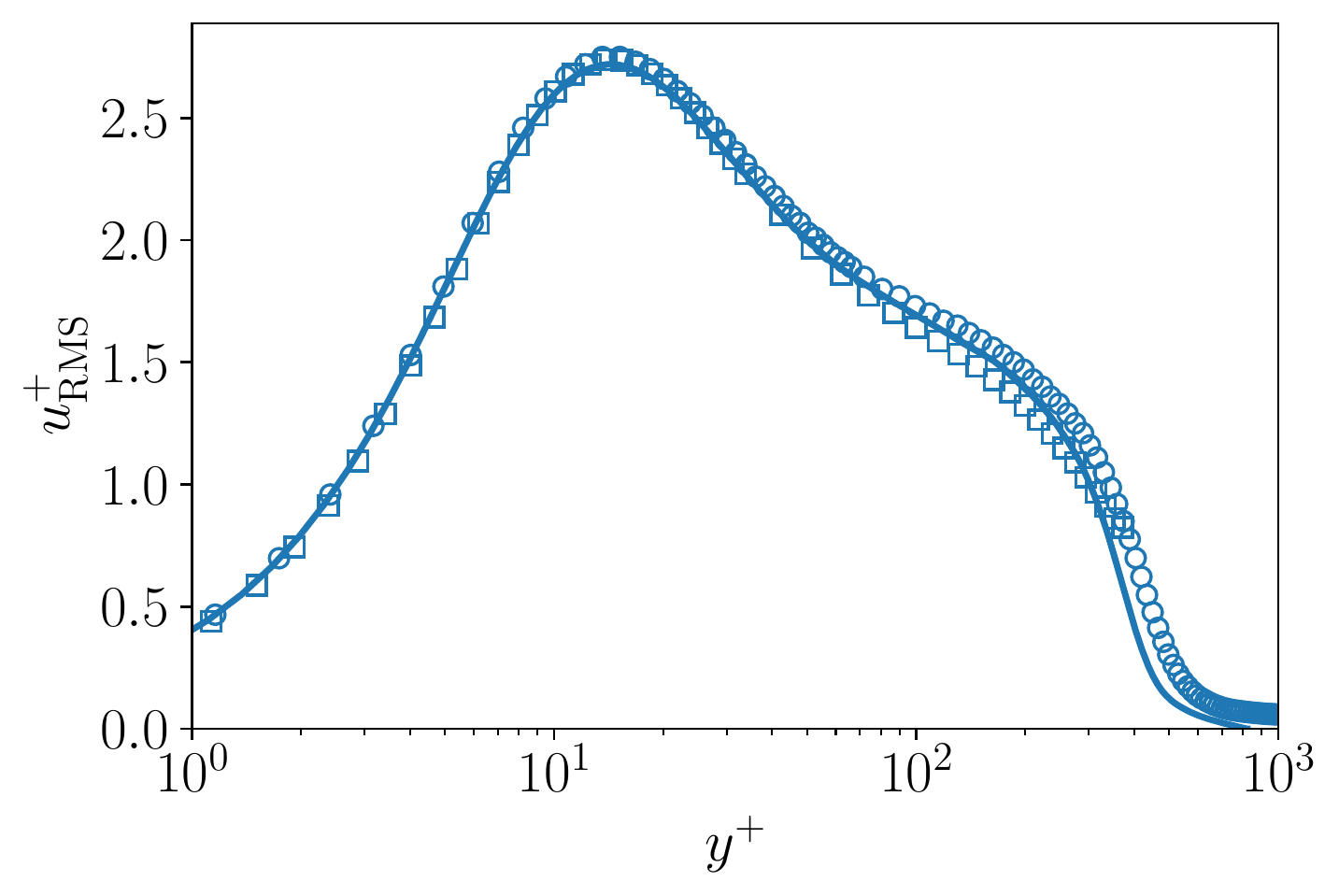}}
\resizebox*{0.45\linewidth}{!}{\includegraphics{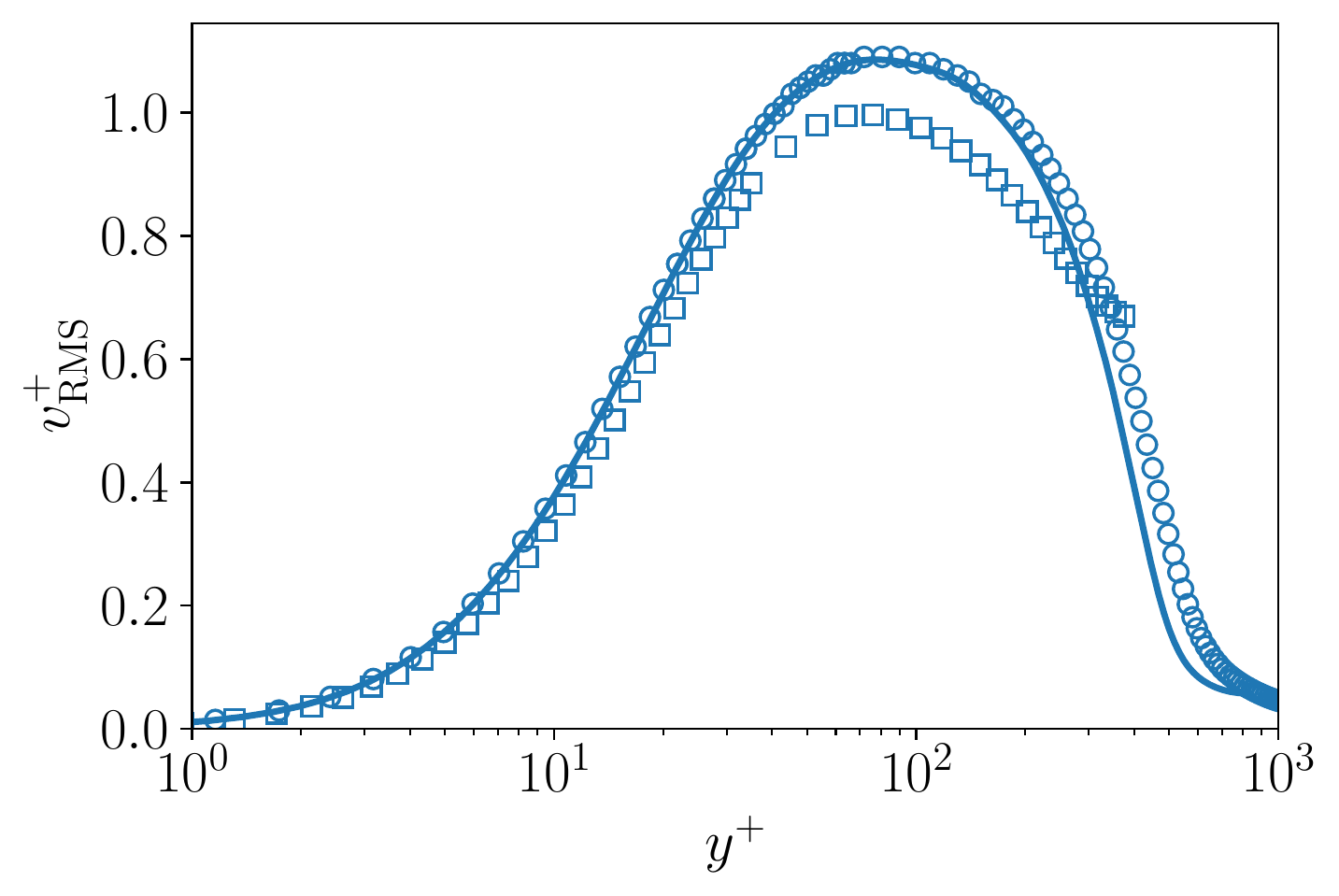}}
\resizebox*{0.45\linewidth}{!}{\includegraphics{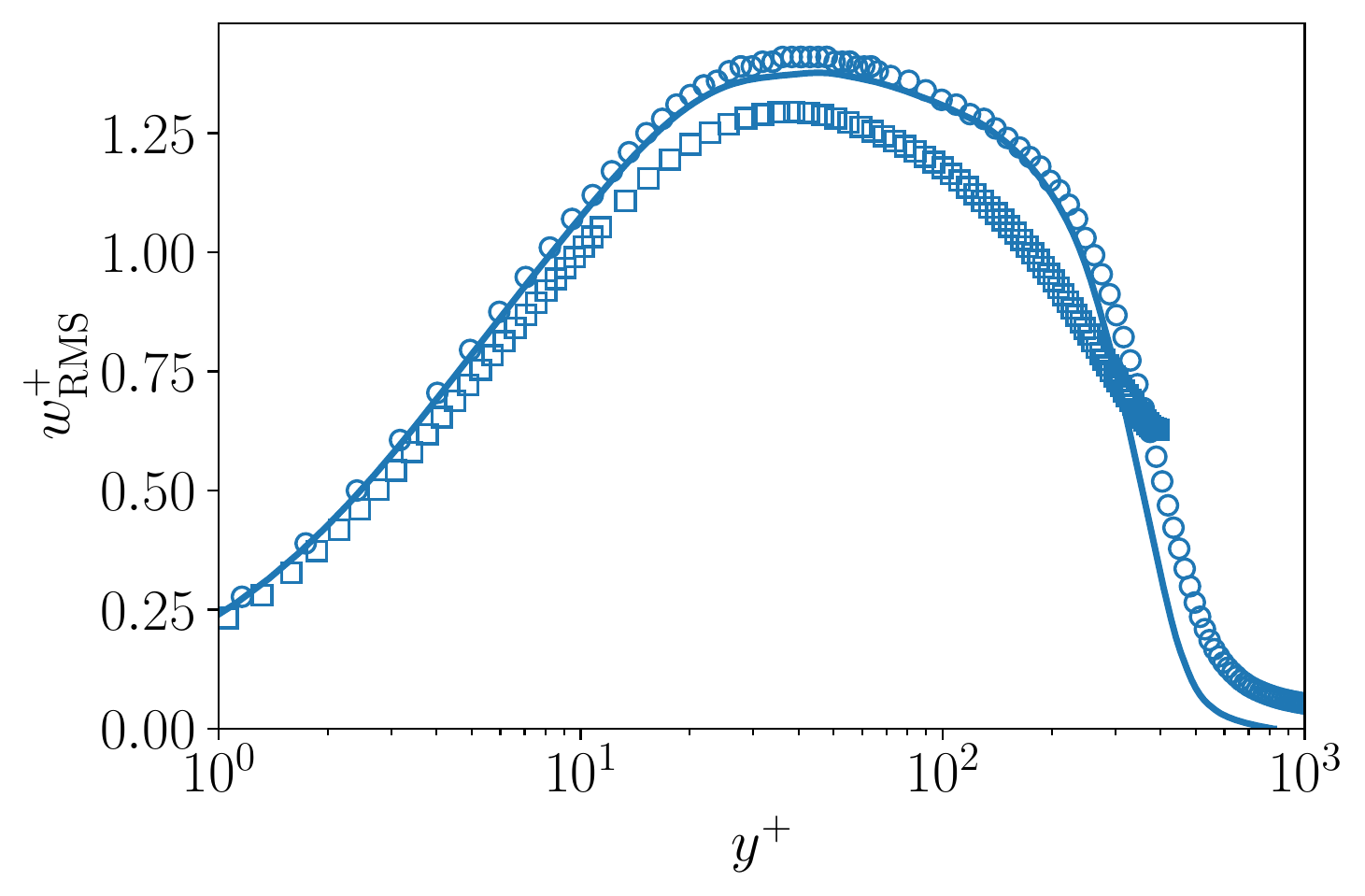}}
\resizebox*{0.45\linewidth}{!}{\includegraphics{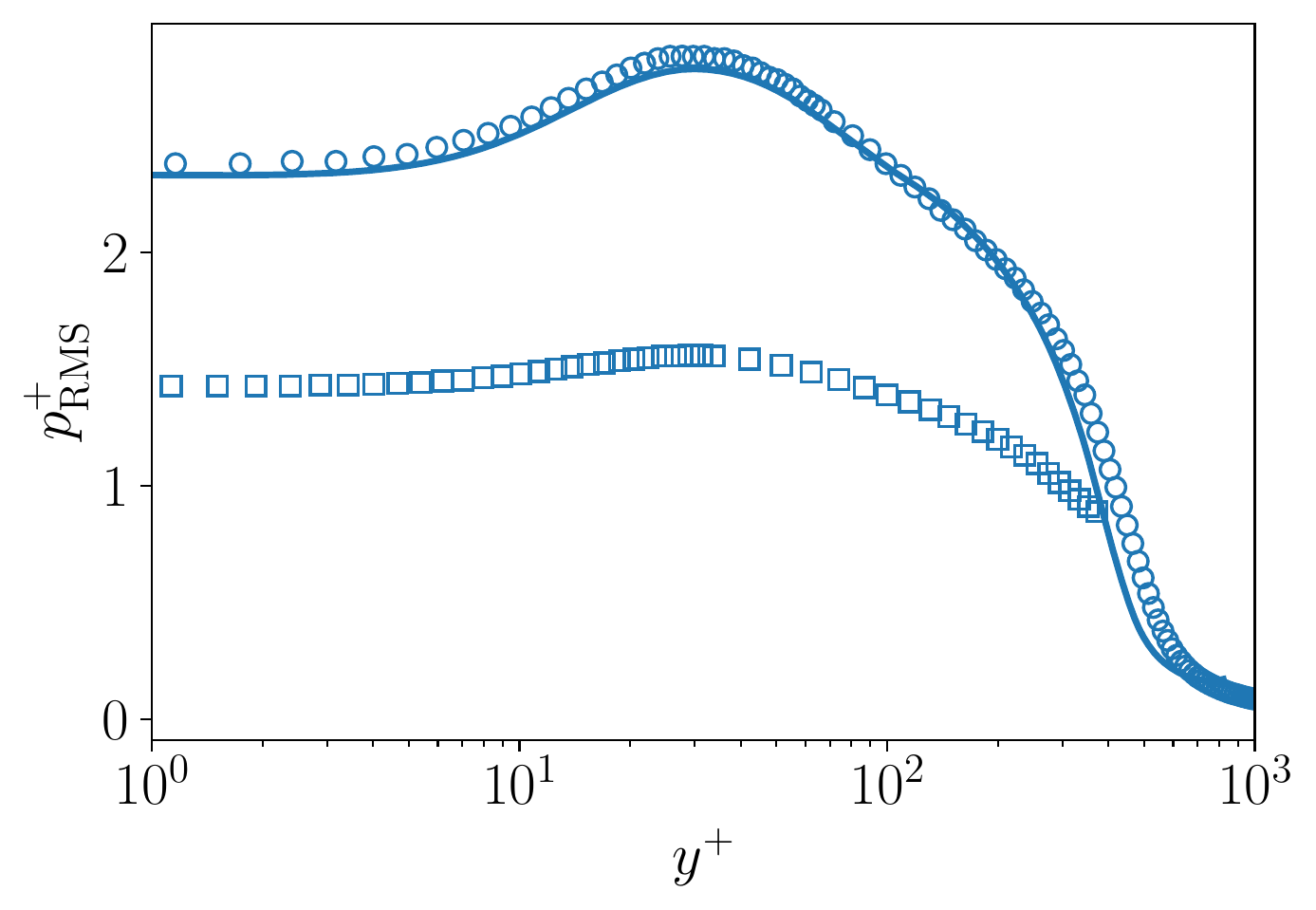}}
\caption[]{RMS velocity fluctuations and RMS pressure fluctuation at $Re_{\theta} = 1070$. {\plotA}~Present DNS, {\plotAmarker}~\cite{jiminez2010} at $Re_{\theta} = 1100$, {\plotAmarkers}~channel DNS data of~\cite{abe} at $Re_{\tau}=395$.} 
\label{fig_mean_vel_profile}
\end{figure}

As shown in figure~\ref{fig_mean_vel_profile}, the present velocity-fluctuation root-mean-squared (RMS) data shows a trend similar to that of the results by~\cite{jiminez2010}. The RMS profiles of the three velocity components are in good agreement in the inner region, while a minor difference can be observed in the outer region. Nonetheless, the peaks of the velocity fluctuations match in both position and magnitude. There is a slight offset in the plots of $p_{\rm RMS}$, which can be attributed to the small difference in the considered $Re_{\theta}$ for comparison. The RMS of the velocity components are also compared with the channel DNS data provided by~\cite{abe}. The streamwise RMS agrees well with the present DNS results and the near-wall  peak value coincides with the present observations. As expected, there is a difference observed in the outer region of flow, since channel and boundary layer flows are fundamentally different farther from the wall. Additionally, the RMS of  the pressure fluctuations observed in the boundary layer is different compared with the channel flow. From the present DNS data, the peak of the streamwise velocity fluctuation is found at $y^{+} = 14$, corresponding in outer units to $y/\delta_{99} = 0.035$.

\begin{figure}
\centering
\resizebox*{0.7\linewidth}{!}{\includegraphics{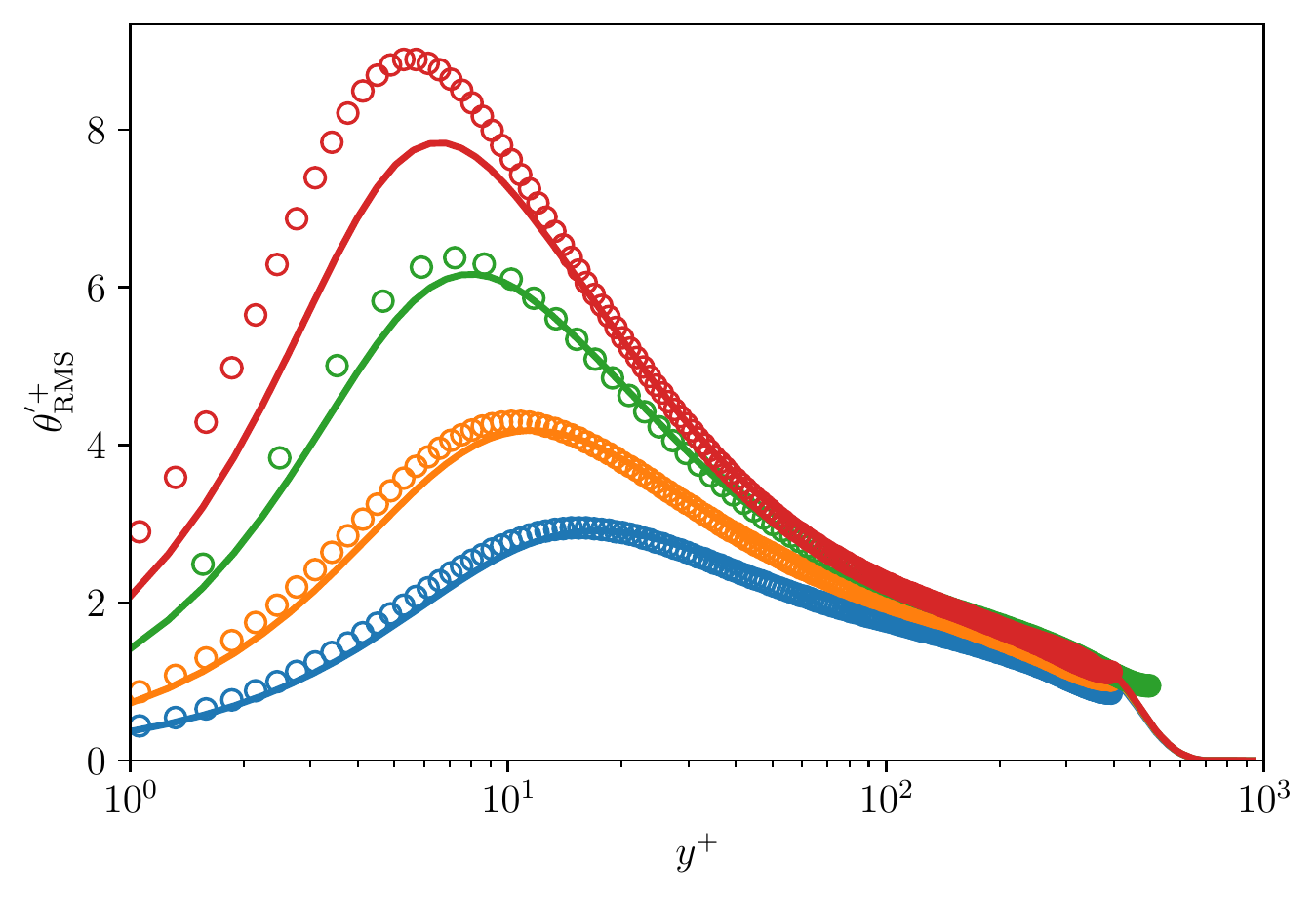}}
\caption[]{Scalar fluctuation RMS at $Re_{\theta} = 1070$. Present DNS data for {\plotA}~$Pr = 1$, {\plotB}~$Pr = 2$, {\plotC}~$Pr = 4$, {\plotD}~$Pr = 6$, channel DNS data by~\cite{kozuka} at $Re_{\tau} = 395$ and {\plotAmarker}~$Pr=1$, {\plotBmarker}~$Pr=2$, {\plotDmarker}~$Pr=7$, {\plotCmarker}~channel DNS data by~\cite{alcantra18} at $Re_{\tau} = 500$ and $Pr = 4$.} 
\label{fig_scalar_fluctuation}
\end{figure}

The RMS of the scalars at different Prandtl numbers are plotted in figure~\ref{fig_scalar_fluctuation}. The scalar RMS profile at $Pr = 1$ is similar to the streamwise velocity RMS and has a higher (roughly 5\%) near-wall peak comparatively, as expected. The comparison of scalar-fluctuation profiles at $Pr = 1,2$ with the channel DNS data from~\cite{kozuka} shows a good agreement in the inner and logarithmic region in addition to a good match of the peak value and wall-normal location. Despite the small difference in $Re_{\tau}$, the profiles at $Pr = 4$ as obtained by~\cite{alcantra18} shows a reasonably good agreement with the present results. With increasing Prandtl number, the peak value of the scalar fluctuation RMS increases and is located closer to the wall. The scalar fluctuations decay to zero at the wall due to the isothermal boundary condition and they also decay to zero outside the boundary layer due to the absence of disturbances in the free-stream.

\begin{figure}
\centering
\resizebox*{0.7\linewidth}{!}{\includegraphics{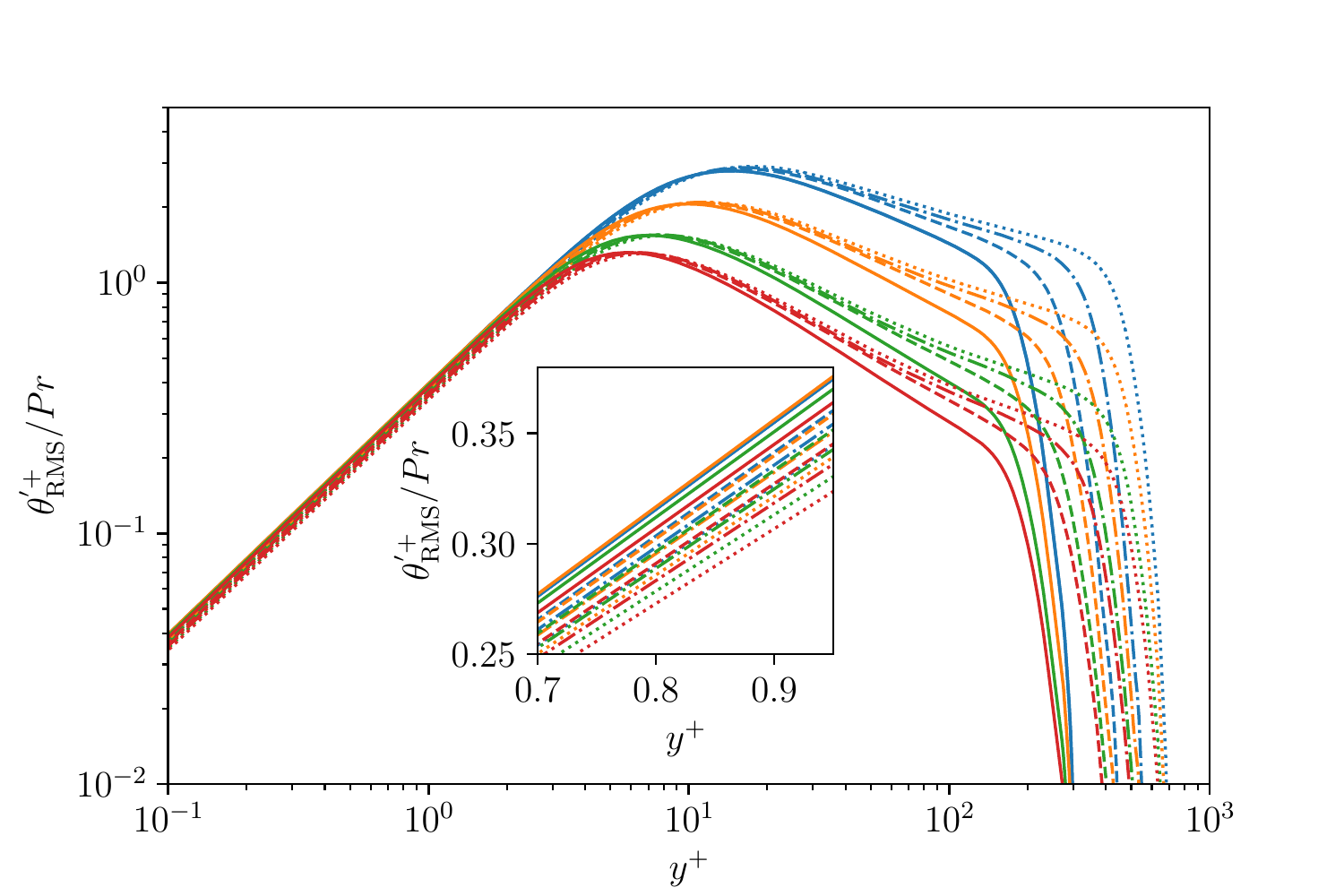}}
\caption[]{Scalar fluctuations scaled by Prandtl number at {\plotA}~$Pr = 1$, {\plotB}~$Pr = 2$, {\plotC}~$Pr = 4$, {\plotD}~$Pr = 6$ and {\plotblacksolid}~$Re_{\theta} = 420$, {\plotblackdashed}~$Re_{\theta} = 628$, {\plotblackdashdot}~$Re_{\theta} = 830$, {\plotblackdotted}~$Re_{\theta} = 1070$ at the corresponding Prandtl numbers.} 
\label{fig_scalar_fluctuation_scaled}
\end{figure}

\begin{figure}
\centering
\resizebox*{0.7\linewidth}{!}{\includegraphics{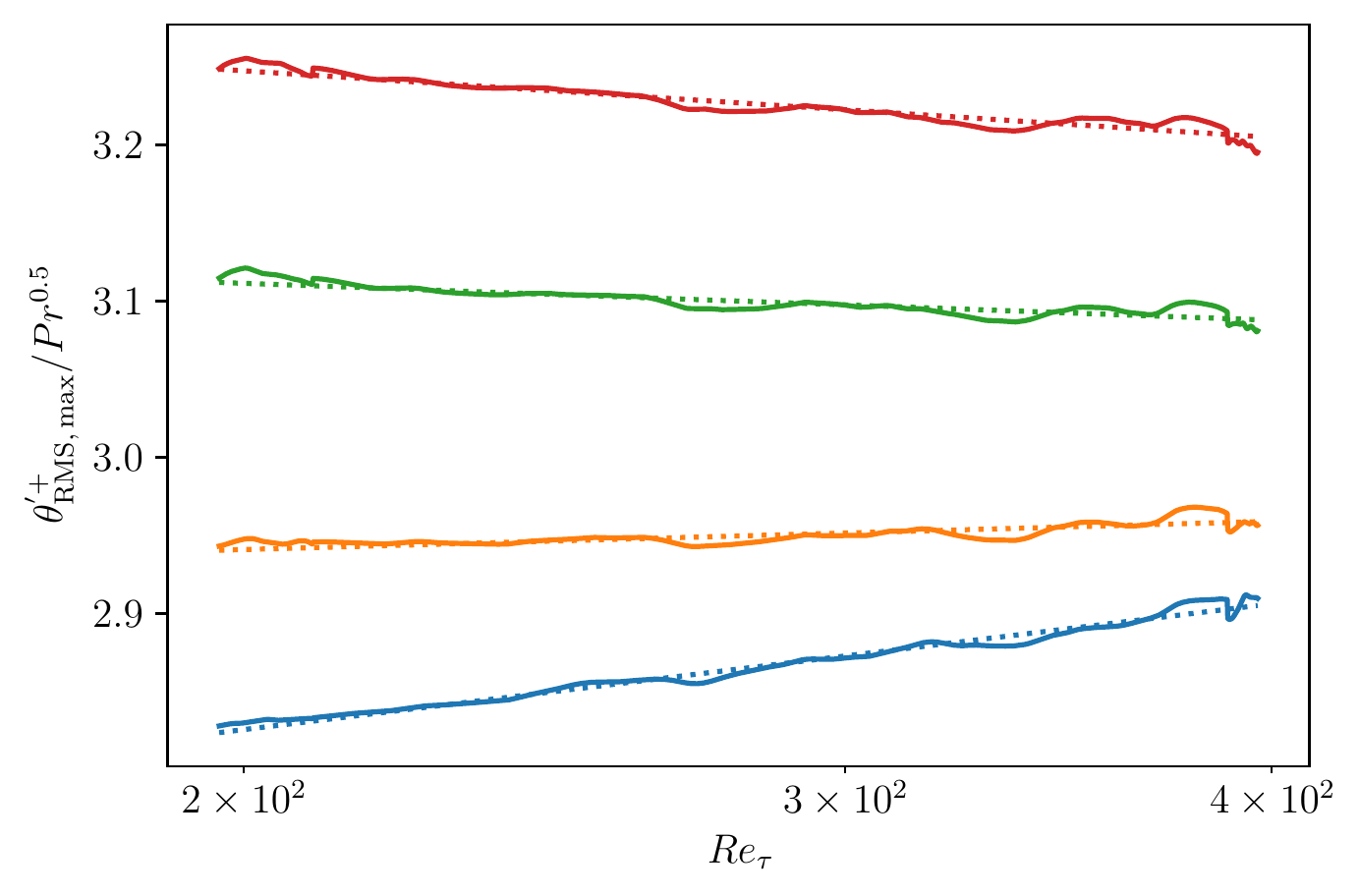}}
\caption[]{Variation of the peak of scalar fluctuation RMS with respect to $Re_{\tau}$, for {\plotA}~$Pr = 1$, {\plotB}~$Pr = 2$, {\plotC}~$Pr = 4$, {\plotD}~$Pr = 6$ including the corresponding empirical fits {\plotAdotted}~$0.116 \log{\left(Re_{\tau}\right)} + 2.210$ with $R^{2} = 0.976$, {\plotBdotted}~$0.037 \log{\left(Re_{\tau}\right)} + 3.963$ with $R^{2} = 0.620$, {\plotCdotted}~$-0.068 \log{\left(Re_{\tau}\right)} + 6.583$ with $R^{2} = 0.726$, {\plotDdotted}~$-0.151 \log{\left(Re_{\tau}\right)} + 8.753$ with $R^{2} = 0.894$.} 
\label{fig_scalar_fluctuation_correlation}
\end{figure}

\begin{figure}
\centering
\resizebox*{0.7\linewidth}{!}{\includegraphics{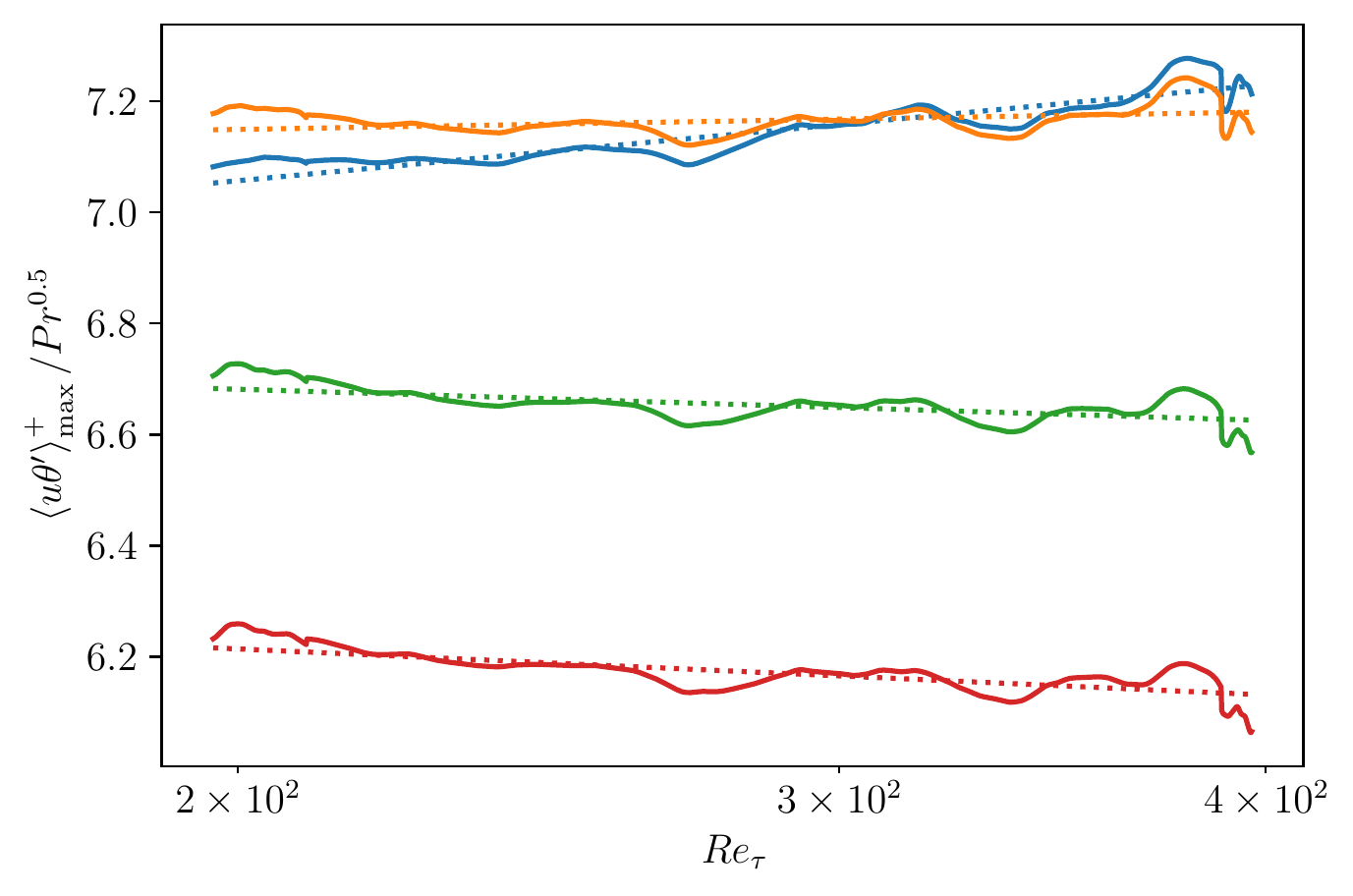}}
\caption[]{Variation of the peak of streamwise heat flux with respect to $Re_{\tau}$, for {\plotA}~$Pr = 1$, {\plotB}~$Pr = 2$, {\plotC}~$Pr = 4$, {\plotD}~$Pr = 6$ including the corresponding empirical fits {\plotAdotted}~$0.250 \log{\left(Re_{\tau}\right)} + 5.732$ with $R^{2} = 0.808$, {\plotBdotted}~$0.064 \log{\left(Re_{\tau}\right)} + 9.769$ with $R^{2} = 0.119$, {\plotCdotted}~$-0.1626 \log{\left(Re_{\tau}\right)} + 14.223$ with $R^{2} = 0.328$, {\plotDdotted}~$-0.293 \log{\left(Re_{\tau}\right)} + 16.773$ with $R^{2} = 0.513$.} 
\label{fig_heat_flux_correlation}
\end{figure}

The obtained scalar-fluctuation RMS profiles are scaled with the respective Prandtl numbers and is plotted in figure~\ref{fig_scalar_fluctuation_scaled}. We observe that the lines of $\theta^{\prime +}_{\rm RMS}$ for different scalars at different Reynolds numbers are parallel and not coinciding. Similar observations were also made by~\cite{alcantra21jfm} for a particular scalar at $Pr = 0.71$ and attributed the differences in the viscous-diffusion term at the wall to the increase in slope of $\theta^{\prime +}_{\rm RMS}$. The slope of $\theta^{\prime +}_{\rm RMS}$ changes with Reynolds number because the peak of $\theta^{\prime +}_{\rm RMS}$ increases and the location of such a peak moves farther from the wall whereas with respect to Prandtl number, the location of $\theta^{\prime +}_{\rm RMS}$ moves closer to the wall as discussed earlier. A correlation for $\theta_{\rm RMS,\, max}^{\prime +}$ was obtained as in~\cite{alcantra21jfm} for each passive scalar considered in the present study as shown in the figure~\ref{fig_scalar_fluctuation_correlation}. It should be noted that if the maximum value is obtained directly at the collocation point, some variations are observed along $Re_{\tau}$ due to the grid resolution. In order to minimize the high-frequency variation along $Re_{\tau}$ as plotted in figures~\ref{fig_scalar_fluctuation_correlation}  and \ref{fig_heat_flux_correlation}, a simple convolution operation is performed which does not alter the obtained empirical fits. Clearly, for higher Prandtl numbers of $4,6$, the peak of the scalar-fluctuation RMS tends to decrease with increasing $Re_{\tau}$. When the present DNS data with $Re_{\theta} \in [470,1070]$ and $Pr = 1,2,4,6$ is used to find an overall variation of the peak in the scalar-fluctuation RMS we find
\begin{equation}
    \theta_{\rm RMS, \,max}^{\prime +} = 2.969 Re_{\tau}^{-0.00858} Pr^{0.571}\,,
    \label{eqn_theta_max_corr}
\end{equation}
with $R^2 = 0.99$. The observed correlation shows a weak dependence on $Re$ compared with that on $Pr$. Further, from equation~(\ref{eqn_theta_max_corr}) a decaying trend of $\theta_{\rm RMS,\, max}^{\prime +}$ with $Re$ is obtained although such a trend has not been observed in the literature except for $\theta_3,\; \theta_4$ in the present study. The studies by~\cite{pirozzoli} and~\cite{alcantra21jfm} have reported the increasing trend of $\theta_{\rm RMS,\,max}^{\prime +}$ with respect to $Re$ but for a lower $Pr$ than in the present work.~\cite{pirozzoli} have suggested that the attached-eddy arguments support the increase of inner peak of streamwise-velocity RMS with respect to $Re_\tau$ due to the effect of overlying attached eddies and they assumed that the same argument applies to the passive scalars. On the other hand, we find that the inner peak of the passive-scalar RMS at high $Pr$ does not follow the previous argumentation. It should also be noted that the range of $Re_\tau$ in our present simulation is narrow compared with those of the works by~\cite{pirozzoli} and~\cite{alcantra21jfm} and a more detailed investigation of this topic is necessary to make any conclusive statements.  

The above procedure was also performed for streamwise heat flux as shown in figure~\ref{fig_heat_flux_correlation}. A similar behaviour was observed with the overall variation in streamwise heat flux given by
\begin{equation}
    \left<u\theta^{\prime}\right>^{+}_{\rm max} = 7.74 Re_{\tau}^{-0.006} Pr^{0.401}\,,
\end{equation}
with $R^{2} = 0.9913$.

\subsection{Integral quantities and non-dimensional numbers}

\begin{figure}
\centering
\resizebox*{0.7\linewidth}{!}{\includegraphics{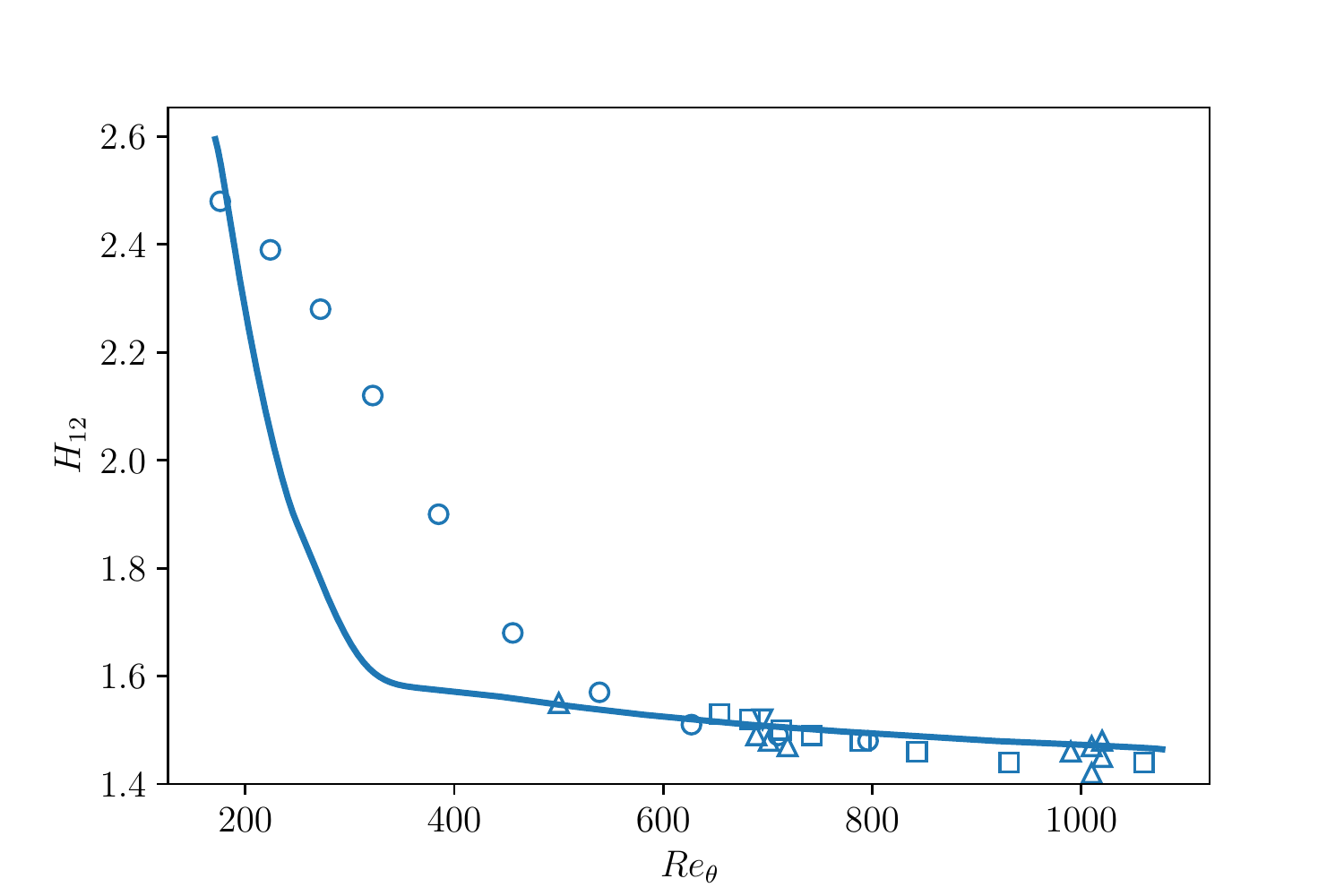}}
\caption[]{Streamwise evolution of the shape factor $H_{12}$. {\plotA}~Present DNS, {\plotAmarkers}~\cite{simens}, experimental data by {\plotAmarker}~\cite{roach}, {\plotAmarkerut}~\cite{erm}, {\plotAmarkerdt}~\cite{purtell}.} 
\label{fig_h12}
\end{figure}

The shape factor $H_{12}$, which measures the ratio between the displacement thickness $\delta^{*}$ and the momentum thickness $\theta$ is plotted in figure~\ref{fig_h12}. The shape factor is lower in the turbulent region with increasing $Re_{\theta}$ and agrees well with the experimental and numerical data for $Re_{\theta} > 600$ where the turbulent boundary layer is fully developed.

\begin{figure}
\centering
\resizebox*{0.7\linewidth}{!}{\includegraphics{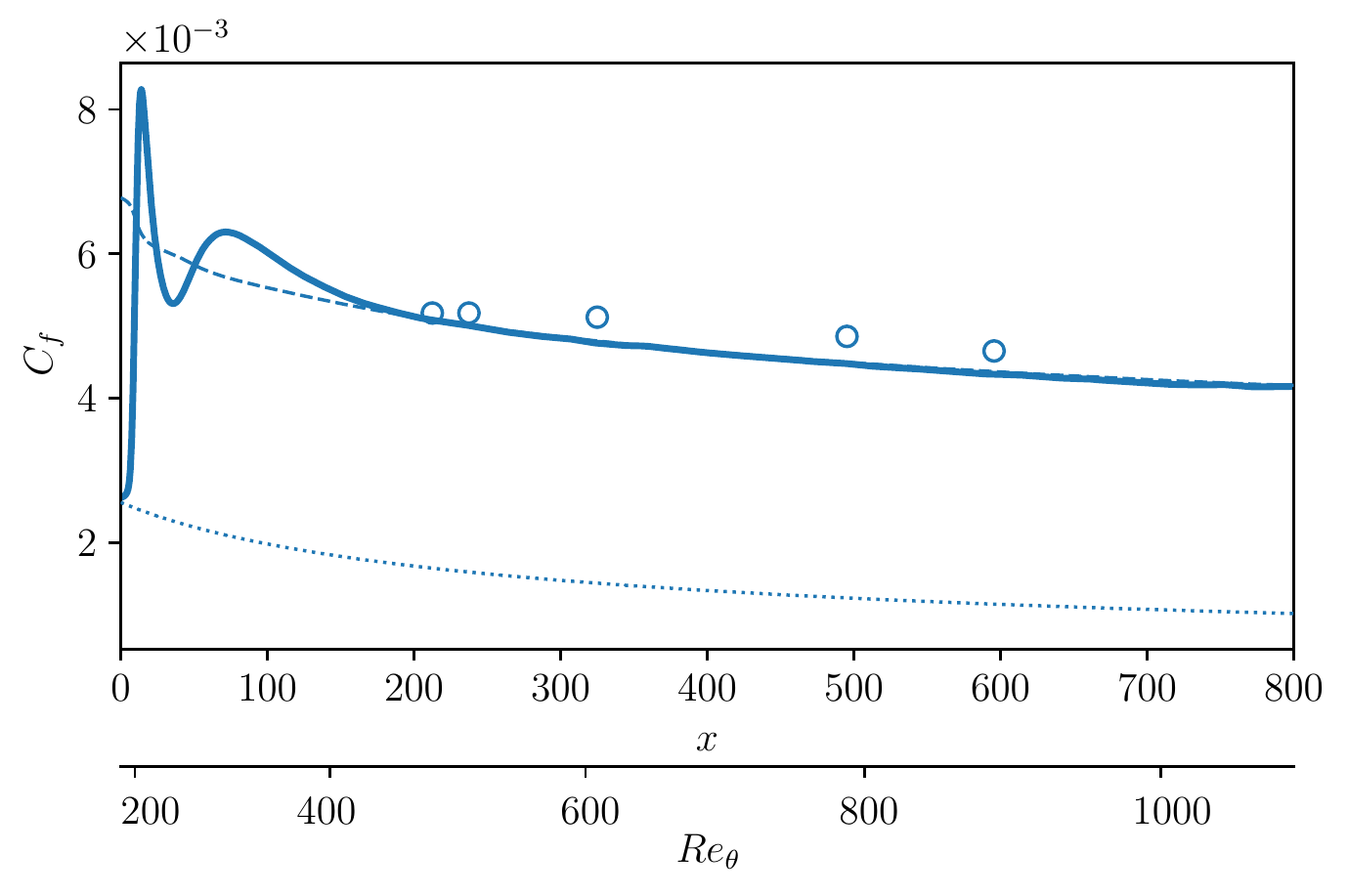}}
\caption[]{Variation of skin-friction coefficient along the streamwise direction. {\plotA}~Present DNS, {\plotAdotted}~theoretical laminar skin-friction solution, {\plotAdashed}~turbulent solution given by~\cite{schoenherr}, {\plotAmarker}~experimental data by~\cite{erm} with wire tripping.} 
\label{fig_cf_x}
\end{figure}

Figure~\ref{fig_cf_x} depicts the streamwise variation of the skin-friction coefficient $C_{f}$ for the present simulation and indicates that the obtained result is in good agreement with the turbulent skin-friction solution provided by~\cite{schoenherr}, which is given by
\begin{equation}
    C_{f} = 0.31\left[\ln^2{\left(2Re_{\theta}\right)} + 2\ln{\left(2Re_{\theta}\right)}\right]^{-1}\,.
\end{equation}
The computed skin-friction coefficient is also in good agreement with the correlation proposed by~\cite{smith83}, which is given as:
\begin{equation}
    C_{f} = 0.024 Re_{\theta}^{-1/4}\,.
\end{equation}
The trip location is at $x/\delta_0^*=10$ with a strong peak in the skin-friction coefficient followed by the transition to turbulence before $x/\delta_0^*=200$. The experimental data provided by~\cite{erm} with wire tripping also closely corresponds to the calculated turbulent skin-friction coefficient. It should be noted that \cite{erm} also repeated the experiments with different tripping devices and found the influence of tripping to persist until $Re_{\theta} \approx 1500$. Due to this, \cite{jiminez2010} found the experimental data to be scattered for $Re_{\theta} <1070$  and also showed the scatter to decrease with $Re_{\theta} > 1600$.

\begin{figure}
\centering
\resizebox*{0.7\linewidth}{!}{\includegraphics{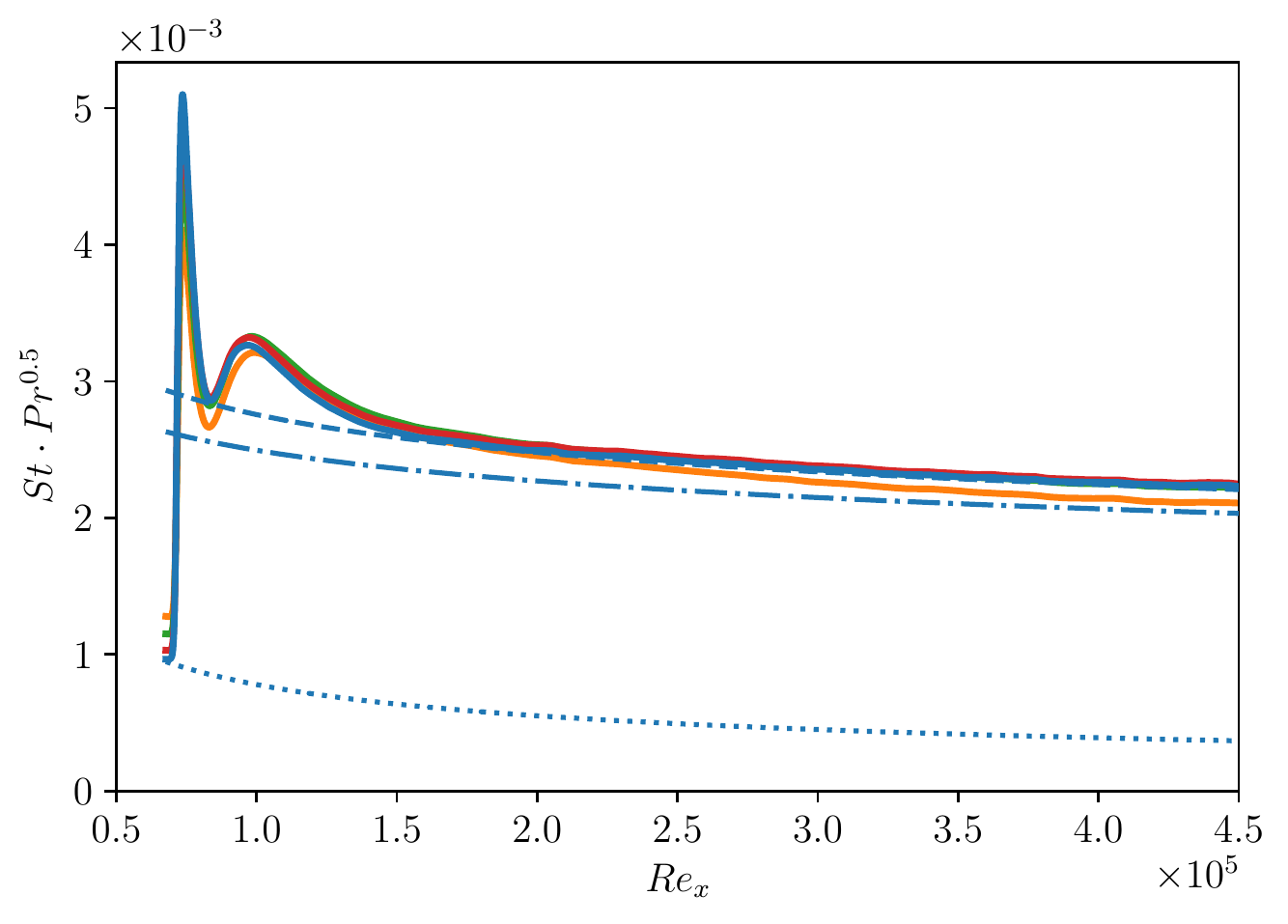}}
\caption[]{Variation of Stanton number along the streamwise direction for different passive scalars. {\plotB}~$\theta_{1}\,$, {\plotC}~$\theta_{2}$, {\plotD}~$\theta_{3}$, {\plotA}~$\theta_{4}$, {\plotAdotted}~\cite{kays_crawford93} correlation corresponding to~$\theta_4$ and {\plotAdashed}~Reynolds--Colburn analogy as given by~\cite{rcolburn} plotted for $\theta_{4}$, {\plotAdashdot}~interpolation developed by~\cite{hollingsworth} plotted for $\theta_{4}$.} \label{fig_stanton}
\end{figure}

\label{ReyAnaly}
The Stanton number normalizes the convective heat transfer into the fluid with respect to the thermal capacity of the fluid. The spatial evolution of the Stanton numbers for different passive scalars scaled with the square-root of $Pr$ (as plotted in figure~\ref{fig_stanton}) is very similar to the skin-friction profiles plotted in figure~\ref{fig_cf_x}. There is a difference observed in the laminar Stanton-number at $x=0$ with the present scaling. However, the individual Stanton-number profiles are well-corresponding to the laminar solution given by~\cite{kays_crawford93} which is
\begin{equation}
    St = \frac{0.332}{\sqrt{Re_{x}}Pr^{2/3}}\,,
\end{equation}
and with the turbulent solution obtained from the Reynolds--Colburn analogy as given by~\cite{rcolburn}. For $\theta_{1}$, due to Reynolds analogy the Stanton profile matches with the skin-friction profile scaled by a factor of two. For the passive scalars at higher Prandtl numbers, a generalization of the Reynolds--Colburn analogy can be obtained, as reported in the study by~\cite{lienhard}:
\begin{equation}
    St = \frac{Nu}{Re_{x}Pr} = \frac{C_{f}/2}{a_{1}+a_{2}\left(Pr^{a_{3}}-1\right)\sqrt{C_{f}/2}}\,,
\end{equation}
with the values of $a_{1}=1, a_{2} = 12.8$ and $a_{3}=0.68$ provided in~\cite{rcolburn} and $C_f$ being expressed as
\begin{equation}
    C_{f} = \frac{0.455}{\ln^2{\left(0.06Re_{x}\right)}}\,.
\end{equation}
The Stanton-number plot for the passive scalar at $Pr = 6$ agrees well with the turbulent solution obtained from the Reynolds--Colburn analogy as shown in figure~\ref{fig_stanton}. On the other hand, reference curves for Kays and Crawford and the Reynolds--Colburn analogy are only reported for $\theta_{4}$ for clarity.

Using the data for water with $Pr = 5.9$, \cite{hollingsworth} developed an interpolation for Prandtl numbers from 0.7 to 5.9 assuming the critical thickness of the sub-layer to be a simple function of Prandtl number. The empirical expression is given by
\begin{equation}
    St = 0.02426 Pr^{-0.895} Re_{x}^{-0.1879 Pr^{-0.18}}\,,
\end{equation}
which is plotted for the scalar at $Pr = 6$ in figure~\ref{fig_stanton}. We find that the relationship proposed by~\cite{hollingsworth} under predicts the Stanton number at $Pr = 6$. However, the interpolation relation provides a good match with the calculated Stanton number for $\theta_{1}$, better than the Reynolds--Colburn analogy, which is not indicated in the plot for clarity.

Using the data obtained in the present simulation, a correlation between the Nusselt, Prandtl and Reynolds numbers can be obtained as: 
\begin{equation}
    Nu_{x} = 0.02 Re_{x}^{0.828} Pr^{0.514}\,,
\end{equation}
which yields $R^{2} = 0.9985$. This is similar to the correlation proposed by~\cite{kays80} but for fully-developed profiles in circular tubes and computed for gases:
\begin{equation}
    Nu = 0.021 Re^{0.8} Pr^{0.5}\,.
\end{equation}

\begin{figure}
\centering
\resizebox*{0.7\linewidth}{!}{\includegraphics{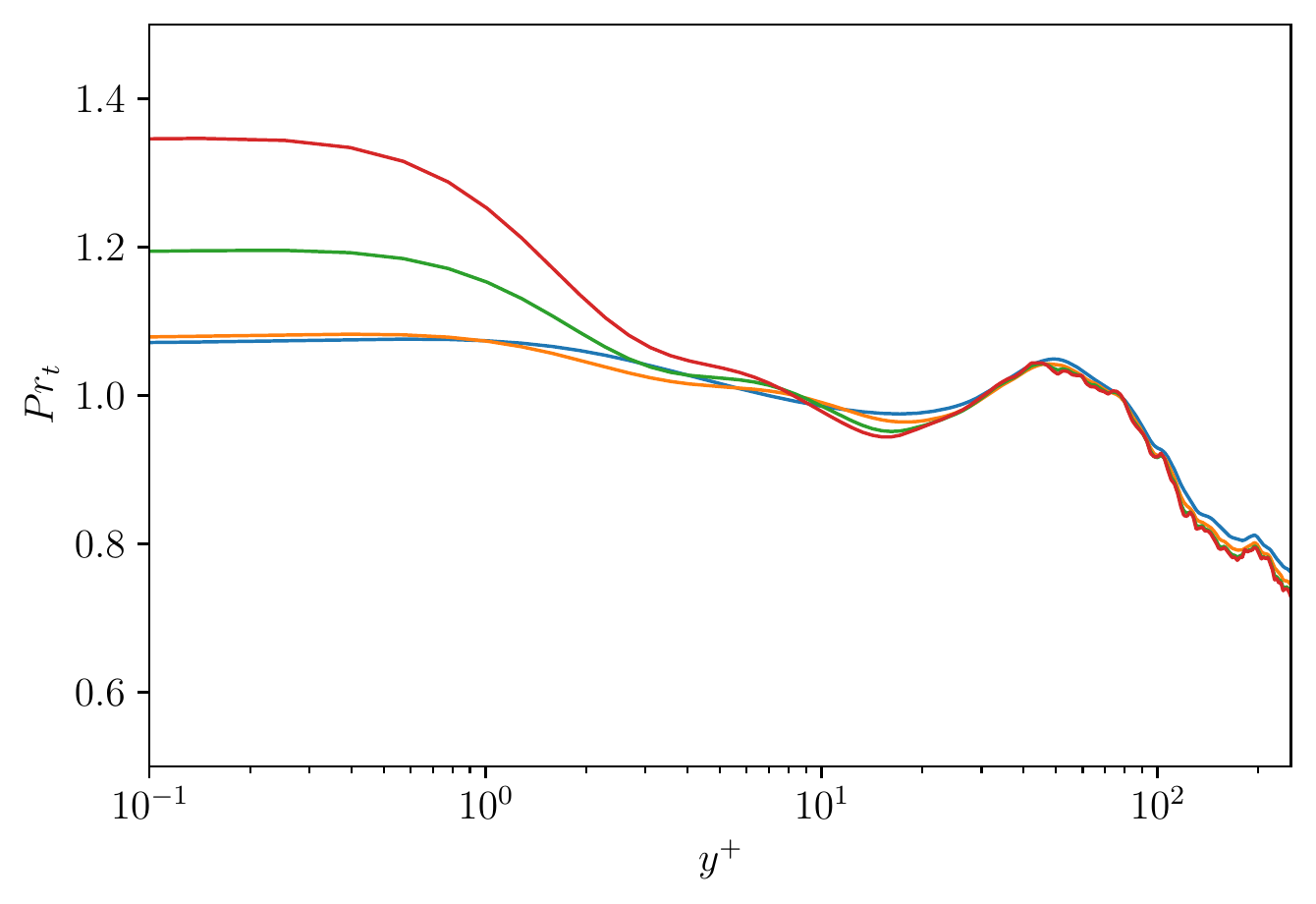}}
\caption[]{Variation of turbulent Prandtl number with respect to wall-normal distance in inner scale at $Re_{\theta} = 1070$ for passive scalar at {\plotA}~$Pr = 1$, {\plotB}~$Pr = 2$, {\plotC}~$Pr = 4$, {\plotD}~$Pr = 6$.} \label{fig_prt_y}
\end{figure}

\begin{figure}
\centering
\resizebox*{0.7\linewidth}{!}{\includegraphics{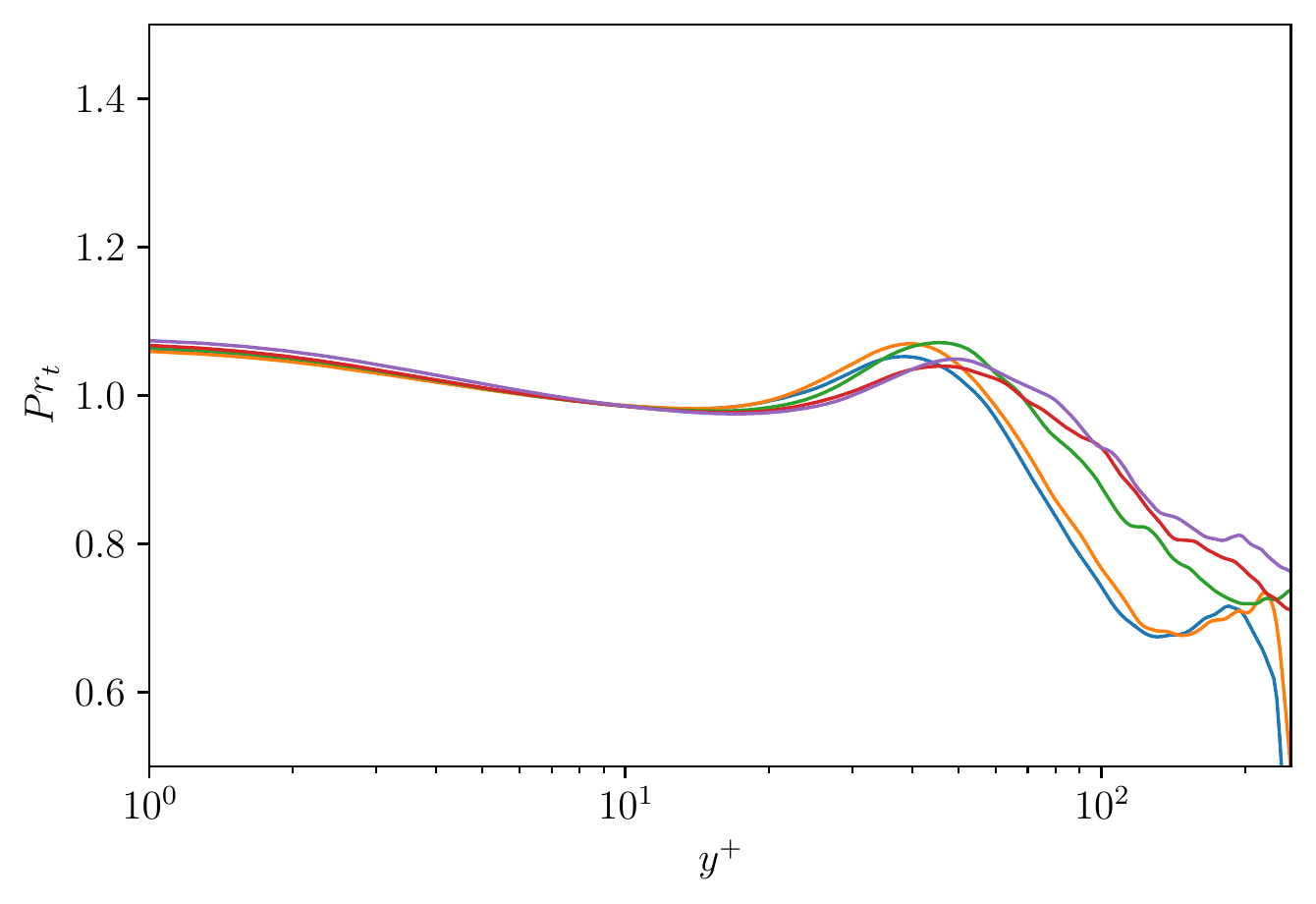}}
\caption[]{Variation of turbulent Prandtl number with respect to wall-normal distance in inner scale at {\plotA}~$Re_{\theta} = 396$, {\plotB}~$Re_{\theta}=420$, {\plotC}~$Re_{\theta}=628$, {\plotD}~$Re_{\theta}=830$, {\plotE}~$Re_{\theta}=1070$ for a passive scalar at $Pr = 1$.} \label{fig_prt_retheta}
\end{figure}

An important parameter for scalar transport is the turbulent Prandtl number $Pr_{t}$, which is defined as the ratio between turbulent eddy viscosity and turbulent eddy diffusivity:
\begin{equation}
  Pr_{t} = \frac{\nu_{t}}{\alpha_{t}}\,.
\end{equation}
The eddy viscosity and the eddy diffusivity arise from the Boussinesq hypothesis~\cite{bous} for modelling turbulent stresses and the heat-flux vector, respectively. For parallel flows (the ones in which the velocity profile does not vary in the streamwise direction), the turbulent eddy viscosity and the turbulent eddy diffusivity are used to describe the turbulent momentum transfer and heat transfer with respect to the mean-flow conditions, in particular the mean velocity strain and temperature gradients, respectively. They are defined as:
\begin{equation}
    \nu_{t} = -\frac{\left<u^{\prime}v^{\prime}\right>}{{\partial \left<u\right>}/{\partial y}}
\end{equation}
and 
\begin{equation}
    \alpha_{t} = -\frac{\left<v^{\prime}\theta^{\prime}\right>}{{\partial \left<\theta\right>}/{\partial y}}\,.
\end{equation}
It should be noted that the eddy viscosity or diffusivity does not represent a physical property of the fluid, like the molecular viscosity, rather a property of the flow. 
The Reynolds analogy introduces the similarity between the turbulent momentum exchange and turbulent heat transfer in a fluid. \cite{reynolds90} noted that, for a fully-turbulent field, both the momentum and heat are transferred due to the motion of turbulent eddies. This yields a simpler model for the turbulent Prandtl number, where the turbulent eddy viscosity for the momentum exchange and turbulent eddy diffusivity for the scalar transport are equal, such that $Pr_{t} = 1$.

Substitution of the eddy viscosity and diffusivity into the definition of turbulent Prandtl number results in
\begin{equation}
    Pr_{t} = \frac{\left<u^{\prime}v^{\prime}\right>}{\left<v^{\prime}\theta ^{\prime}\right>}\frac{{\partial \left<\theta\right>}/{\partial y}}{{\partial \left<u\right>}/{\partial y}}\,.
\end{equation}

\begin{figure}
\centering
\resizebox*{0.45\linewidth}{!}{\includegraphics{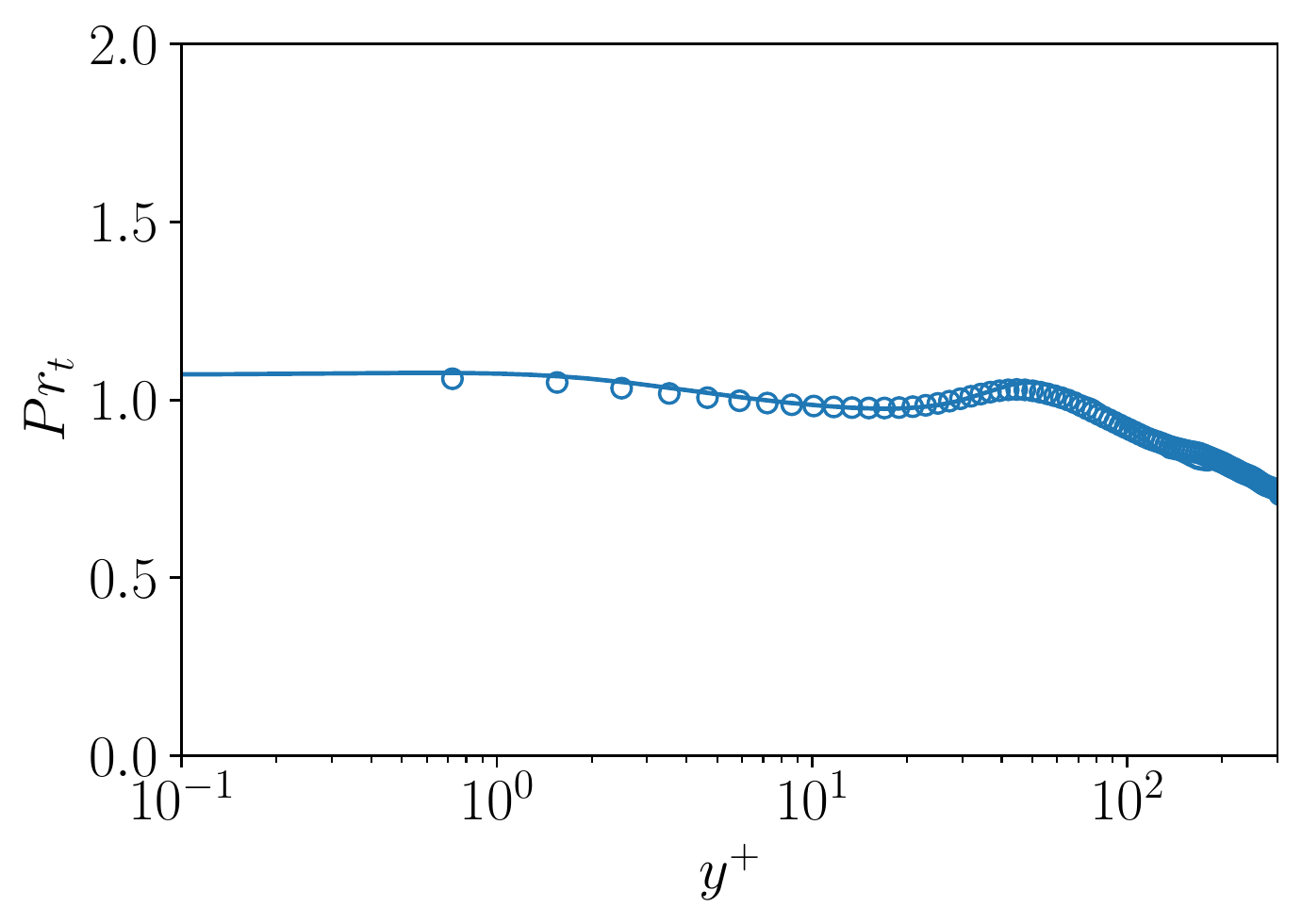}}
\resizebox*{0.45\linewidth}{!}{\includegraphics{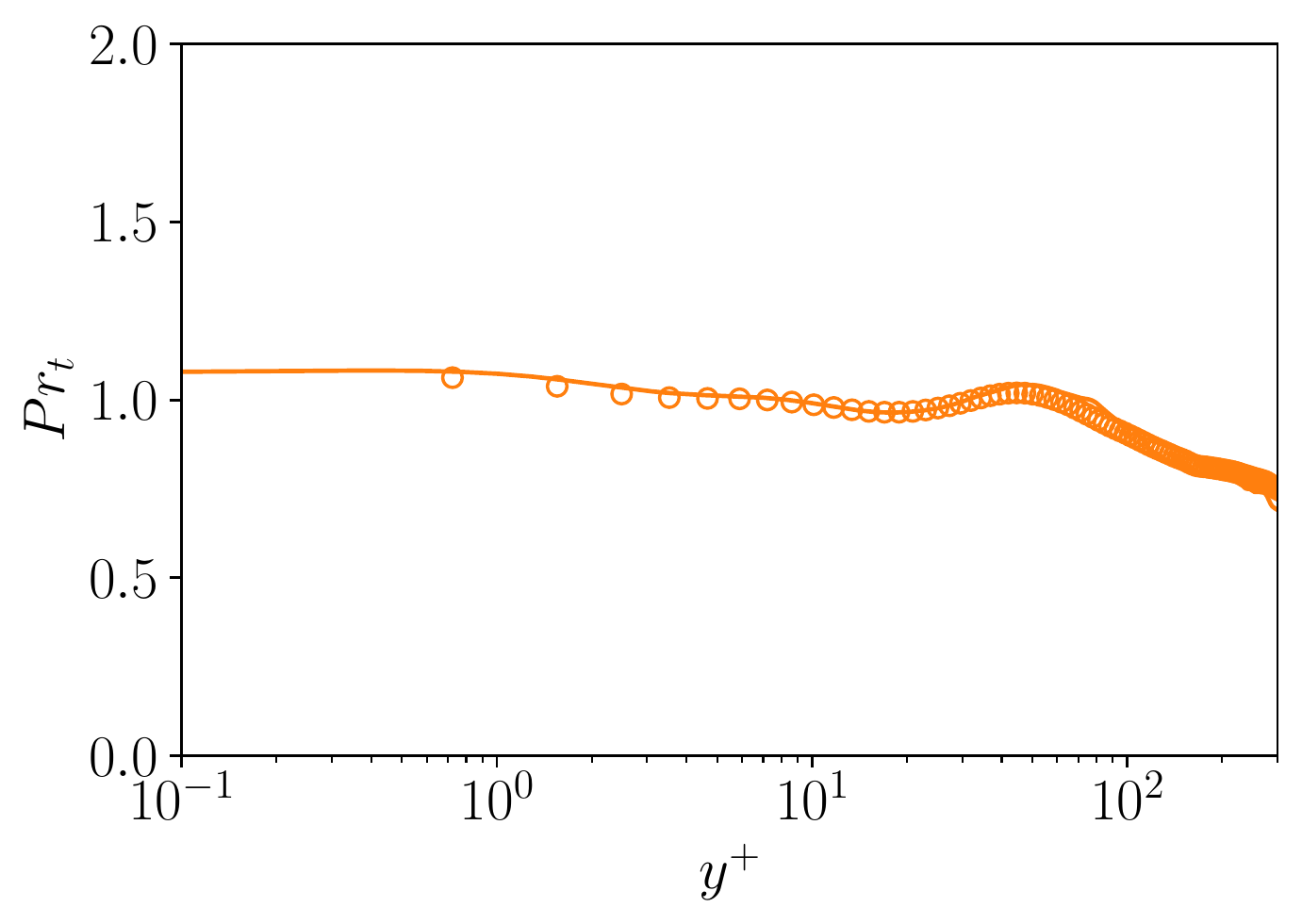}}
\resizebox*{0.45\linewidth}{!}{\includegraphics{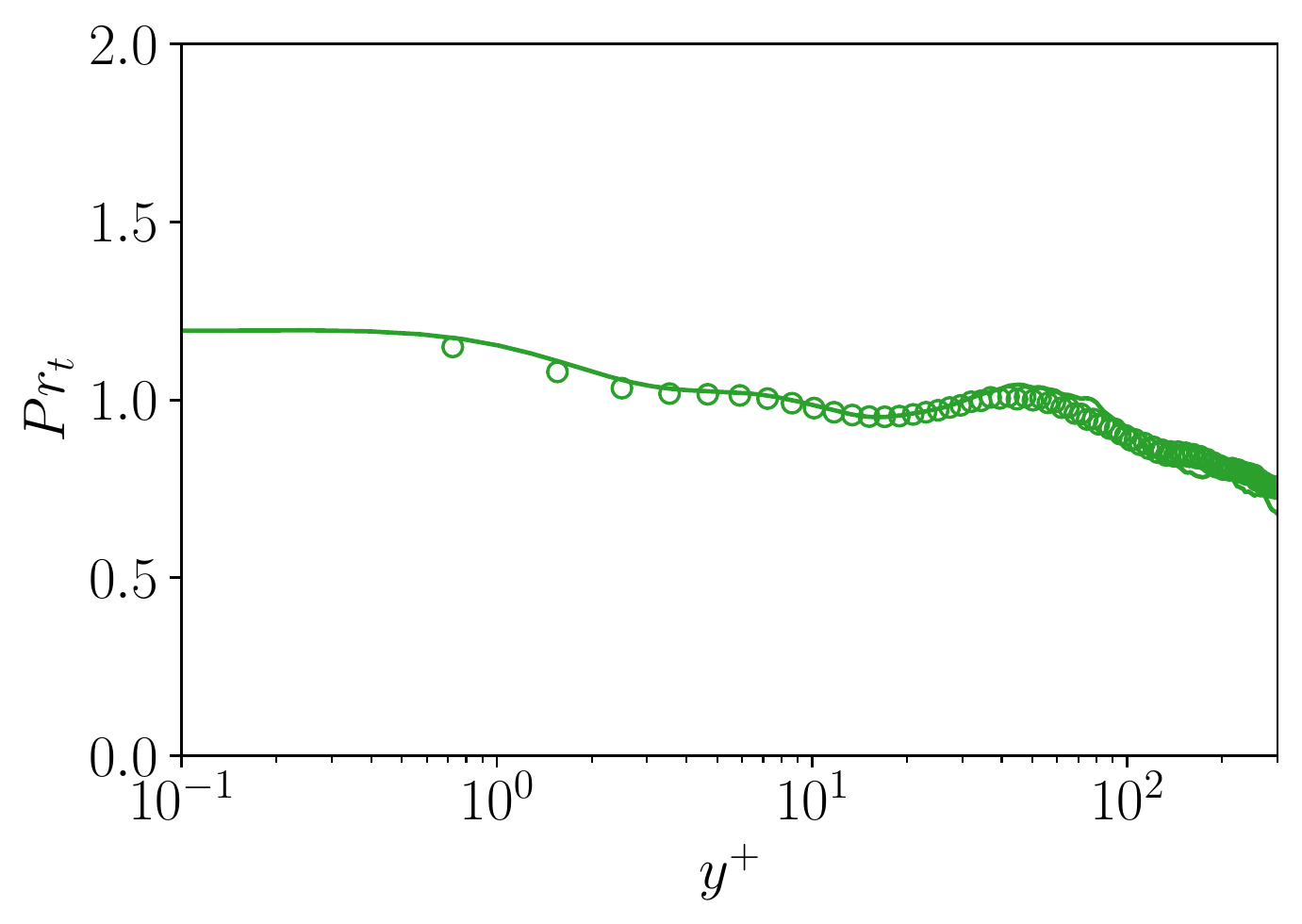}}
\resizebox*{0.45\linewidth}{!}{\includegraphics{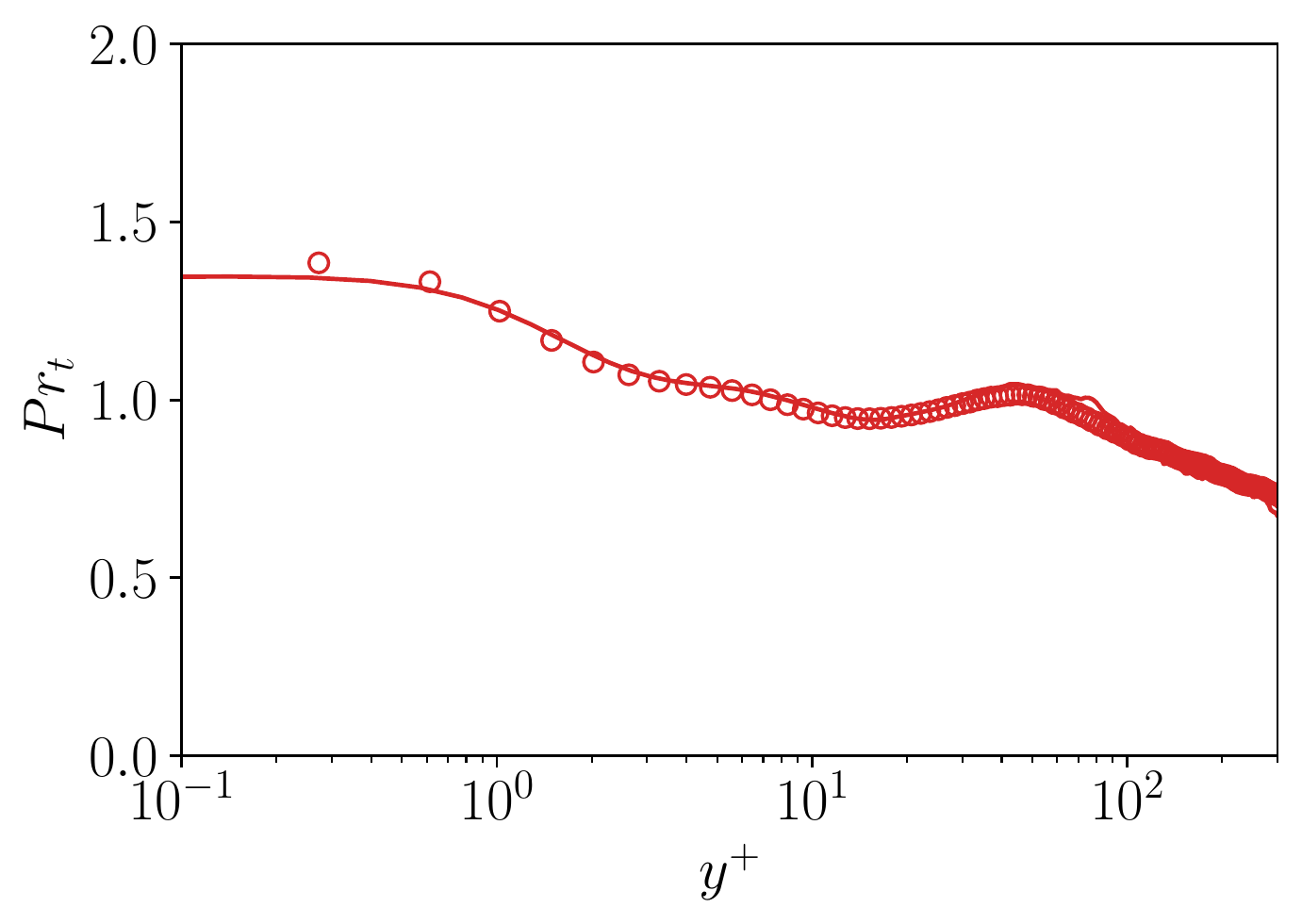}}
\caption[]{Comparison of turbulent Prandtl number against the channel DNS data at $Re_\theta = 1070$ for different molecular Prandtl numbers. Present DNS for {\plotA}~$Pr = 1$, {\plotB}~$Pr = 2$, {\plotC}~$Pr = 4$, {\plotD}~$Pr = 6$, channel DNS data by~\cite{alcantra21} at $Re_{\tau} = 497$ and {\plotAmarker}~$Pr = 1$, {\plotBmarker}~$Pr = 2$, {\plotCmarker}~$Pr = 4$, {\plotDmarker}~$Pr = 6$.}
\label{fig_prt_comp}
\end{figure}

From this definition, evaluating the turbulent Prandtl number at any point in the boundary layer would require the turbulent shear stress, turbulent heat transfer, velocity gradient and temperature gradient. Experimental investigations have limited accuracy in the simultaneous measurement of the Reynolds shear stress and wall-normal turbulent heat flux, in particular close to the wall~\citep{araya12}. For this reason, experimental investigations like~\cite{blom}, \cite{antonia80}, \cite{kays_crawford93} exhibit significant scatter in the data.

The variation of turbulent Prandtl number with the wall-normal distance in inner units at a given $Re_{\theta} = 1070$ is reported in figure~\ref{fig_prt_y}. We observe that the turbulent Prandtl number varies at the wall and increases with respect to molecular Prandtl number of the scalar. From figure~\ref{fig_prt_retheta}, we also see the turbulent Prandtl number decays for $y^+ > 15$ and this decay becomes steeper as $Re$ decreases. The turbulent Prandtl number is usually assumed to be a constant and is independent of molecular Prandtl number and wall-normal distance~\citep{qiang_thesis}. Studies such as the ones by~\cite{kestin} and \cite{blom} have analyzed experimentally the turbulent Prandtl number in order to assess the validity of this assumption. Such experimental investigations have reported a constant value of turbulent Prandtl number as the wall is approached, see figure~\ref{fig_prt_y}. However, different experimental campaigns provided data that could not show a conclusive interpretation of the behaviour of turbulent Prandtl number close to the wall.

For the different passive scalars considered in the present study, the turbulent Prandtl number approaches a constant value $>1$ close to the wall in the viscous sub-layer. The plots of $Pr_{t}$ exhibit a significant difference closer to the wall with respect to the various molecular Prandtl numbers. There is a slight decrease in $Pr_{t}$ up to $y^{+} \approx 20$ and then the increase is maintained farther from the wall up to $y^{+} \approx 50$, the point after which the trend steadily decreases. It is pointed out in~\cite{kays94} that the peak between $y^{+} \approx 20$ and $100$ is not observed in the experimental data, an observation which was attributed to the high Reynolds number of the experiments, while DNS data is not available for comparison. Our data is consistent with these experimental observations.

Following the discussions about a constant $Pr_{t}$ in the logarithmic region, \cite{kays_crawford93} proposed a correlation for turbulent Prandtl number that is applicable to air. In this correlation, the value of $Pr_{t}$ approaches a constant value of 0.85 in the logarithmic region. In the studies by~\cite{hollingsworth}, a correlation was proposed based on the measurement of the temperature profile of water at $Pr \approx 6$. Again, the value of $Pr_{t}$ approaches a value of 0.85 as $y^{+}$ is increased. This observation of constant $Pr_{t}$ in the logarithmic region is not clearly observed with the present DNS data. This can be due to the low-Reynolds number range considered in the present study. 
Based on the correlation proposed by \cite{hollingsworth}, \cite{kays94} suggested a constant value of $Pr_{t} = 1.07$ for $0<y^{+}<5$. Indeed, if we consider only the passive scalars at $Pr = 1,2$, the turbulent Prandtl number approaches a constant value of 1.07 closer to the wall. It appears that the turbulent Prandtl number is independent of the molecular Prandtl number as shown in the studies by~\cite{qiang}. This independence of the turbulent Prandtl number at the wall with respect to the molecular Prandtl number has also been reported in many other studies like~\cite{kong} and~\cite{jacobs} for TBL flow, as well as in~\cite{kim89} and~\cite{kasagi92} for turbulent channel flow. However, from the present study, we find that the turbulent Prandtl number indeed depends on the molecular Prandtl number and this observation is based on the increasing value of $Pr_t$ at the wall with respect to the scalars with $Pr=4,6$ as shown in figure~\ref{fig_prt_y}.

In order to verify the plausibility of the present observations at $Pr = 4,6$, the obtained DNS data was compared with the DNS channel data reported by~\cite{alcantra21}. Figure~\ref{fig_prt_comp} shows the comparison of $Pr_{t}$ at different molecular Prandtl numbers where the present DNS data was at $Re_{\theta} = 1070$, corresponding to $Re_{\tau} = 395$, and the data from~\cite{alcantra21} was at $Re_{\tau} = 500$. Despite these differences, the turbulent Prandtl numbers close to the wall are in good agreement, confirming that the $Pr$-scaled wall-normal heat flux decreases with increase in $Pr$ for $Pr \gtrapprox 4$, as stated by~\cite{alcantra21}. Thus, the present observation confirms the constant behaviour of the turbulent Prandtl number very close to the wall and highlights its dependence on the molecular Prandtl number, which has been often ignored in turbulent heat-transfer calculations.

A brief discussion of the Reynolds stress budget is provided in Appendix~\ref{appA}.

\subsection{Higher order statistics, shear stress and heat flux}

\begin{figure}
\centering
\resizebox*{0.45\linewidth}{!}{\includegraphics{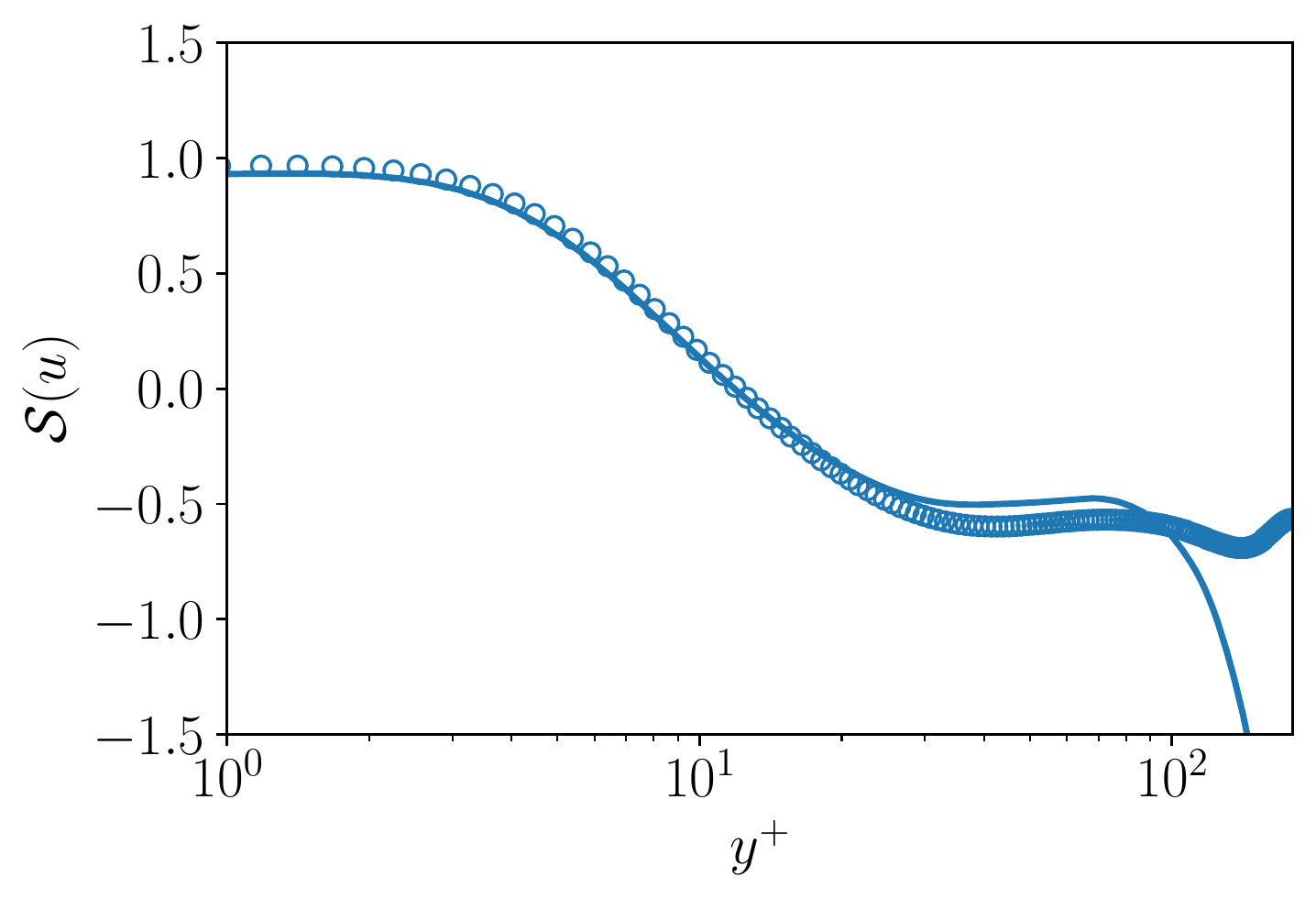}}
\resizebox*{0.42\linewidth}{!}{\includegraphics{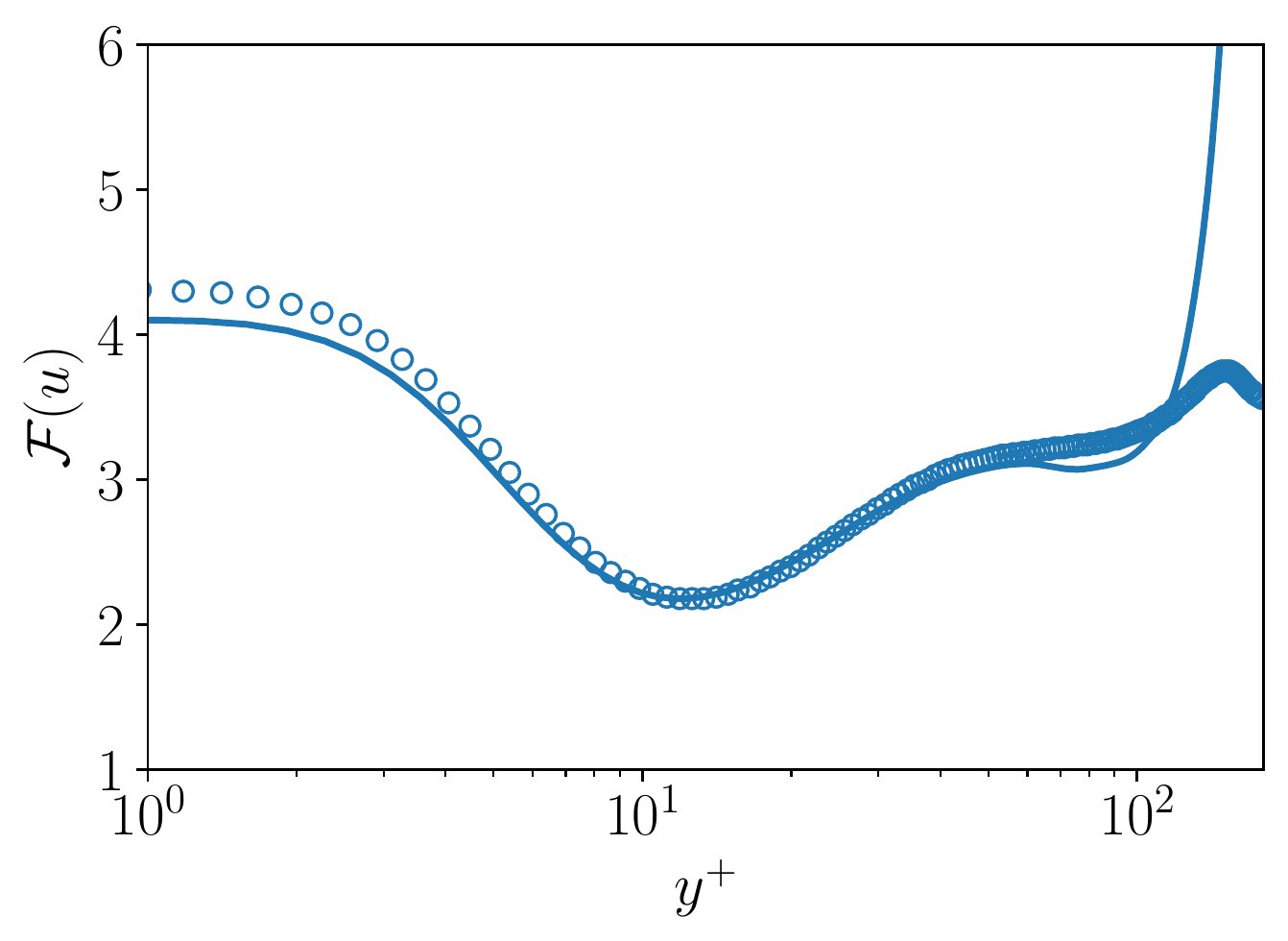}}
\resizebox*{0.45\linewidth}{!}{\includegraphics{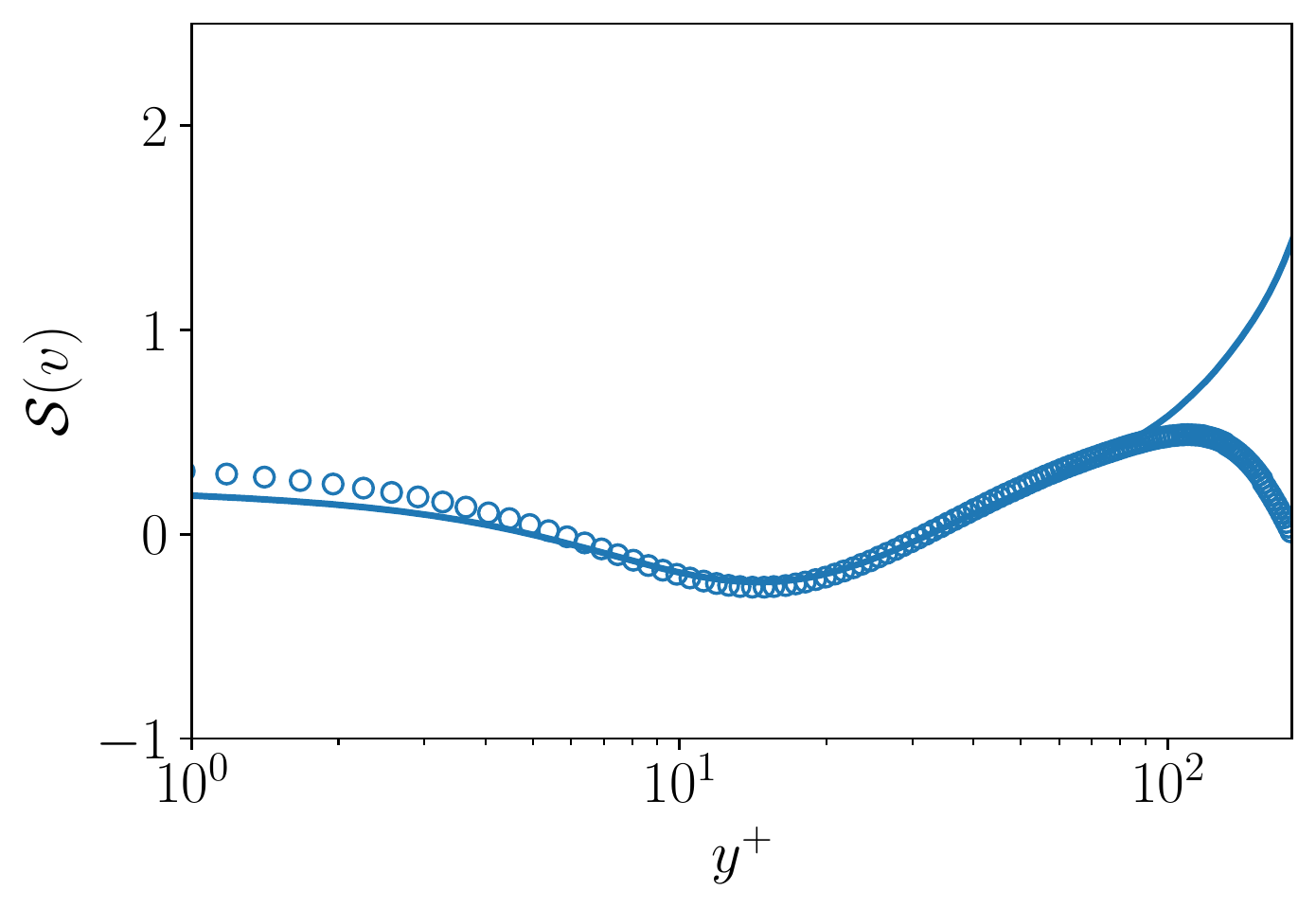}}
\resizebox*{0.445\linewidth}{!}{\includegraphics{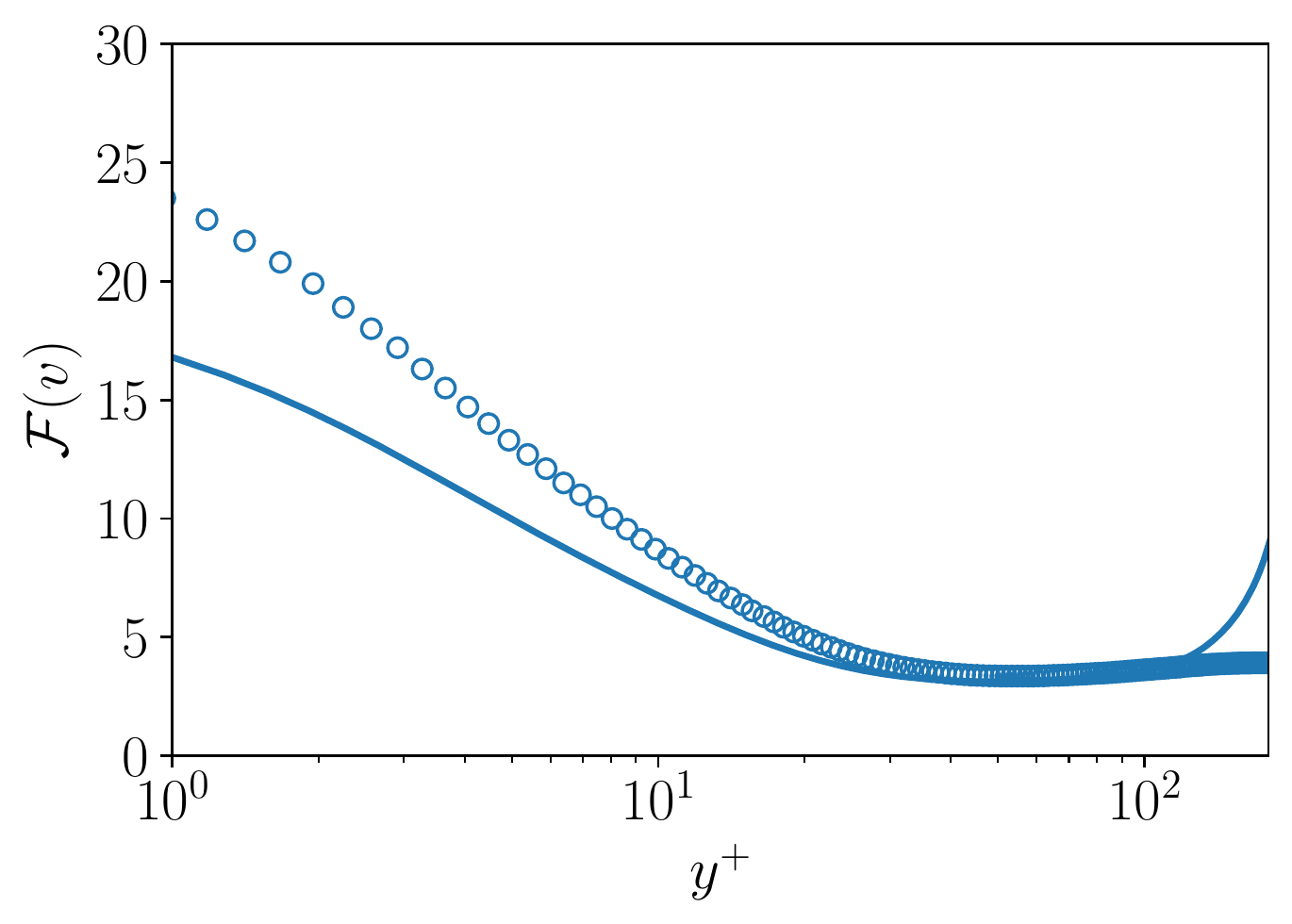}}
\resizebox*{0.45\linewidth}{!}{\includegraphics{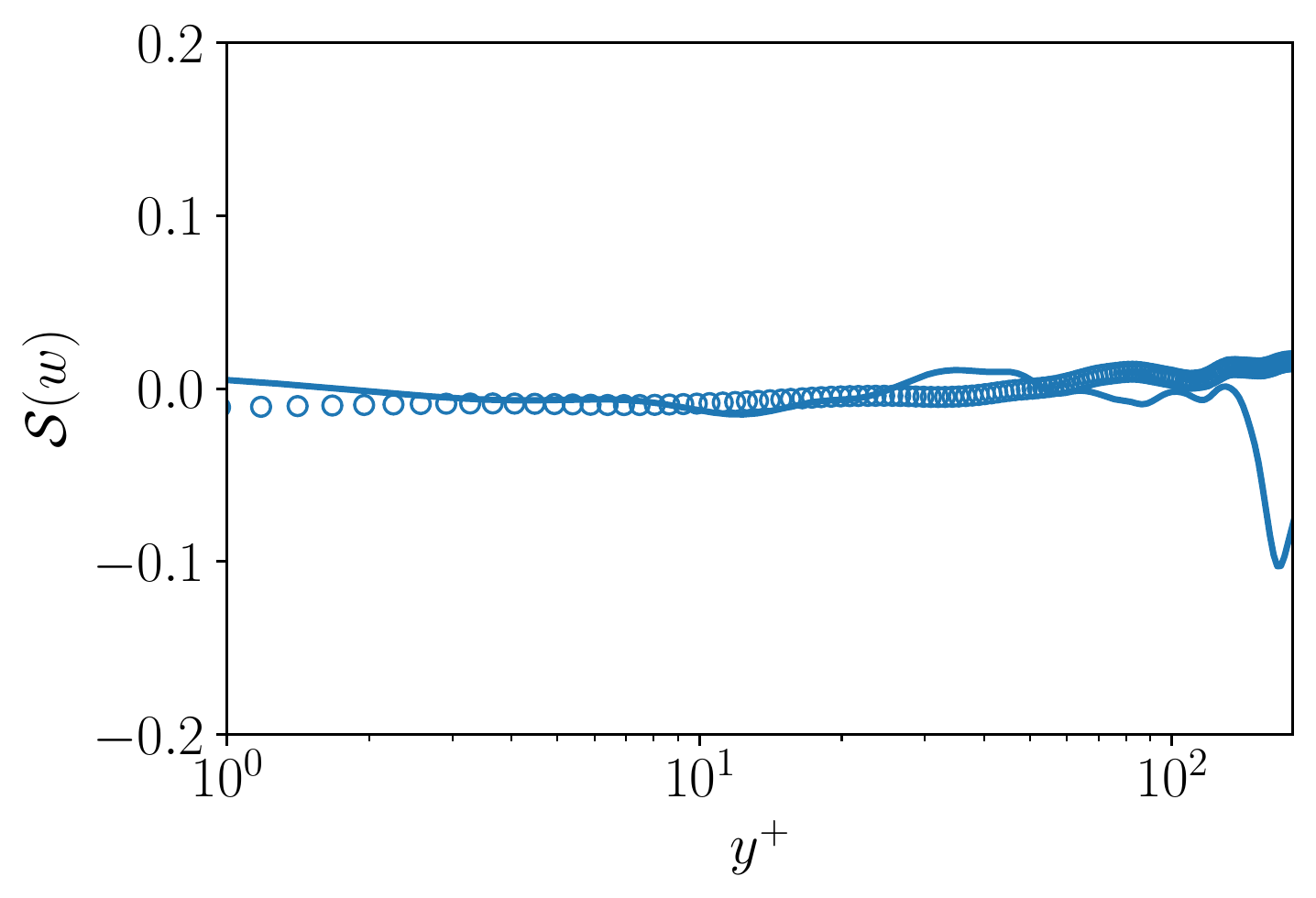}}
\resizebox*{0.43\linewidth}{!}{\includegraphics{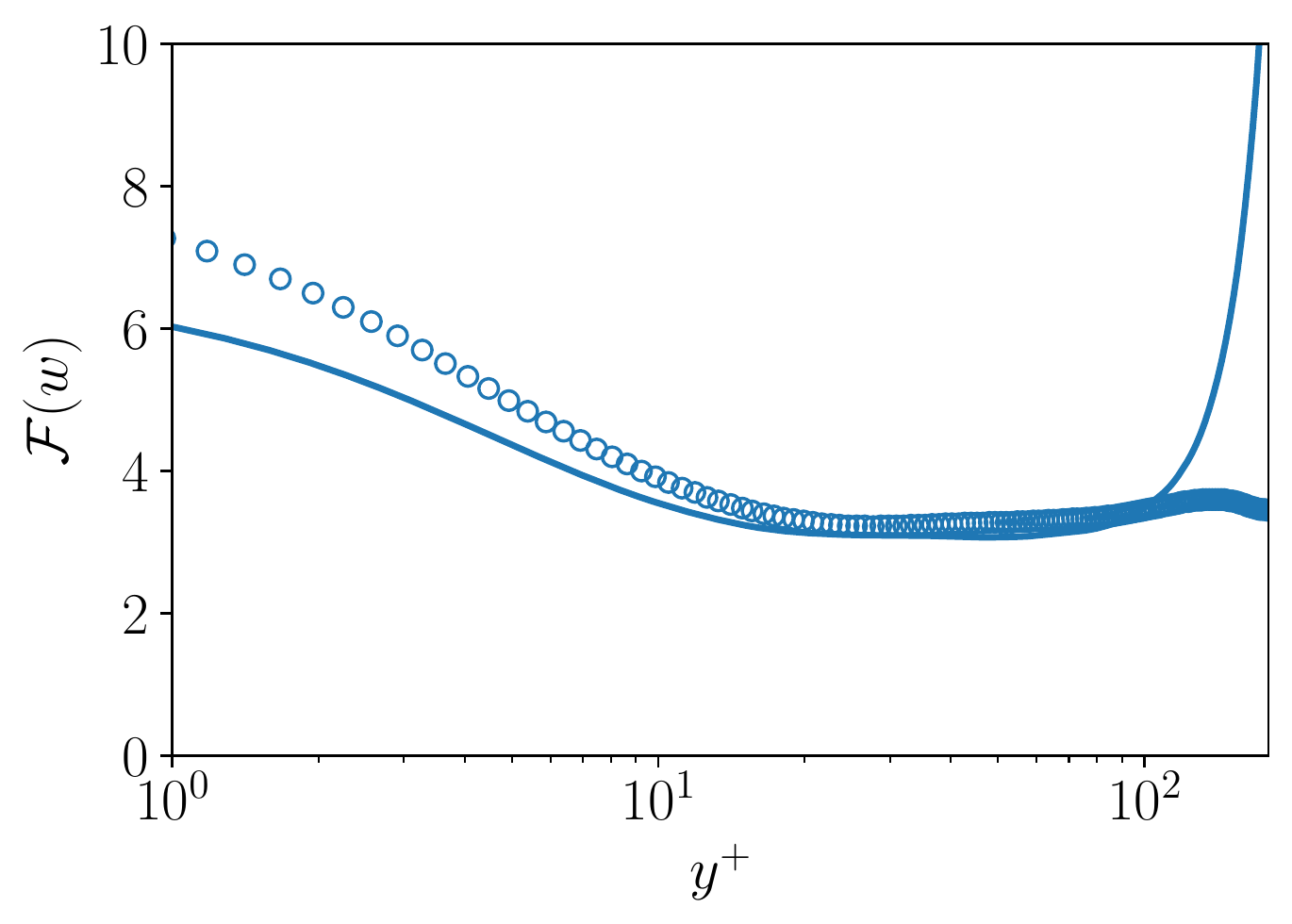}}
\resizebox*{0.45\linewidth}{!}{\includegraphics{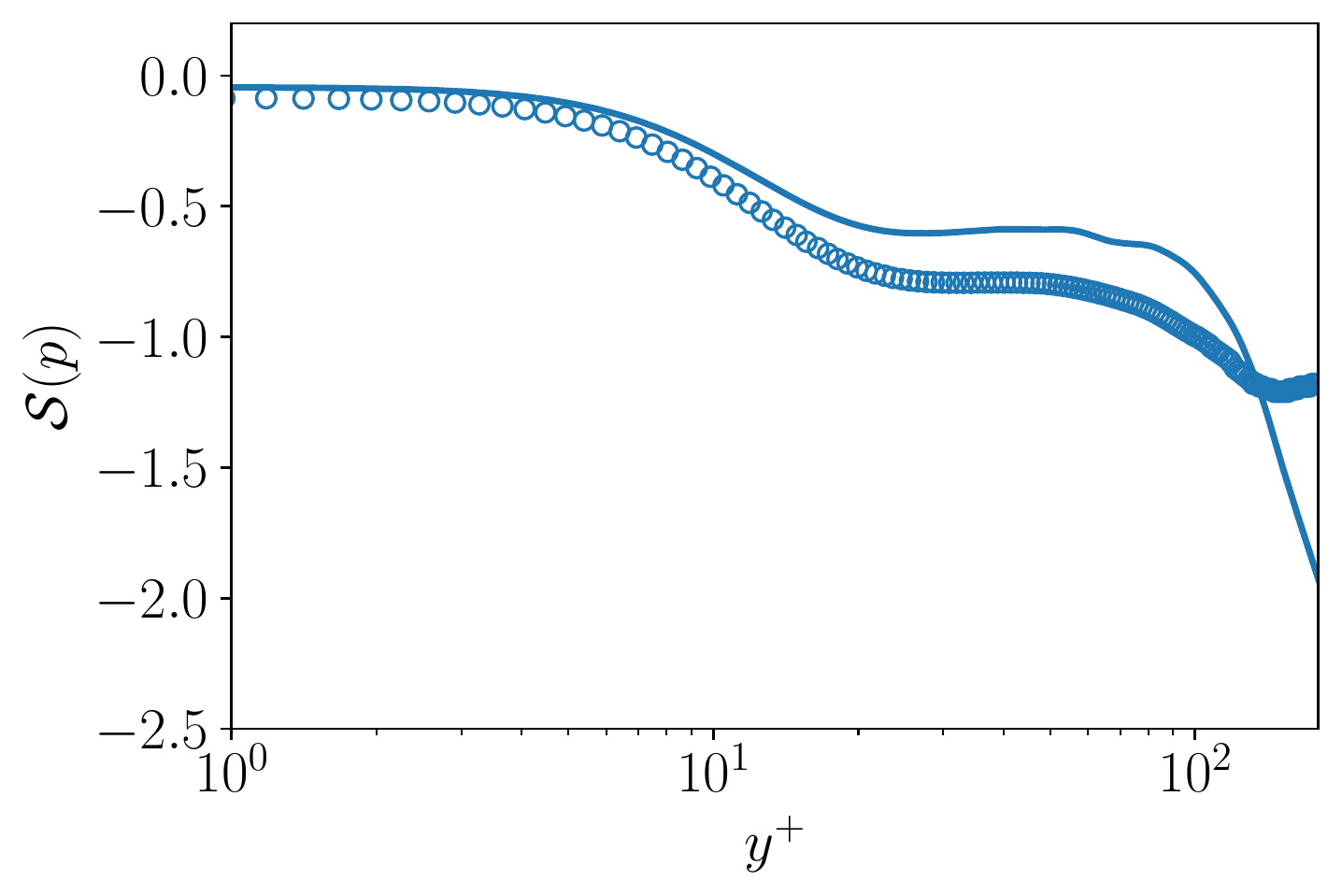}}
\resizebox*{0.43\linewidth}{!}{\includegraphics{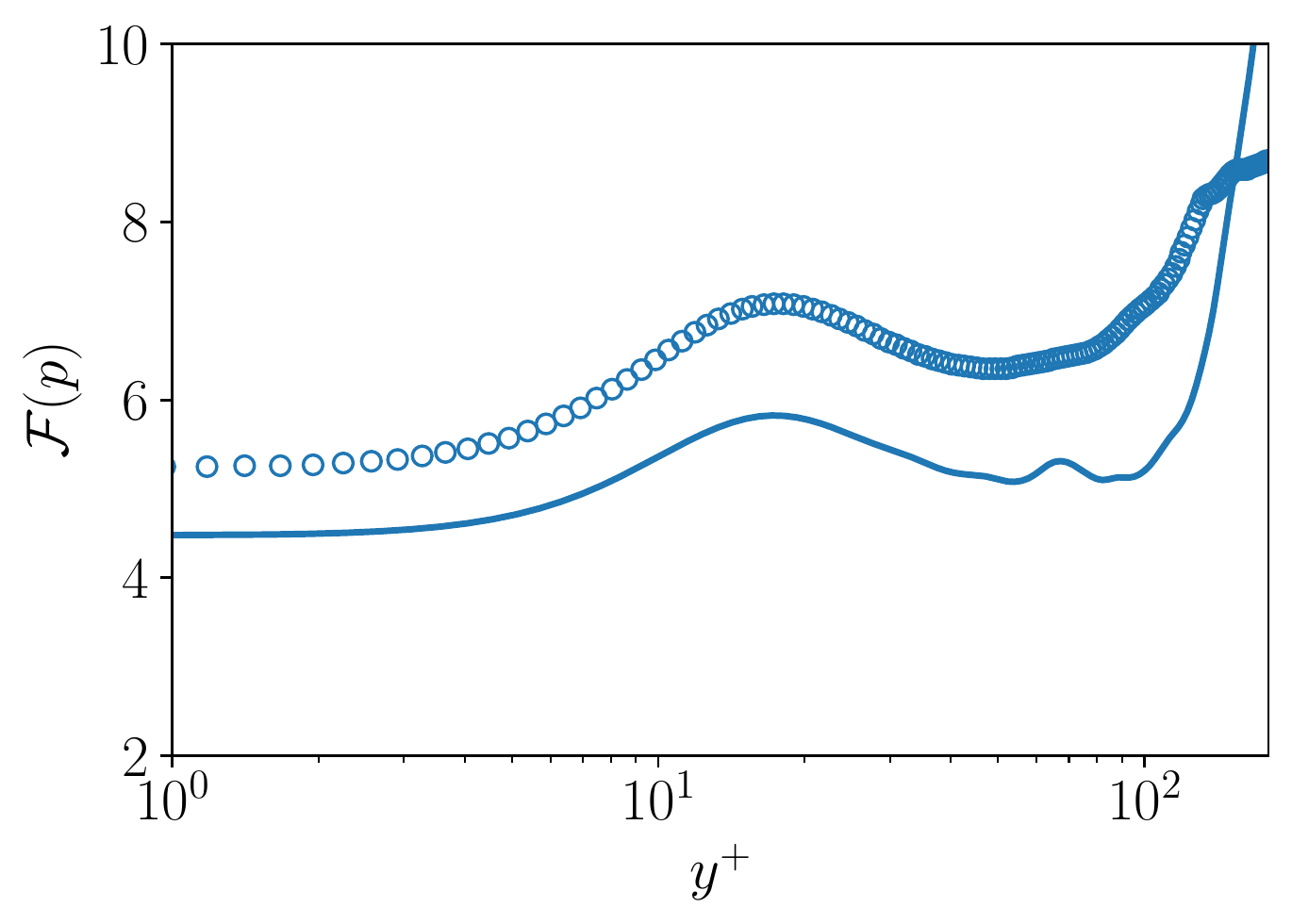}}
\caption[]{Skewness and flatness for the streamwise, wall-normal and spanwise velocity components at $Re_{\theta} = 420$ corresponding to $Re_{\tau} = 180$. {\plotA}~Present DNS, {\plotAmarker}~channel DNS data by~\cite{vreman} at $Re_{\tau} = 180$.} \label{fig_high_statistics}
\end{figure}

The higher-order statistics (in specific, the values of third and fourth-order moments of a quantity) provides information on the non-Gaussian behaviour of turbulence. The third and fourth-order moments are also called as skewness and flatness, respectively and for a statistically stationary variable $m$, it is defined as
\begin{align}
    \mathcal{S}(m) &= \left<m^3\right>/\left<m^2\right>^{3/2}\,, \\
    \mathcal{F}(m) &= \left<m^4\right>/\left<m^2\right>^{2}\,.
\end{align}
If the stochastic variable $m$ were to follow a Gaussian probability distribution, then $\mathcal{S}(m) = 0$ and $\mathcal{F}(m) = 3$.
The deviation from such values for the skewness and flatness of the streamwise and wall-normal velocity components is shown in figure~\ref{fig_high_statistics} whereas, for the spanwise velocity we observe a Gaussian behaviour in the overlap region as expected. The higher order moments obtained by~\cite{vreman} for the turbulent channel flow is compared with the present TBL results at $Re_{\tau} = 180$ (corresponding to $Re_\theta = 420$), showing a reasonable agreement. From figure~\ref{fig_cf_x}, it should be highlighted that $Re_\theta = 420$ is roughly in the beginning of the turbulent regime and the deviation of the profiles for $y^{+}>100$ as observed in figure~\ref{fig_high_statistics} is due to the intermittent wake region which decay to the values for Gaussian distribution as free-stream is approached. There are some differences observed in the overlap region where, the skewness of $u$ is roughly $10\%$ higher than that reported by~\cite{vreman}. Although the plots indicate an overall good agreement, some drastic differences are observed in the flatness of wall-normal and spanwise velocity components close to the wall, where the turbulence is highly intermittent. The flatness of wall-normal velocity component approaches a value of $\approx 29$ close to the wall for the data provided by~\cite{vreman} whereas in the TBL, it converges to $\approx 20$. This is still lower than the value of $22$ obtained by~\cite{kim87}. The skewness of pressure fluctuations is roughly 20\% higher in the overlap region of TBL compared to the channel. Further, the intermittency (flatness) of pressure is higher than the velocity components as reported by~\cite{kim87} whereas the data obtained with TBL shows an offset of 15\% throughout and is lower than the flatness behaviour observed in the channel. The flatness factor for pressure fluctuation at the wall approaches a value of 4.5 in the present simulations as compared to the values of 4.7~\citep{qiang} and 4.9~\citep{schewe}, whereas a value of $\approx$5.2 is reported by~\cite{vreman}.

 \label{sec_heat_flux}
\begin{figure}
\centering
\resizebox*{0.7\linewidth}{!}{\includegraphics{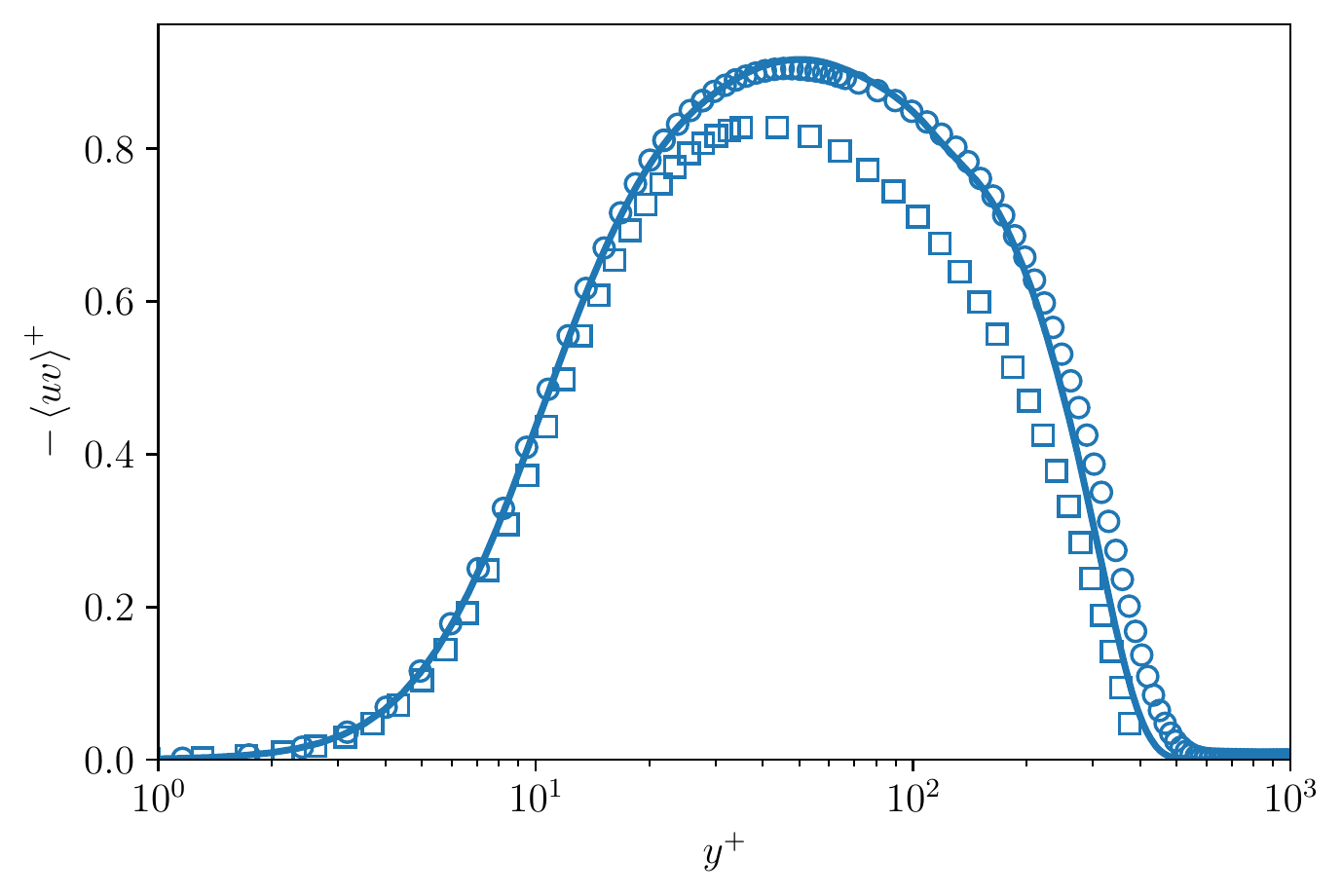}}
\caption[]{Inner-scaled Reynolds shear stress as a function of the wall-normal direction at $Re_{\theta}=1070$. {\plotA}~Present DNS, {\plotAmarker}~\cite{jiminez2010} at $Re_{\theta} = 1100$, {\plotAmarkers}~channel DNS data by~\cite{kozuka} at $Re_{\tau} = 395$.} \label{fig_shear_stress}
\end{figure}

The Reynolds shear stress is plotted in figure~\ref{fig_shear_stress} and it is compared with the turbulent boundary layer data provided by~\cite{jiminez2010} and the channel DNS data by~\cite{kozuka}. Even if there is a good agreement between the data in the inner region, there is a clear difference in the overlap region between the channel and the turbulent boundary layer. Such differences were also observed in the plots of the fluctuations in the streamwise and wall-normal directions shown in figure~\ref{fig_mean_vel_profile}. The peak of the Reynolds shear stress is higher for the turbulent boundary layer compared to a channel flow, which indicates a higher momentum transfer by the fluctuating velocity field in the turbulent boundary layer. Looking at the energy budgets for shear stress component in both the channel and turbulent boundary layer corresponding to $Re_\tau = 395$ (plots not shown here), we find that there is higher production in the turbulent boundary layer compared to channel flows (with uniform heat flux wall conditions) in the overlap region whereas the dissipation was of similar magnitude.

\begin{figure}
\centering
\resizebox*{0.7\linewidth}{!}{\includegraphics{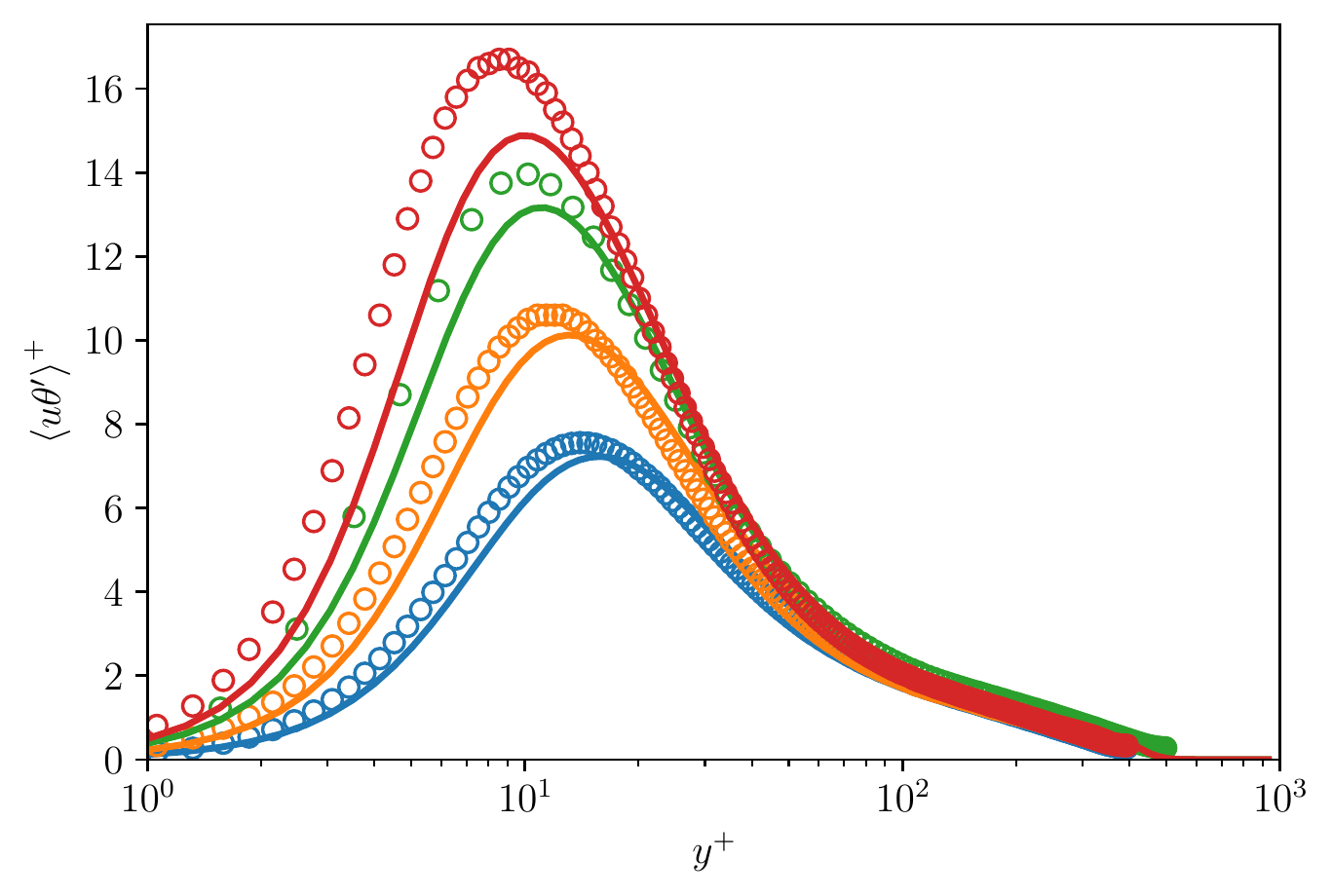}}
\caption[]{Variation of streamwise heat flux along the wall-normal direction at $Re_\theta = 1070$ for different passive scalars. Present DNS data for {\plotA}~$Pr = 1$, {\plotB}~$Pr = 2$, {\plotC}~$Pr = 4$, {\plotD}~$Pr = 6$, channel DNS data by~\cite{kozuka} at $Re_{\tau} = 395$ and {\plotAmarker}~$Pr=1$, {\plotBmarker}~$Pr=2$, {\plotDmarker}~$Pr=7$, {\plotCmarker}~channel DNS data by~\cite{alcantra18} at $Re_{\tau} = 500$ and $Pr = 4$.} \label{fig_u_heat_flux}
\end{figure}

\begin{figure}
\centering
\resizebox*{0.7\linewidth}{!}{\includegraphics{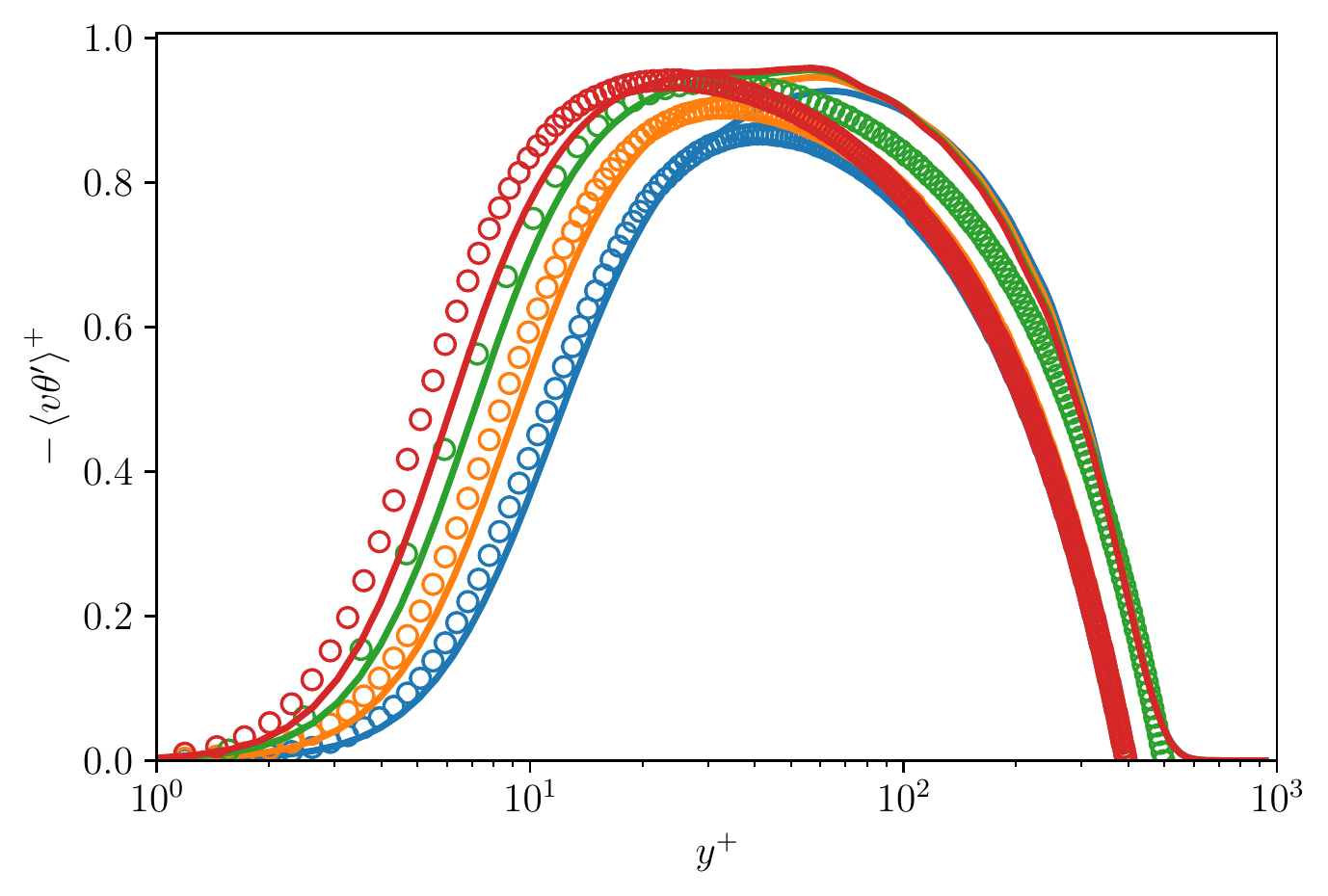}}
\caption[]{Variation of wall-normal heat flux along the wall-normal direction at $Re_\theta = 1070$ for different passive scalars. Present DNS data for {\plotA}~$Pr = 1$, {\plotB}~$Pr = 2$, {\plotC}~$Pr = 4$, {\plotD}~$Pr = 6$, channel DNS data by~\cite{kozuka} at $Re_{\tau} = 395$ and {\plotAmarker}~$Pr=1$, {\plotBmarker}~$Pr=2$, {\plotDmarker}~$Pr=7$, {\plotCmarker}~channel DNS data by~\cite{alcantra18} at $Re_{\tau} = 500$ and $Pr = 4$.} \label{fig_v_heat_flux}
\end{figure}

The turbulent streamwise heat flux shows a reasonable agreement with the channel data available at $Pr = 1, 2$, as shown in figure~\ref{fig_u_heat_flux} although the peaks for the channel-flow data are slightly higher and closer to the wall compared to the present observations. For $Pr = 4$, however, the peak of the streamwise heat flux is higher in the channel data by~\cite{alcantra18}, since the heat flux is reported at a higher Reynolds number than our DNS. Furthermore, the comparison at the highest $Pr$ in our simulation was at a slightly lower $Pr$ than that in the channel and hence there is a difference in the slope of the streamwise heat flux close to the wall.
Figure~\ref{fig_v_heat_flux} shows the wall-normal heat flux for the different simulations. A clear difference can be observed in the outer region when comparing similar Prandtl numbers. On the other hand, the boundary-layer data and the channel data show a better agreement in the inner region.

\begin{figure}
\centering
\resizebox*{0.45\linewidth}{!}{\includegraphics{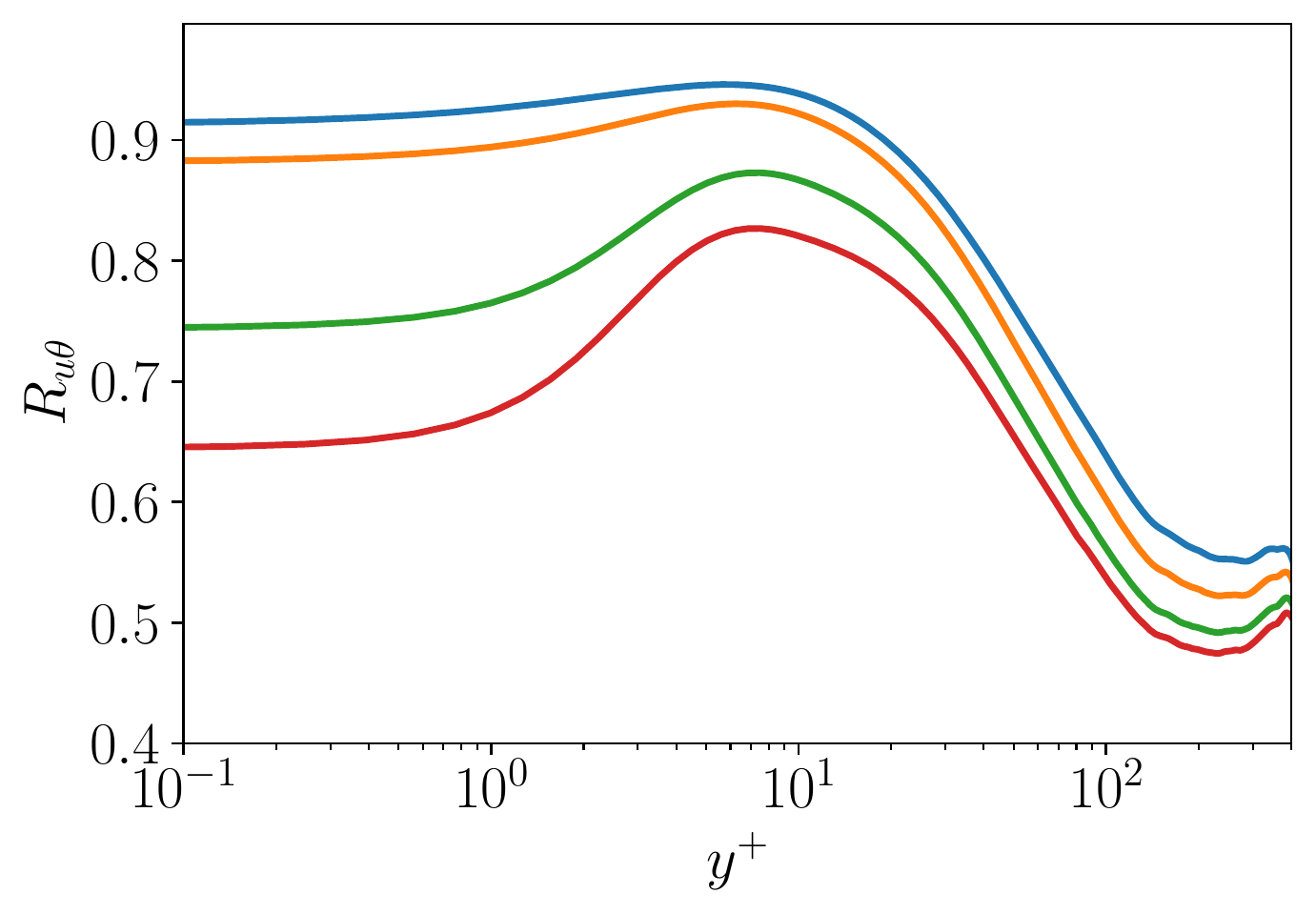}}
\resizebox*{0.45\linewidth}{!}{\includegraphics{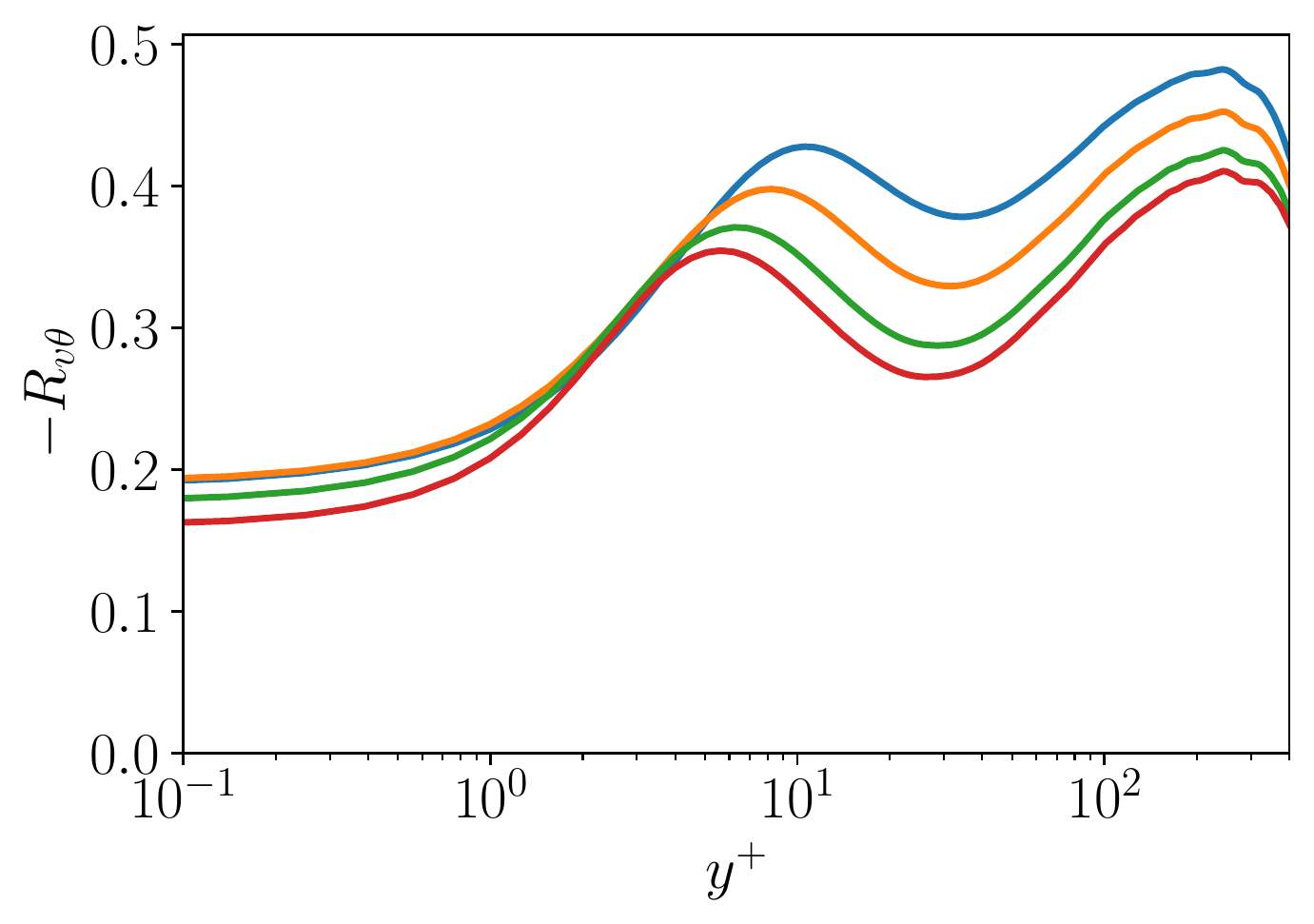}}
\caption[]{Correlation coefficients: (left) $u-\theta$, (right) $v-\theta$. Present DNS data for scalars at {\plotA}~$Pr = 1$, {\plotB}~$Pr = 2$, {\plotC}~$Pr = 4$, {\plotD}~$Pr = 6$ at $Re_\theta = 1070$.} \label{fig_Rutheta}
\end{figure}

The correlation coefficients provide more information on the statistical association between the fields, and here the structure of the flow field and scalar fluctuations are analyzed in terms of
\begin{equation}
\begin{aligned}
    R_{u\theta} &= \frac{\left<u \theta^{\prime}\right>}{u_{\rm{RMS}}\theta^{\prime}_{\rm{RMS}}}\,,\\
    R_{v\theta} &= \frac{\left<v \theta^{\prime}\right>}{v_{\rm{RMS}}\theta^{\prime}_{\rm{RMS}}}\,.
\end{aligned}
\end{equation}
The correlation-coefficient plot in figure~\ref{fig_Rutheta} corresponding to $u-\theta$ shows a strong correlation of streamwise velocity and scalar fluctuations at $Pr = 1$ and it decreases with increase in Prandtl number. This is related to the similarity in the momentum and passive-scalar transport by the turbulent eddies close to the wall. On the other hand, the $v-\theta$ correlation coefficient also shown in figure~\ref{fig_Rutheta} exhibits an increasing trend with the Prandtl number. In the conductive sub-layer the correlation coefficients coincide for the various Prandtl numbers under study and then approach different values at the wall. This highlights a similar behaviour of the turbulent wall-normal momentum and passive-scalar fields with a caveat, {\it i.e. the differences are  present} very close to the wall for different Prandtl numbers.

\begin{figure}
\centering
\resizebox*{0.7\linewidth}{!}{\includegraphics{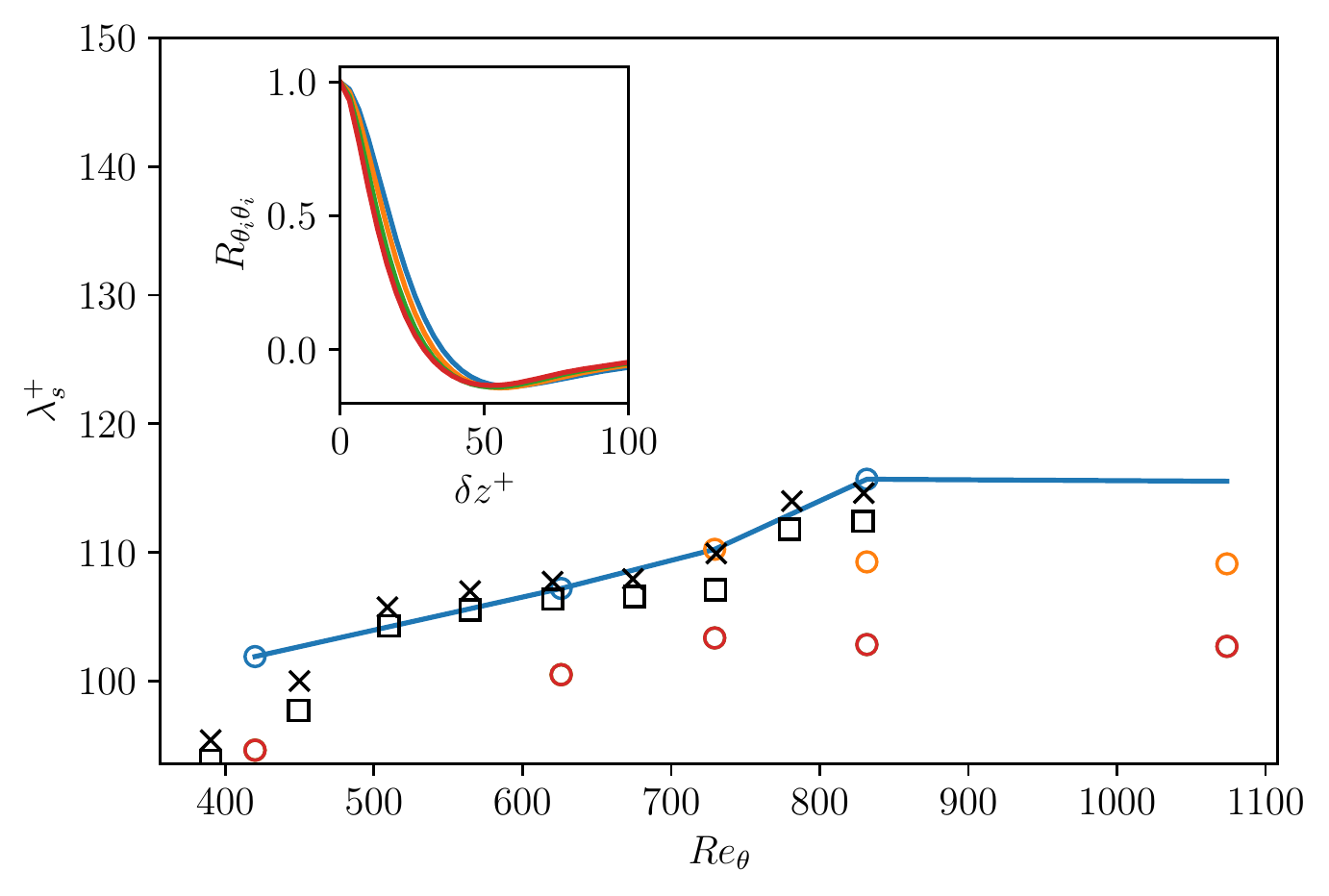}}
\caption[]{Variation of mean spanwise streak spacing along the streamwise direction at $y^+ \approx 7$. {\plotA}~$U$, {\plotAmarker}~$\theta_1$, {\plotBmarker}~$\theta_2$, {\plotCmarker}~$\theta_3$, {\plotDmarker}~$\theta_4$. Comparison data from~\cite{qiang} for~{\plotkmarkercr}~$U$, {\plotkmarkersq}~$\theta_2$. (Inset)~Spanwise two-point correlation for {\plotA}~$\theta_1$, {\plotB}~$\theta_2$, {\plotC}~$\theta_3$, {\plotD}~$\theta_4$ at $Re_\theta \approx 830$.}
\label{fig_Rplot}
\end{figure}

\subsection{Spanwise two-point correlations} \label{sec_Rplot}
The two-point correlations provide some quantitative information of the turbulent structures near the wall. For example, the streak spacing near the wall is of interest and can also be observed in an experimental setting. In order to identify the mean streak spacing, spanwise two-point correlations of the velocity components and passive scalars were obtained at five different positions along the streamwise direction at different wall-normal positions. Overall, the obtained results at $Re_\theta = 830$ were compared with the data reported by~\cite{kim87} and shows good agreement (not shown here). The obtained two-point correlations for different scalars at $Re_\theta = 830$ are shown in figure~\ref{fig_Rplot} to assess the differences for varying $Pr$. The two-point correlation becomes negative and reaches a minimum at an inner-scaled correlation length of $\delta z^+ \approx 50$. The length at which the minimum occurs provides an estimate of the half-mean separation between the streaks in the spanwise direction, {\it i.e.} $\left(\lambda_s^+/2\right)$. The streak spacing is plotted along the streamwise direction at a wall-normal position of $y^+ \approx 7$ as shown in figure~\ref{fig_Rplot}. Overall, the velocity and scalar streak spacing increase with $Re_\theta$ as reported in the works of~\cite{qiang} and~\cite{schlatter09}. Note that the correlations are available only at five streamwise locations, but the comparison with~\cite{qiang} is in reasonably good agreement for the velocity and scalar streaks at $Pr = 2$. The velocity-streak spacing increases from 102 to 115 and appears to saturate for $Re_\theta > 830$. Note that such a saturation of streak spacing was also pointed out by~\cite{qiang} for $Re_\theta > 1500$. From figure~\ref{fig_Rplot}, we also observe that the streak spacing decreases with increasing $Pr$ and that the streak spacings for the scalars at $Pr = 4, 6$ are indistinguishable, although the rate of decay of the two-point correlations for the scalars is different. A higher grid resolution might be necessary to quantify the possible differences in the scalar-streak spacing at higher $Pr$.

\subsection{Scaling of wall-heat-flux fields}
\begin{figure}
\centering
\resizebox*{0.95\linewidth}{!}{\includegraphics{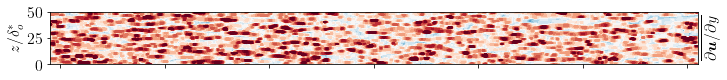}}
\resizebox*{0.95\linewidth}{!}{\includegraphics{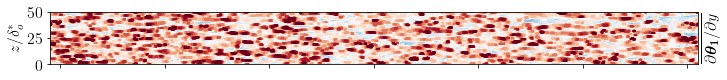}}
\resizebox*{0.95\linewidth}{!}{\includegraphics{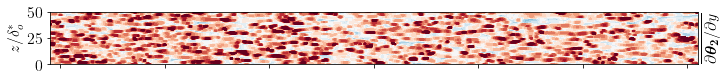}}
\resizebox*{0.95\linewidth}{!}{\includegraphics{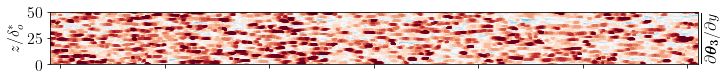}}
\resizebox*{0.95\linewidth}{!}{\includegraphics{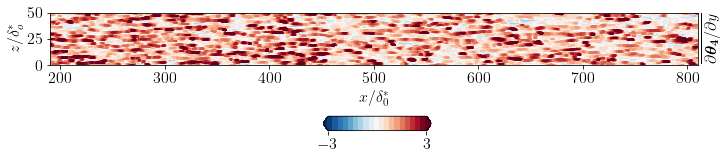}}
\caption[]{Instantaneous normalized streamwise wall-shear and heat-flux fields at different Prandtl numbers.}
\label{fig_Nu_field_comp}
\end{figure}

\begin{figure}
\centering
\resizebox*{0.45\linewidth}{!}{\includegraphics{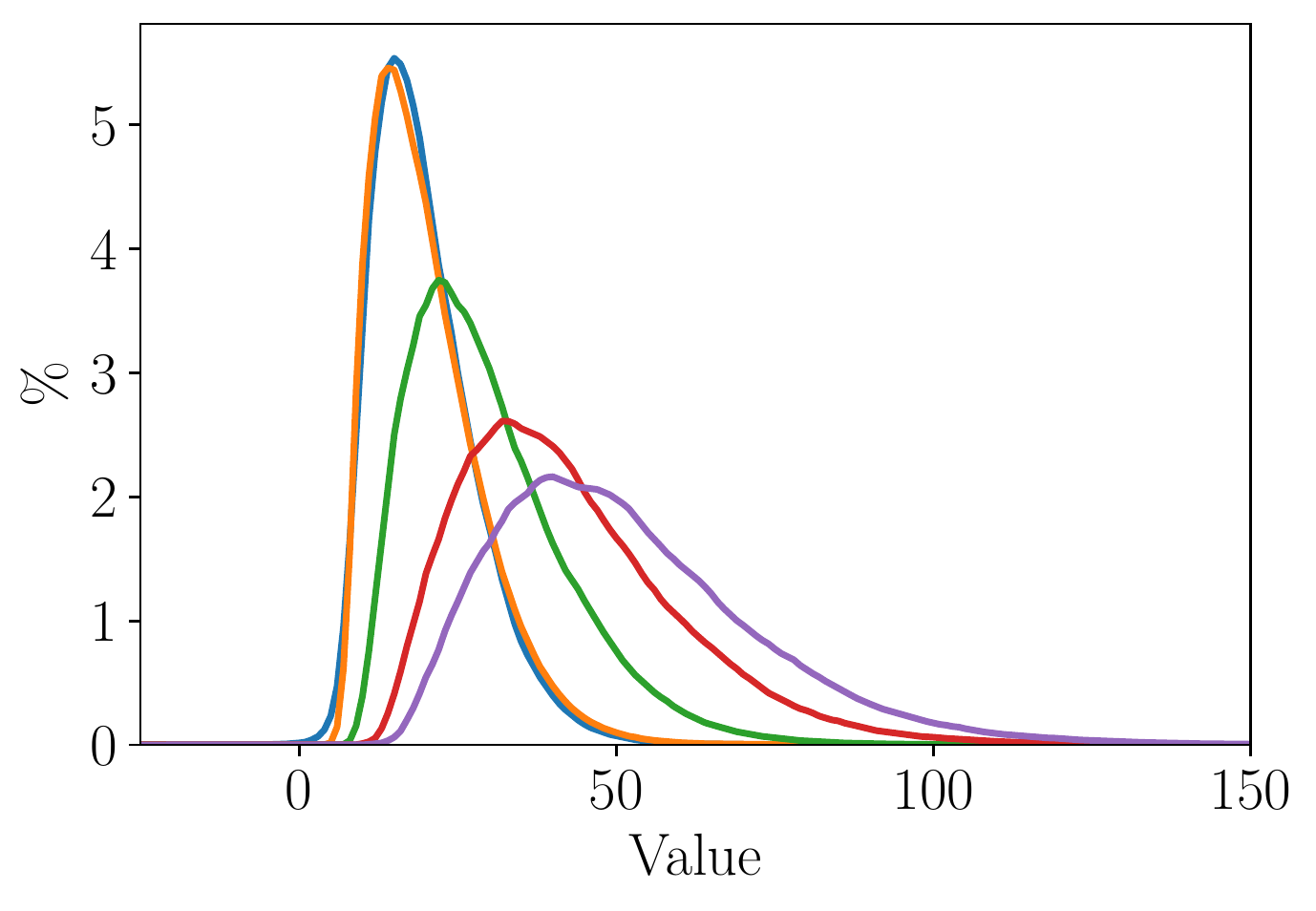}}
\resizebox*{0.45\linewidth}{!}{\includegraphics{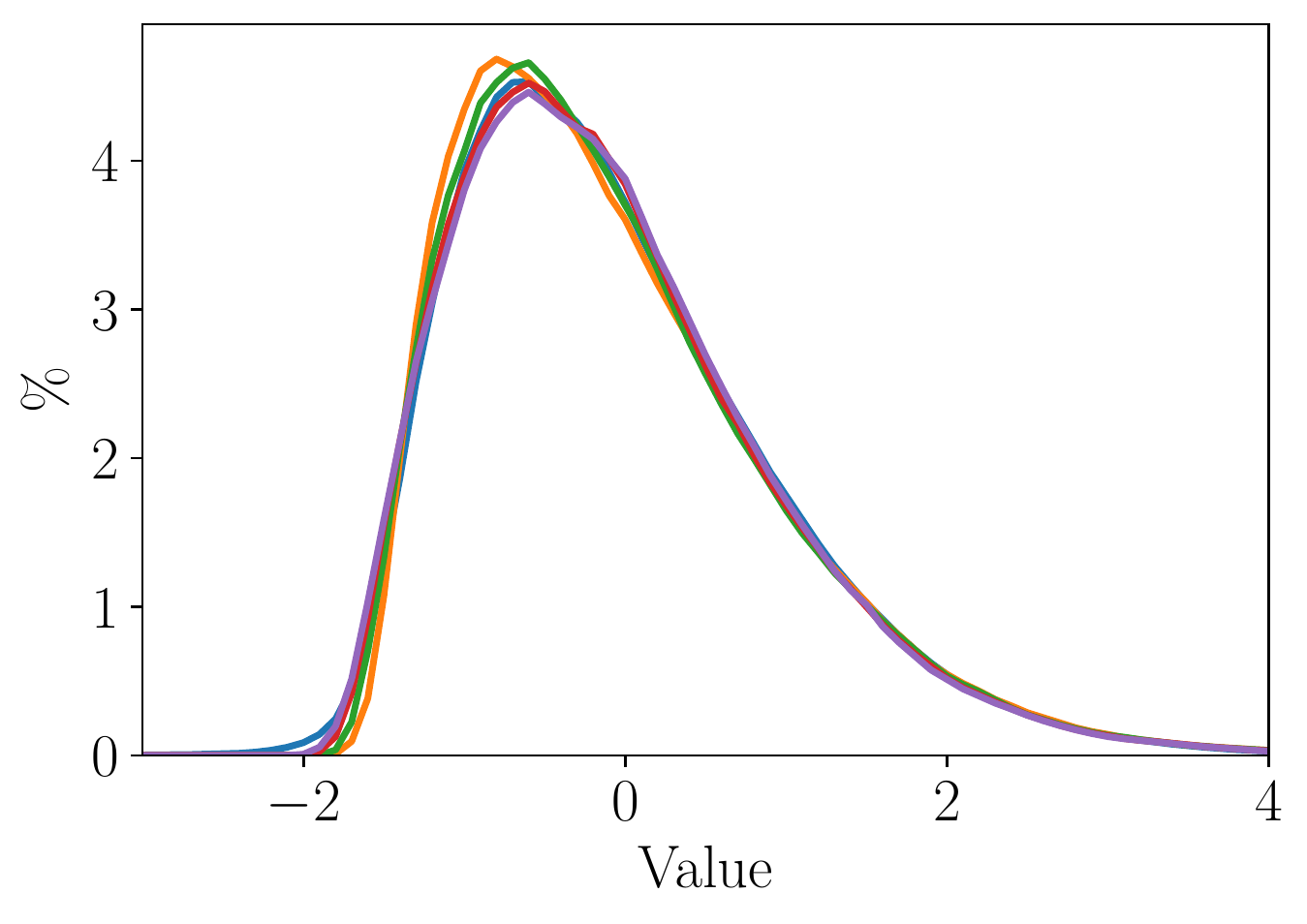}}
\caption[]{Probability density function of wall-shear and heat-flux fields at different Prandtl numbers. (Left) distribution of absolute data in the fields with a bin size of 1 unit, (right) distribution of the normalized data in the fields with a bin size of 0.1 unit, {\plotA}~${\partial u}/{\partial y}$ or $\overline{{\partial u}/{\partial y}}$, {\plotB}~${\partial \theta_{1}}/{\partial y}$ or $\overline{{\partial \theta_{1}}/{\partial y}}$, {\plotC}~${{\partial \theta_{2}}/{\partial y}}$ or $\overline{{\partial \theta_{2}}/{\partial y}}$, {\plotD}~${{\partial \theta_{3}}/{\partial y}}$ or $\overline{{\partial \theta_{3}}/{\partial y}}$, {\plotE}~${{\partial \theta_{4}}/{\partial y}}$ or $\overline{{\partial \theta_{4}}/{\partial y}}$.}
\label{fig_Nu_distribution}
\end{figure}

The wall-shear and heat-flux fields at different Prandtl numbers are normalized by subtracting the mean and dividing by the corresponding RMS quantities. The normalization of a quantity $n$ is calculated as:
\begin{align}
    \overline{n} = \frac{n-\left<n\right>}{n_{\rm RMS}}\,.
\end{align}
Figure~\ref{fig_Nu_field_comp} shows the instantaneous normalized streamwise wall-shear and heat-flux fields at different Prandtl numbers. The normalized fields appear qualitatively similar and this result is confirmed by the distribution of the data in the streamwise shear and heat flux fields obtained from 3,700 samples, shown in figure~\ref{fig_Nu_distribution}. Though the distribution of data is different in the various fields, after normalization the distributions becomes identical. This indicates the uniformity in the distribution of the fluctuations of shear and heat flux when scaled with the corresponding RMS quantities. This observation is useful for certain applications, for instance in the prediction of fluctuating flow quantities from the wall, as discussed in the study by~\cite{luca_tsfp}.

%% file: sec4_PSD.tex
\section{Spectral analysis}\label{sec_4}

The analysis of thermal-boundary-layer statistics reveals that the scalar fluctuations and heat flux are strongly affected by the Prandtl number of the scalar in the flow and the corresponding scalings were reported in~$\S$\ref{sec_3}. Additional insight can be obtained by analyzing the energy distribution at the different lengthscales for the scalars at the simulated Prandtl numbers. In this regard, time series of the wall-shear, wall-heat flux, streamwise velocity and different scalars were sampled at different wall-normal locations with a sampling time ($\sim$$\Delta t^+_s = 1$ with the reference friction velocity at $x/\delta_0^* = 500$) corresponding to 12,000 time units $\left({\delta_{0}^{*}}/{U_{\infty}}\right)$. The two-dimensional (2D) premultiplied power-spectral density (PSD) $k_z k_t \phi\left(\lambda_z^+,\lambda_t^+\right)$ is obtained in the spanwise direction and in time based on the time series, for a total sampled time of about 11.5 flow-through times. Note that here $k$ denotes wavenumber and $\phi$ is the power-spectral density defined for the particular quantity under study.

\begin{figure}
\centering
\resizebox*{0.349375\linewidth}{!}{\includegraphics{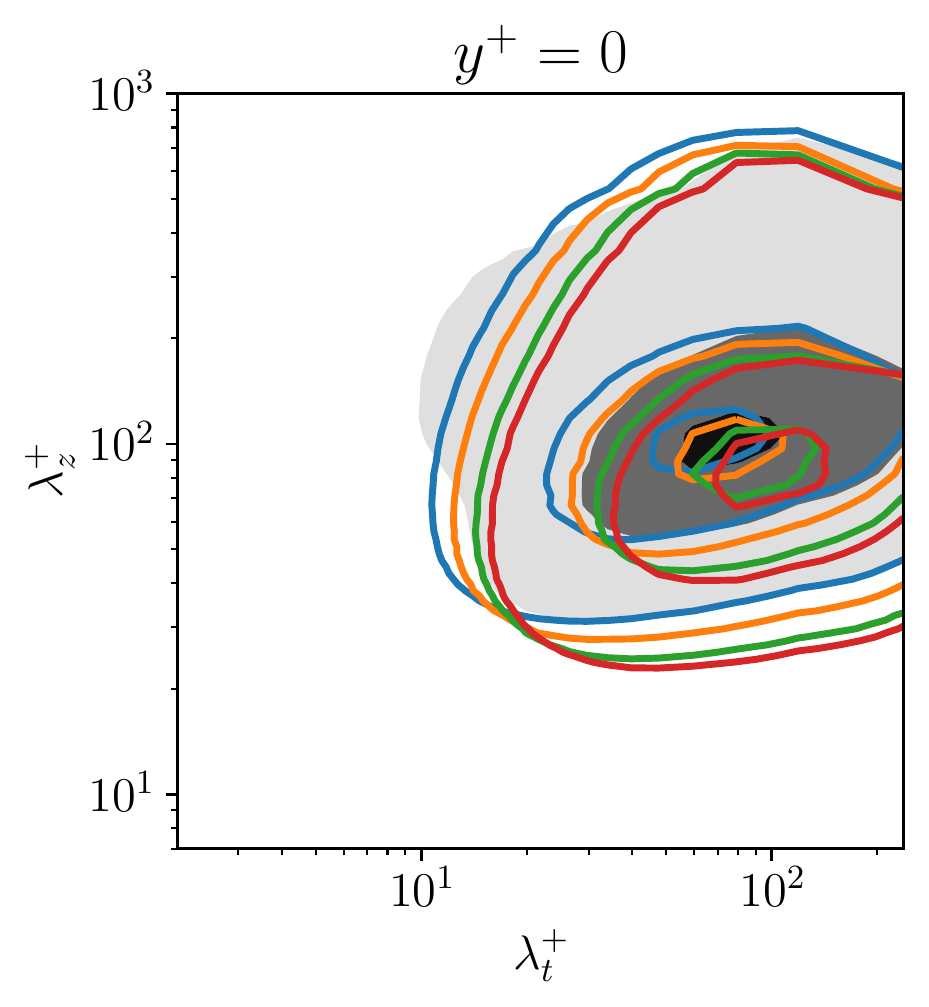}}
\resizebox*{0.325\linewidth}{!}{\includegraphics{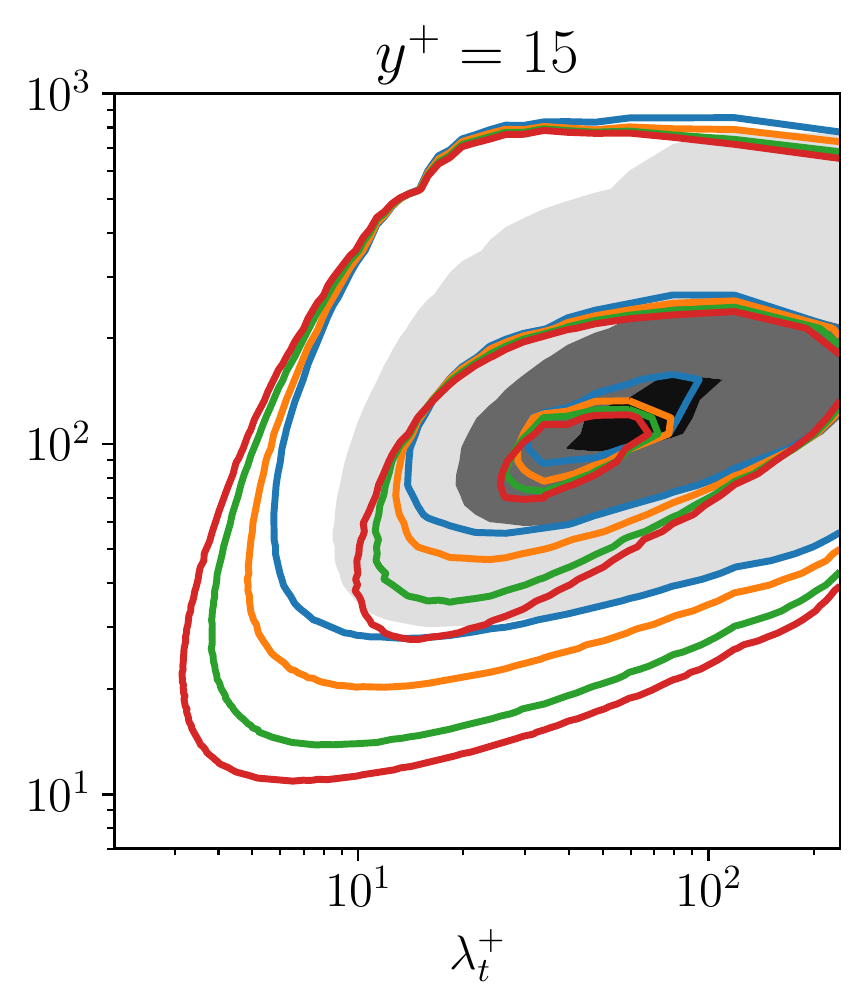}}
\caption[]{Two-dimensional premultiplied power-spectral density of (left) streamwise wall-shear and wall-heat-flux fields and (right) streamwise velocity and scalar at $y^+ = 15$, for the ZPG thermal TBL with $Re_\theta$ ranging from 467 to 1072. The three contour lines correspond to 10\%, 50\% and 90\% of the maximum energy density. We show: $k_t k_z \phi_{u_y u_y}$ or $k_t k_z \phi_{u u}$~(shaded contours), {\plotA}~$k_t k_z \phi_{{\theta_1}_y {\theta_1}_y}$ or $k_t k_z \phi_{{\theta_1} {\theta_1}}$, {\plotB}~$k_t k_z \phi_{{\theta_2}_y {\theta_2}_y}$ or $k_t k_z \phi_{{\theta_2} {\theta_2}}$, {\plotC}~$k_t k_z \phi_{{\theta_3}_y {\theta_3}_y}$ or $k_t k_z \phi_{{\theta_3} {\theta_3}}$, {\plotD}~$k_t k_z \phi_{{\theta_4}_y {\theta_4}_y}$ or $k_t k_z \phi_{{\theta_4} {\theta_4}}$, correspondingly. Here, the spatial coordinate in subscript denotes partial derivative with respect to that coordinate.}
\label{sec4_pdf_1}
\end{figure}

For the calculation of power-spectral density, the procedure outlined in~\cite{ramon} is utilized. After the mean subtraction of the sampled time series to obtain  the turbulent quantities, first a one-dimensional spectrum $E\left[t,x\right]\left(\lambda_z^+\right)$ is obtained by performing a fast Fourier transform (FFT) in the spanwise direction, due to the periodicity condition imposed along $z$. As a result, a spectral decomposition of the energy content into different wavenumbers $k_z$ is obtained with the corresponding wavelengths $\lambda_z$ given by $2\pi/k_z$. Note that the local friction velocity is used to obtain the inner-scaled quantities. It should be pointed out that the flow is developing along the streamwise direction and hence the FFT along the streamwise direction is not applicable. In the present spectral analysis, we consider the $Re_{\theta}$ range between 470 and 1070 and use Welch's overlapping-window method to address the non-periodicity in the temporal signal. As a next step, the spectrum in time is obtained using Welch's method with 15 bins in total, where 8 of them are independent. A Hamming window is used for imposing the periodicity in the bins, and the 2D spectrum is obtained as $E\left[x\right]\left(\lambda_z^+,\lambda_t^+\right)$ by using FFT along $z$ and Welch method in $t$. The obtained spectrum is divided by $\Delta k_t \Delta k_z$ and premultiplied with $k_t k_z$. Finally, an averaging along $x$ is performed to yield the 2D premultiplied power-spectral density $k_t k_z \phi\left(\lambda_z^+,\lambda_t^+\right)$.
\begin{figure}
\centering
\resizebox*{0.349375\linewidth}{!}{\includegraphics{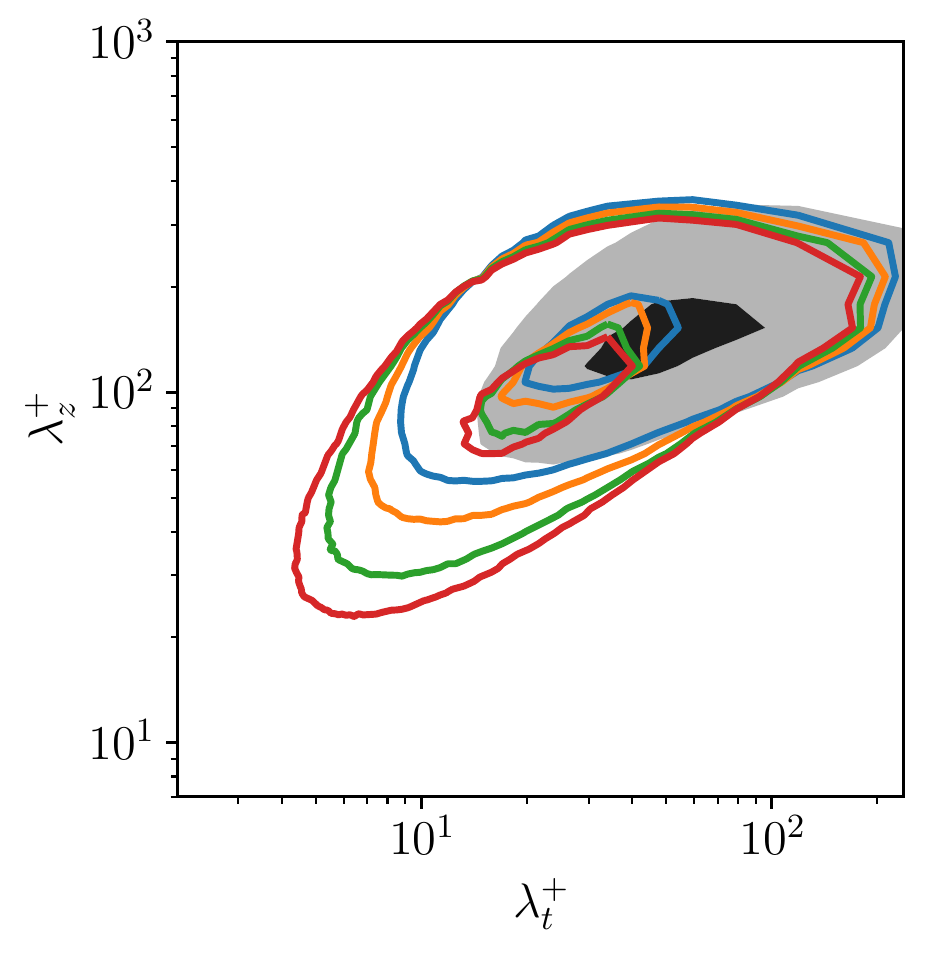}}
\resizebox*{0.325\linewidth}{!}{\includegraphics{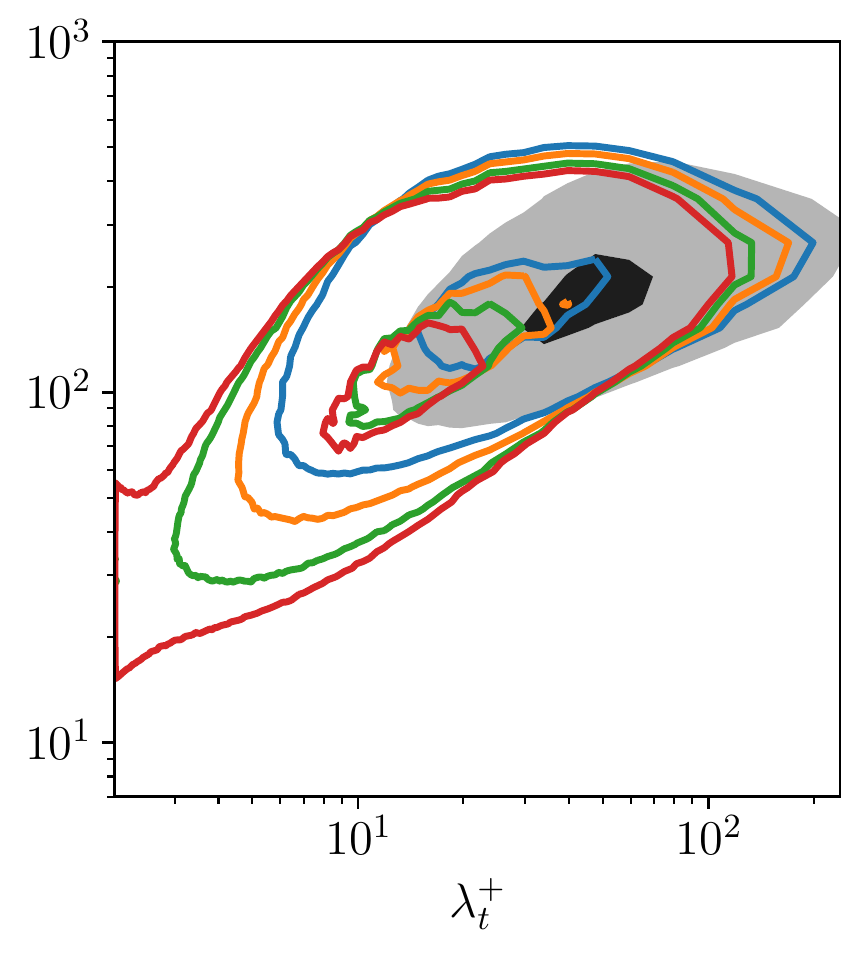}}
\caption[]{Two-dimensional premultiplied power-spectral density of streamwise velocity and scalar fields at (left)~$y^+=30$ and (right)~$y^+=50$. The two contour lines correspond to 10\% and 50\% of the maximum energy density. We show: $k_t k_z \phi_{u u}$~(shaded contours), {\plotA}~$k_t k_z \phi_{{\theta_1} {\theta_1}}$, {\plotB}~$k_t k_z \phi_{{\theta_2} {\theta_2}}$, {\plotC}~$k_t k_z \phi_{{\theta_3} {\theta_3}}$, {\plotD}~$k_t k_z \phi_{{\theta_4} {\theta_4}}$.}
\label{sec4_pdf_2}
\end{figure}

The 2D power-spectral densities of the streamwise wall-shear and wall-heat-flux at different Prandtl numbers are provided in figure~\ref{sec4_pdf_1}, along with the power-spectral density of the streamwise velocity and scalar fluctuations at $y^+ = 15$. The  obtained 2D power-spectral density at $y^+=15$ agrees well with the results reported by~\cite{ramon}, although the latter are at higher Reynolds number. At the wall, the spectral peak is observed at $\lambda_z^+ \approx 100$ and $\lambda_t^+ \approx 100$. Furthermore, at $y^+ = 15$, we observe the maximum of spectral-energy distribution is at $\lambda_z^+ \approx 120$, which corresponds to the characteristic streak spacing in wall turbulence~\citep{smith83n}. It is observed that the power-spectral density for the streamwise wall-shear stress is very similar to that for the wall-heat-flux at $Pr=1$, which is an expected result for the reasons outlined in~$\S$\ref{ReyAnaly}. However, with increasing $Pr$, the power-spectral density shifts to the right, indicating that the energy is not spread over a wider range of scales, and instead is concentrated on longer temporal structures. One could argue that, with a shorter boundary layer, the structures at $Pr=6$ can become larger than the one we develop at $Pr=1$. Because of this, the larger structures have a different footprint at the wall. Additionally, the plots in the figure~\ref{sec4_pdf_1} also exhibit a slight trend downwards for higher $Pr$, a fact that indicates the presence of smaller spanwise scales, in agreement with the discussion in~$\S$\ref{sec_Rplot}. Overall, the temporal wavelength range at which we have the most energetic structures in the wall-heat-flux fields increases for larger Prandtl numbers. 
From the above observations, considering the dominant energetic structures to be composed of streaks at the wall, the scalar at $Pr=6$ (in general for higher $Pr$) might exhibit longer and thinner scalar streak structures at the wall compared with the case at $Pr = 1$. 

The power-spectral densities calculated at $y^+ = 30$ and $50$ are shown in figure~\ref{sec4_pdf_2}. From this figure, we observe that the similarity in the distribution of energy for the scalar at $Pr = 1$ and the streamwise velocity is lost as we move farther from the wall. In contrast to the observation of large streamwise structures at the wall, we find an increasing concentration of energy in smaller scales as we increase the Prandtl number. Further, the range of scales in which the energy is distributed also increases with $Pr$. Focusing on the most energetic structures, we find that these are concentrated in a region of smaller temporal and spanwise scales with increasing $Pr$ at both $y^+=30$ and $50$. For the scalar at $Pr = 6$, the spectral peak is observed at $\lambda_z^+ \approx 100$ and $\lambda_t^+ \approx 20$ for $y^+ = 30$ and $50$.

%% file: sec5_conclusions.tex
\section{Summary and conclusions} \label{sec_5}
In the present study, a direct numerical simulation of the thermal turbulent boundary layer is performed up to $Re_{\theta}=1080$ with passive scalars at $Pr=1,2,4$ and $6$, which to authors' knowledge is the highest Prandtl number simulated for the thermal boundary layer. 
Various statistical quantities for the flow and scalars were computed and compared against the reported channel and TBL data in the literature. Overall, the statistical quantities are in good agreement with the existing data whenever a comparison is possible. For higher $Pr$, we also observed that the peak of the scalar fluctuations decreases when $Re_\tau$ increases, which is different from the trend reported in the literature. Further, we showed that the variation of the peak in scalar fluctuation has a weak dependence with $Re_\tau$ compared to Prandtl number. Similarly, the peak in the heat flux also exhibits a weak dependence with $Re_\tau$ compared to $Pr$ of the scalar and the heat flux scales as~${\sim}Pr^{0.4}$.

In the present study, we also highlighted the behaviour of the turbulent Prandtl number $Pr_t$, which does not approach a constant value of 1.07 as the wall is approached for higher Prandtl numbers. In addition, we also found the corresponding $Pr_t$ to increase with $Pr$, confirming the findings of~\cite{alcantra21}. Finally, a brief description of the energy distribution in the scales for different $Pr$ at different wall-normal locations is presented by analyzing the 2D pre-multiplied power-spectral density.

The analysis and data provided in this work are expected to serve as a database for the research community to assess the validity of new turbulence models and validate other numerical and experimental results.  

%% file: Appendix.tex
\appendix
\section{Reynolds stress budget}\label{appA}

\begin{figure}
\centering
\resizebox*{0.44\linewidth}{!}{\includegraphics{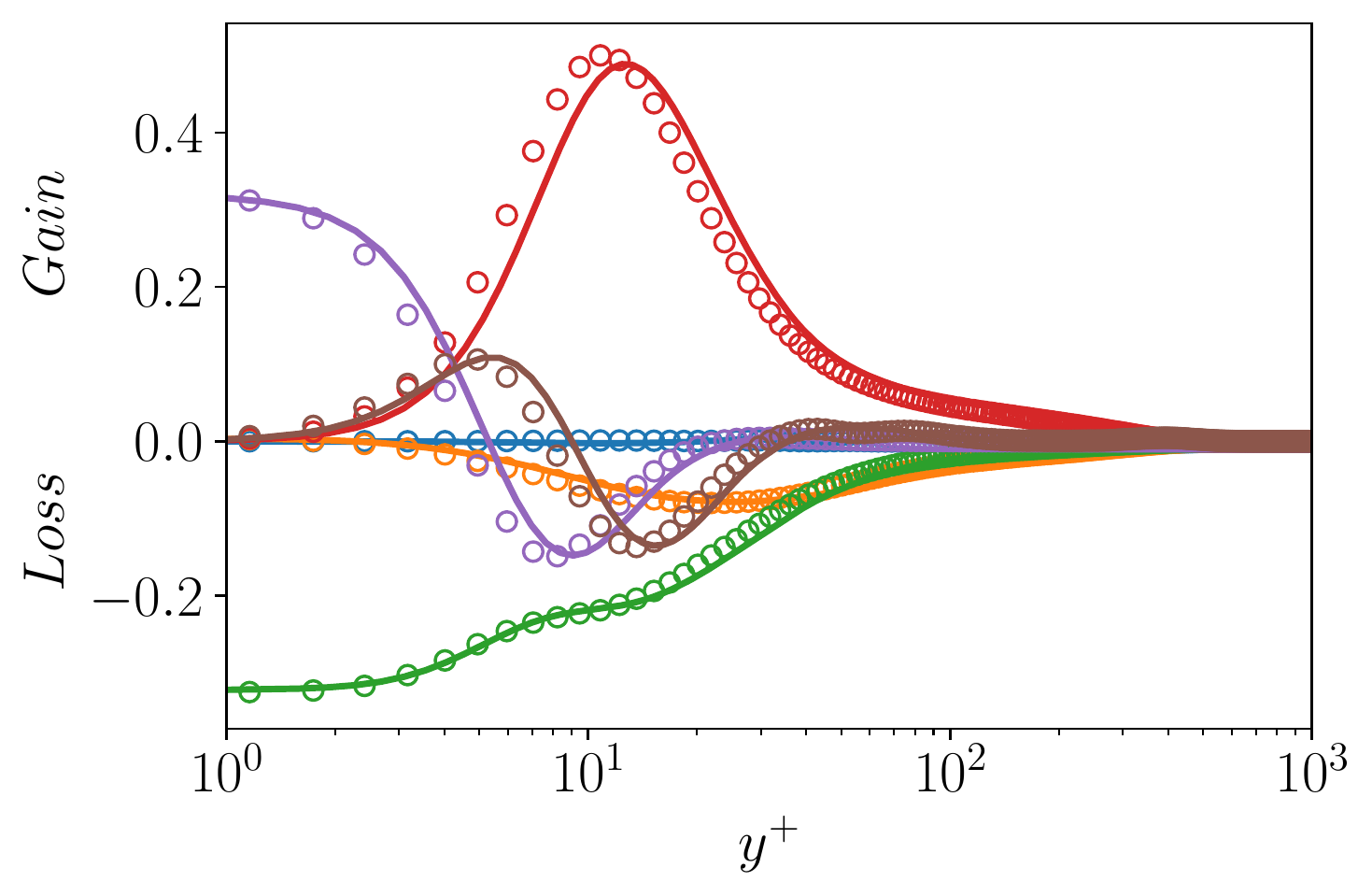}}
\resizebox*{0.45\linewidth}{!}{\includegraphics{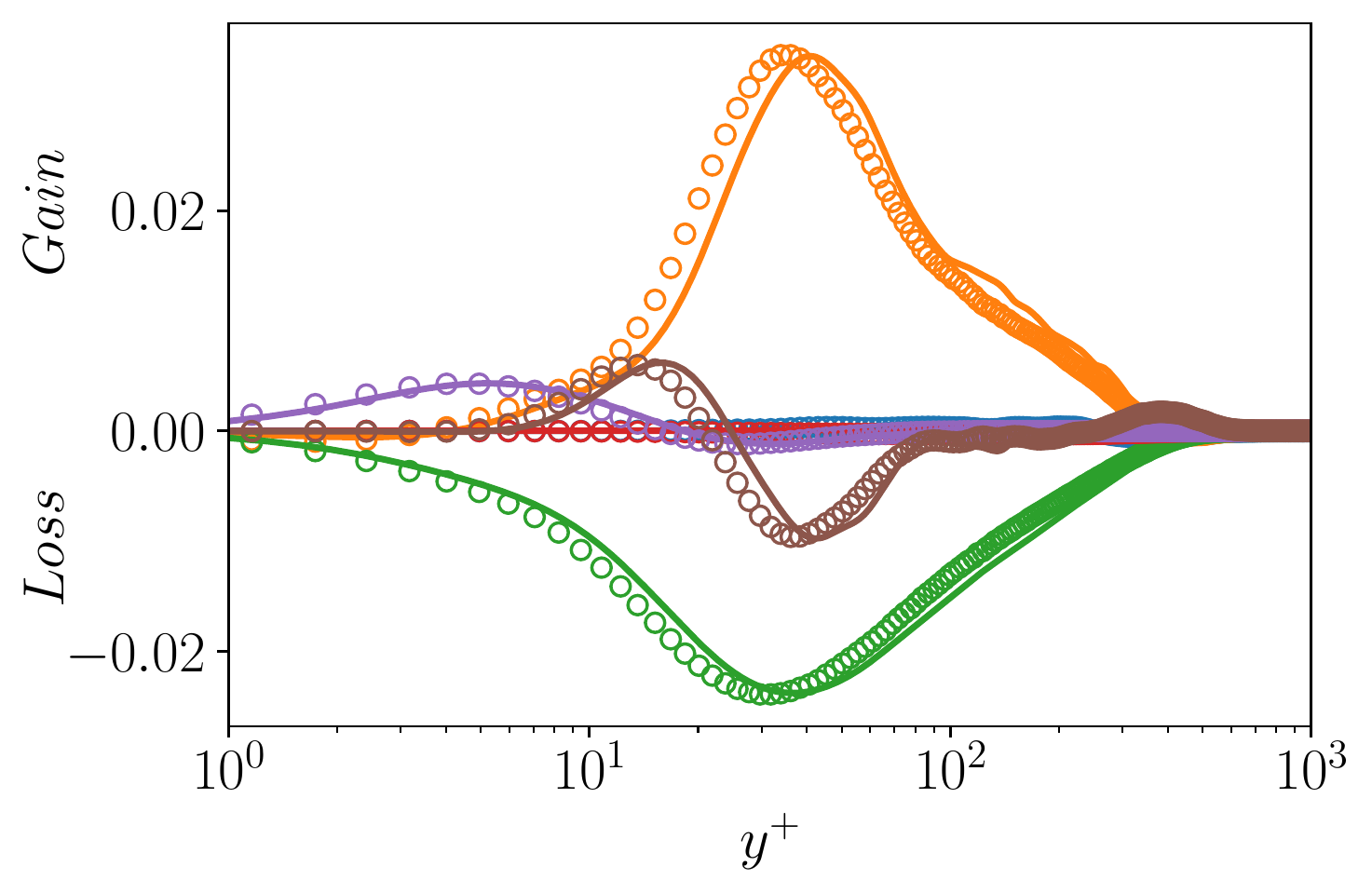}}
\resizebox*{0.44\linewidth}{!}{\includegraphics{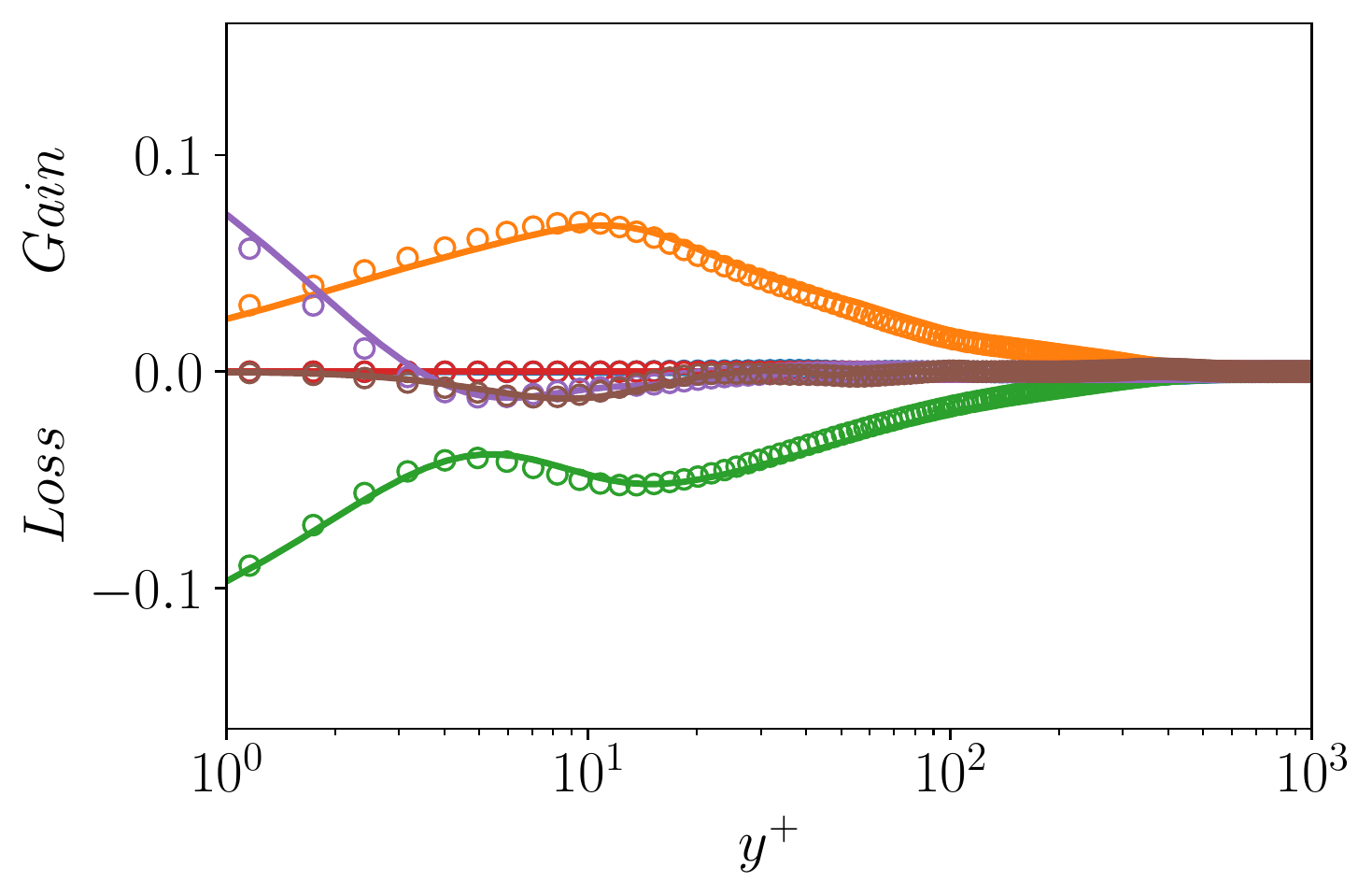}}
\resizebox*{0.45\linewidth}{!}{\includegraphics{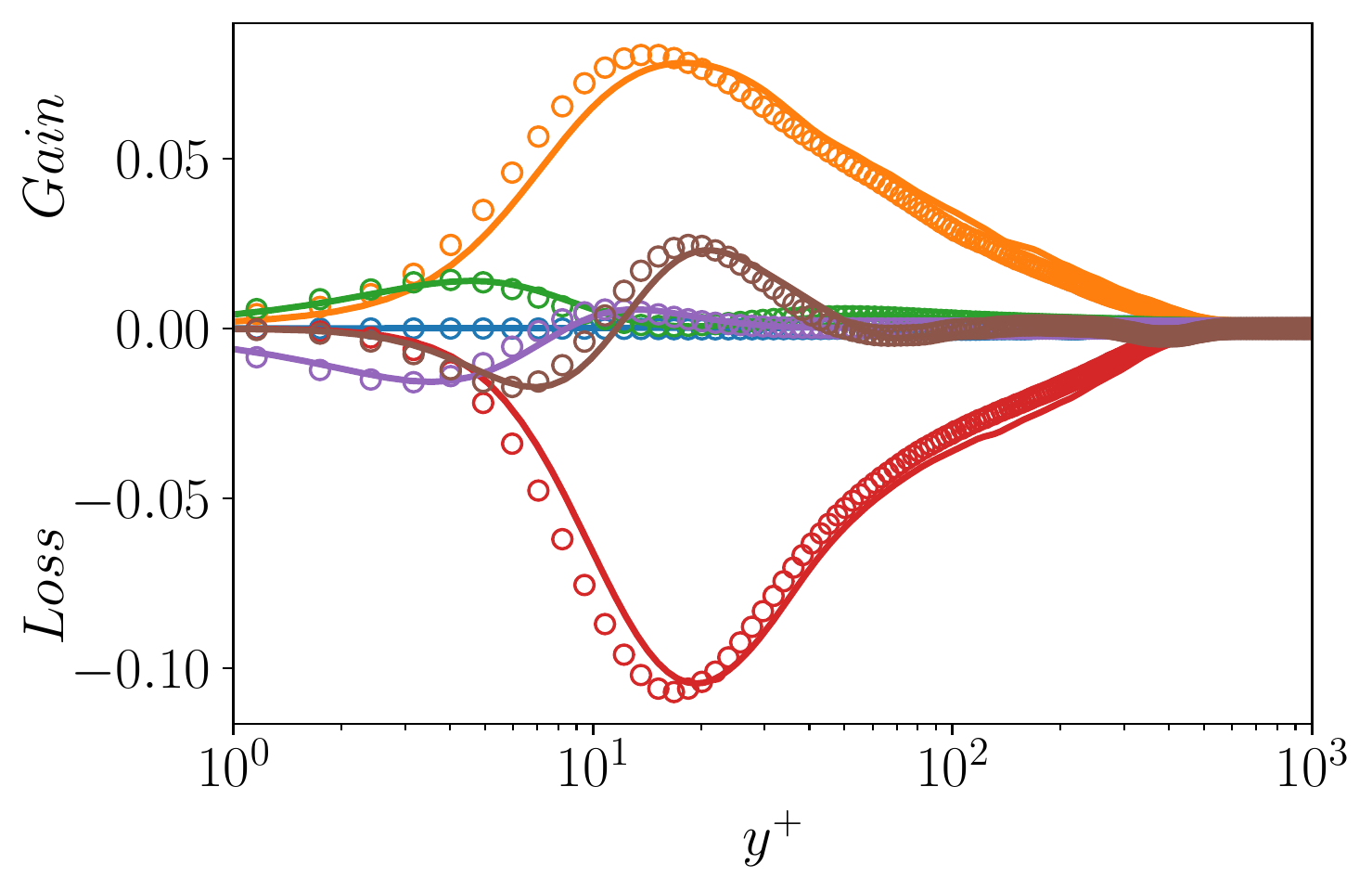}}
\caption[]{Comparison of the Reynolds-stress budgets. (Top-left)~$\left<uu\right>$, (top-right)~$\left<vv\right>$, (bottom-left)~$\left<ww\right>$, (bottom-right)~$\left<uv\right>$, {\plotblacksolid}~present DNS, {\plotblackmarker}~\cite{jiminez2010}, {\plotA}~convection, {\plotB}~velocity-pressure diffusion, {\plotC}~dissipation, {\plotD}~production, {\plotE}~molecular diffusion, {\plotF}~turbulent diffusion.}
\label{fig_vel_budget}
\end{figure}

\begin{figure}
\centering
\resizebox*{0.45\linewidth}{!}{\includegraphics{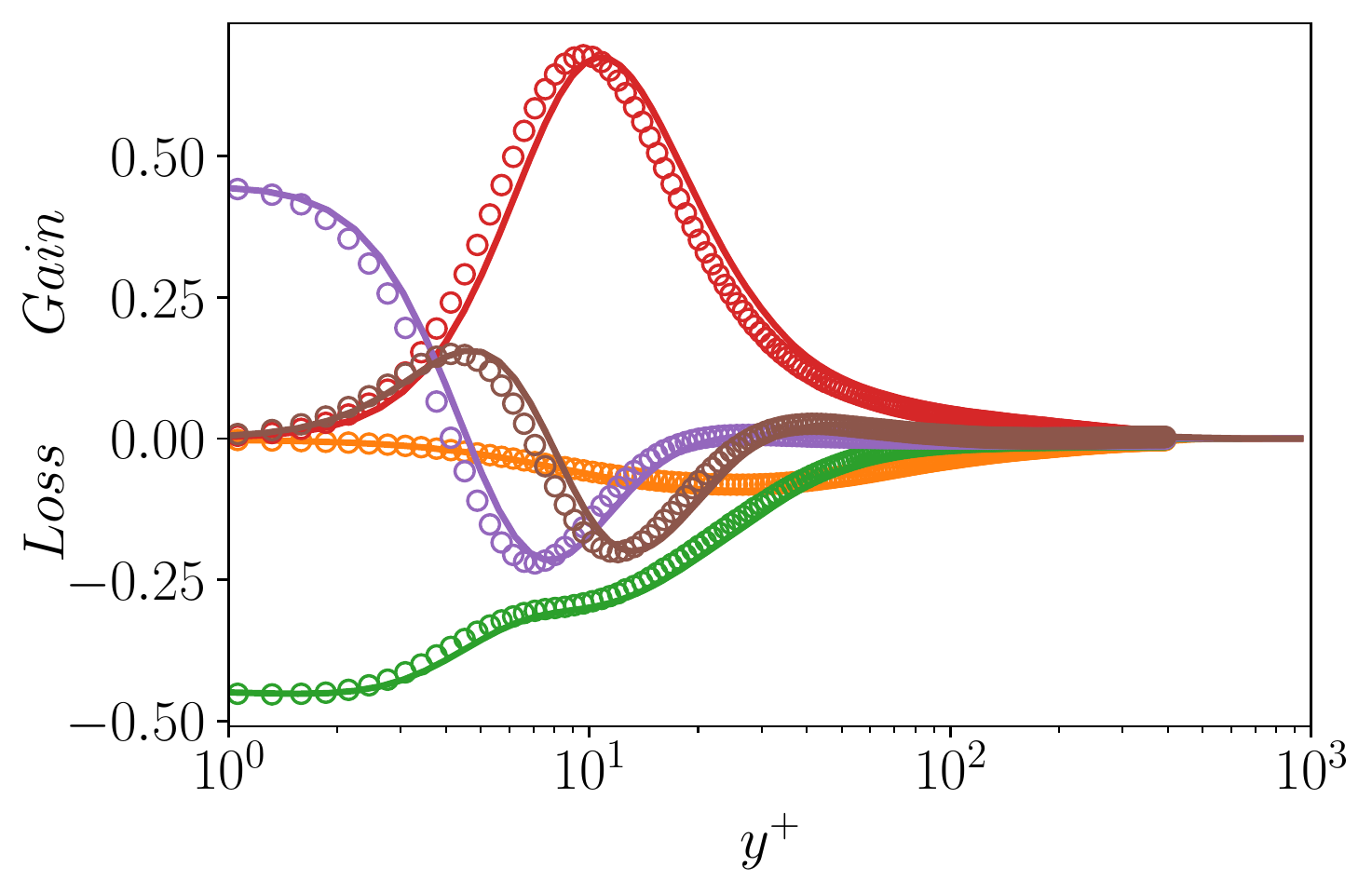}}
\resizebox*{0.45\linewidth}{!}{\includegraphics{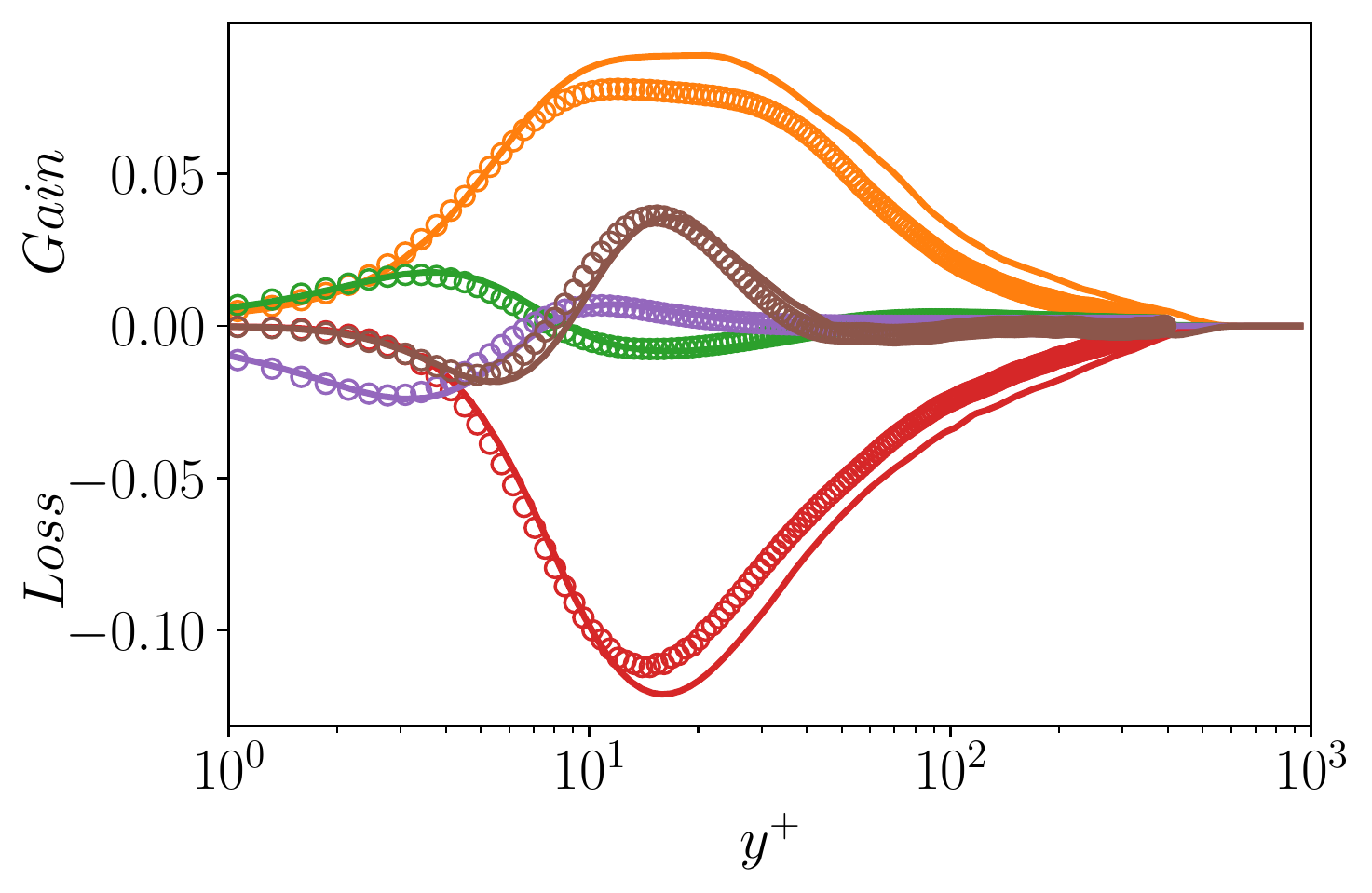}}
\caption[]{Comparison of the scalar-flux budgets. (Left)~$\left<u\theta^{\prime}\right>$, (right)~$\left<v\theta^{\prime}\right>$, {\plotblacksolid}~present DNS at $Pr = 2$, {\plotblackmarker}~\cite{kozuka} at $Pr = 2$, {\plotA}~convection, {\plotB}~scalar-pressure diffusion, {\plotC}~dissipation, {\plotD}~production, {\plotE}~molecular diffusion, {\plotF}~turbulent diffusion.}
\label{fig_s2_budget}
\end{figure}

\begin{figure}
\centering
\resizebox*{0.44\linewidth}{!}{\includegraphics{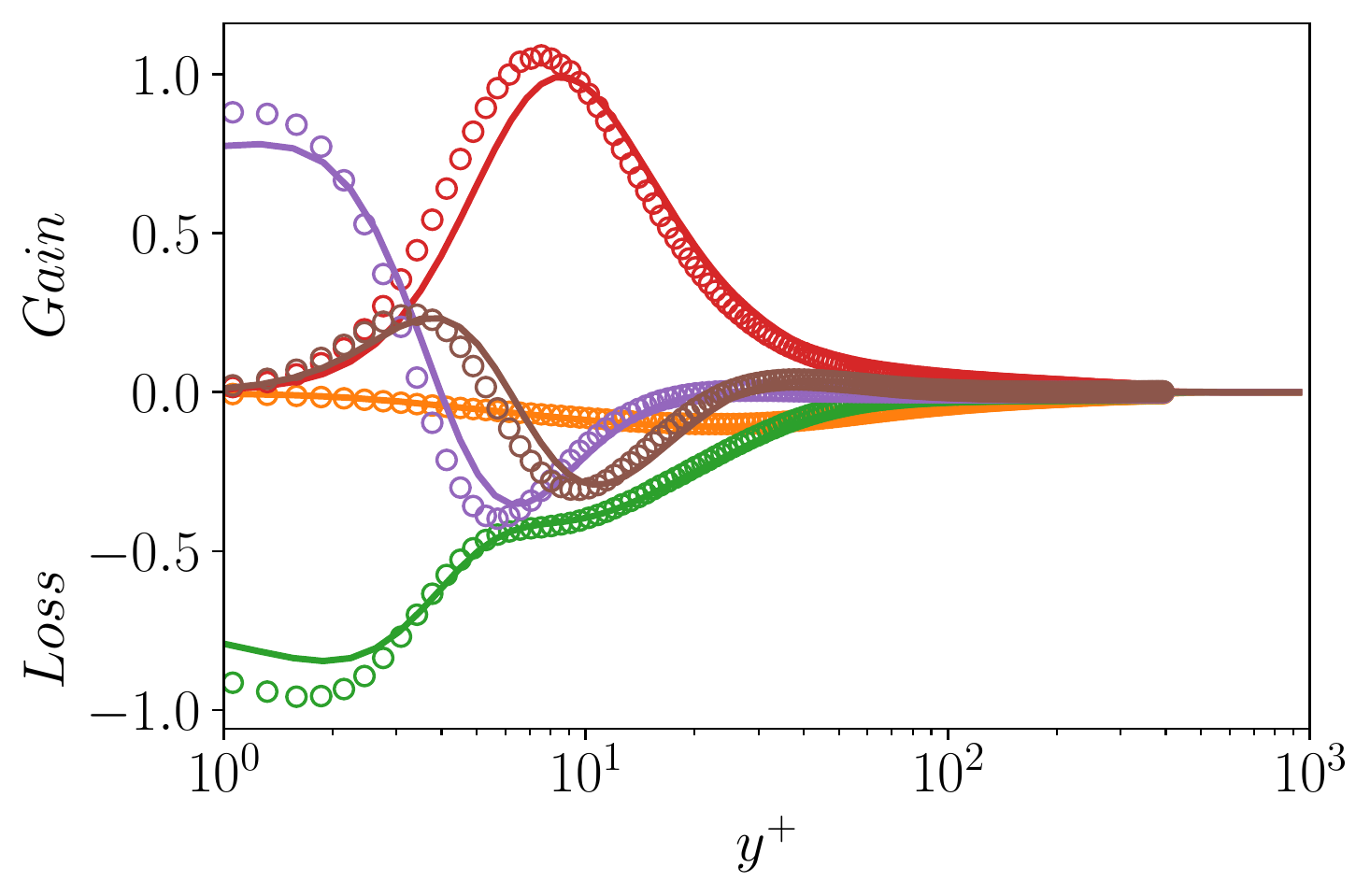}}
\resizebox*{0.45\linewidth}{!}{\includegraphics{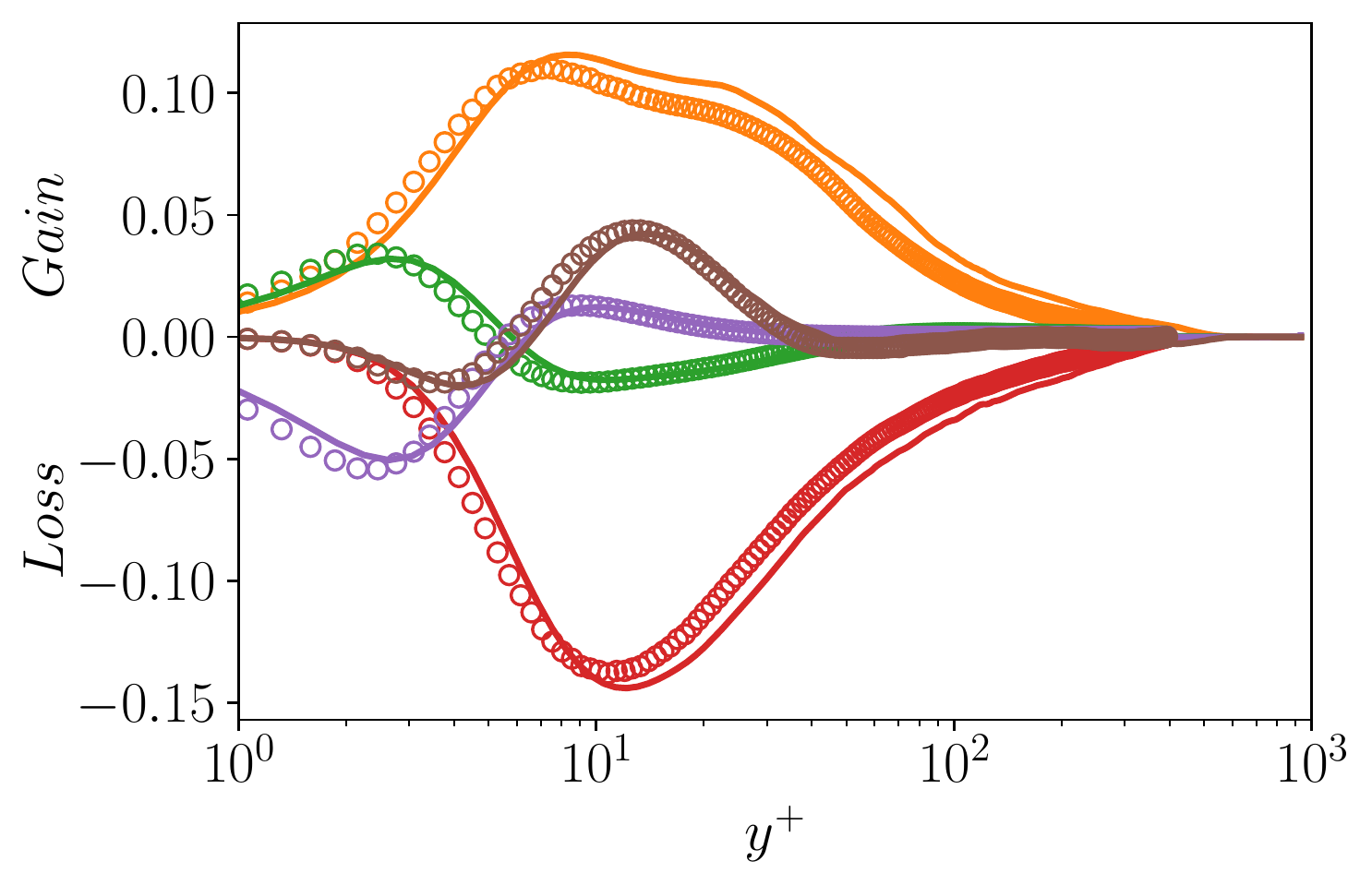}}
\caption[]{Comparison of the scalar-flux budgets. (Left)~$\left<u\theta^{\prime}\right>$, (right)~$\left<v\theta^{\prime}\right>$, {\plotblacksolid}~present DNS at $Pr = 6$, {\plotblackmarker}~\cite{kozuka} at $Pr = 7$, {\plotA}~convection, {\plotB}~scalar-pressure diffusion, {\plotC}~dissipation, {\plotD}~production, {\plotE}~molecular diffusion, {\plotF}~turbulent diffusion.}
\label{fig_s4_budget}
\end{figure}

The Reynolds stress equation is written (in index notation) as:
\begin{align}
    \frac{\Bar{D} \left<u_i u_j\right>}{\Bar{D} t} = \mathcal{P}_{ij} - \epsilon_{ij} + C_{ij} + D_{ij} + T_{ij}\,,
\end{align}
where $\Bar{D}$ represents the material derivative and $\left<u_i u_j\right>$ is the Reynolds stress tensor. Here, $\mathcal{P}_{ij}$ denotes the production term, $\epsilon_{ij}$ is the viscous dissipation rate tensor, $C_{ij}$ is the velocity pressure-gradient term (which can be split into pressure strain term and pressure diffusion term), $T_{ij}$ is the turbulent diffusion and $D_{ij}$ is the molecular diffusion term. The corresponding terms  are written respectively as:
\begin{align}
    \mathcal{P}_{ij} &:= -\left<u_i u_k\right> \frac{\partial \left<U_i\right>}{\partial x_k} - \left<u_j u_k\right>\frac{\partial \left<U_j\right>}{\partial x_k}\,,\\
    \epsilon_{ij} &:= 2\nu \left< \frac{\partial u_i}{\partial x_k}\frac{\partial u_j}{\partial x_k}\right>\,,\\
    C_{ij} &:= -\frac{1}{\rho}\left<u_i \frac{\partial p}{\partial x_j}+u_j\frac{\partial p}{\partial x_i}\right>\,,\\
    T_{ij}  &:= -\frac{\partial }{\partial x_k} \left<u_i u_j  u_k\right>\,,\\
    D_{ij} &:= \nu \frac{\partial^2 }{\partial x_j \partial x_j}\left<u_i  u_j\right>\,.
\end{align}
Detailed description of the above terms can be found in~\cite{pope} and they are non-dimensionalized by $u_{\tau}^4/\nu$.

The obtained Reynolds-stress budgets at $Re_\theta=1080$ (which is located almost at the end of the computational domain before fringe region) are compared with the data by~\cite{jiminez2010} in the turbulent boundary layer at $Re_\theta=1100$ as shown in figure~\ref{fig_vel_budget}. All the different components contributing to the stress terms are in good agreement with the reference data. Due to the lack of TBL data at higher Prandtl numbers, the budgets for the scalar at $Pr = 2$ are compared against the channel DNS data from~\cite{kozuka} as shown in figure~\ref{fig_s2_budget}. Overall, there is a good comparison obtained for the different terms in the scalar flux budgets. Also as discussed in $\S$\ref{sec_heat_flux}, we observe a higher production compared with the channel-flow case for the vertical-heat-flux budget in the overlap region. In addition, the scalar-pressure diffusion term also exhibits the same behaviour as discussed above and it should be noted that the discrepancy not only stems from the different problem setup but also the wall boundary condition for the scalar, which is Dirichlet in the present study and Neumann in the works of~\cite{kozuka}. Further, a comparison of the data obtained at $Pr = 6$ with the data obtained by~\cite{kozuka} at $Pr = 7$ for a channel flow is also provided in figure~\ref{fig_s4_budget}. Overall, the comparison of the stress budgets at different parameter points shows a good agreement with the data available in the literature. Note that the small discrepancy observed in figure~\ref{fig_s2_budget} are due to the different Prandtl numbers. 